\documentclass[preprint,12pt,authoryear]{elsarticle}

\usepackage{amssymb}
\usepackage{graphicx}       
\usepackage{subfigure}
\usepackage{amsthm}
\usepackage{amsmath}
\usepackage{lineno}
\usepackage{array}
\usepackage{booktabs}
\usepackage{multirow}
\usepackage{tabularx}
\usepackage{bm}
\usepackage{float}
\usepackage{color}
\allowdisplaybreaks[4]
\UseRawInputEncoding
\graphicspath{{figures/}}
\usepackage[dvipsnames]{xcolor}
\usepackage[colorlinks=true,
            linkcolor=Maroon,
            urlcolor=blue,
            citecolor=blue]{hyperref}

\journal{}

\begin{document}

\begin{frontmatter}

\title{{A multiple-time-step integration algorithm for particle-resolved simulation with physical collision time}}

\author[xidian]{Zhengping Zhu}
\author[lzu]{Ruifeng Hu}
\author[xidian]{Xiaojing Zheng}

\address[xidian]{Research Center for Applied Mechanics, School of Mechano-Electronic Engineering, Xidian University, Xi'an 710071, China}
\address[lzu]{Center for Particle-Laden Turbulence, Key Laboratory of Mechanics on Disaster and Environment in Western China, Ministry of Education, and Department of Mechanics, College of Civil Engineering and Mechanics, Lanzhou University, Lanzhou 731000, China}

\begin{abstract}
In this paper, we present a multiple-time-step integration algorithm (MTSA) for particle collisions in particle-resolved simulations. Since the time step required for resolving a collision process is much smaller than that for a fluid flow, the computational cost of the traditional soft-sphere model by reducing the time step is quite high in particle-resolved simulations. 
In one state-of-the-art methodology, collision time is stretched to several times the flow solver time step for the fluid to adapt to the sudden change in particle motion. However, the stretched collision time is not physical, the hydrodynamic force may be severely underestimated during a stretched collision, and the simulation of sediment transport may be sensitive to the stretched collision time.
The proposed MTSA adopts different time steps to resolve fluid flow, fluid-particle interaction, and particle collision. 
We assessed the MTSA for particle-wall collisions as well as particle-particle collisions, determined the optimal iteration number in the algorithm, and obtained excellent agreements with experimental measurements and reference simulations. The computational cost of the MTSA can be reduced to about one order of magnitude less than that using the traditional soft-sphere model with almost the same accuracy.
The MTSA was then implemented in a particle-resolved simulation of sediment transport with thousands of particles. {By comparing the results obtained using the MTSA and a version of the stretching collision time algorithm similar to \cite{costa2015collision}, we found that stretching the collision time reduced particle stiffness, weakened particle entrainment, and affected some turbulence and particle statistics.}  
\end{abstract}

\end{frontmatter}


\newpage

\section{Introduction}

Particle-laden two-phase flows are widely {encountered} in industrial engineering and natural environments \cite[]{balachandar2010turbulent,brandt2022particle}, such as aeolian sand and dust movement, sediment transport, and fluidized-bed processes. 
{When simulating particle-laden flows, if the flow field surrounding particles is fully resolved, it is called a particle-resolved simulation. A number of different particle-resolved simulation approaches have been developed, like the immersed boundary (IB) method \cite[]{,mittal2005immersed,uhlmann2005immersed,luo2007full,breugem2012second,kempe2012improved,zhou2014second,tao2018combined,huang2019recent,Wang2019hydrodynamic,Verzicco2022immersed} and the distributed Lagrange multiplier (DLM)/fictitious domain approach \cite[]{glowinski1999distributed,glowinski2001fictitious,yu2007direct,Yu2010direct,Shao2012fully}.
Particle-resolved simulations have been widely used to obtain high-fidelity simulation data, which is greatly helpful for developing reduced-order models \cite[]{bagchi2003effect,eaton2009two,homann2013effect,Tenneti2014particle,Zhou2015direct,Akiki2017pairwise,Seyed2020microstructure,Zhang2020modeling,wang2022direct,Xia2022improved,Xia2022drag} or studying the mechanism of the fluid-particle interactions \cite[]{uhlmann2008interface,kidanemariam2014interface,vreman2016particle,vowinckel2016entrainment,luo2017direct,peng2020flow,Costa2021near,Yu2021modulation}. 

The focus of the present paper is on the collision algorithm for particles in particle-resolved simulations. 
Particle collision modeling is essential in particle-resolved simulations, whether a particle-laden flow is in the dense regime \cite[]{kidanemariam2014interface,vowinckel2014fluid,picano2015turbulent} or the dilute regime \cite[]{costa2020interface}. 
Soft-sphere model \cite[]{Hertz1882beruhrung,mindlin1953elastic} is one of the most popular collision models in particle-resolved simulations when multi-particle contact frequently occurs.
However, the traditional soft-sphere model (\emph{i.e.}, the soft-sphere model with physical collision time) can be inefficient when applied to particle-resolved simulations, {as the time scale associated with particle collision is typically much smaller than that of fluid flow, particularly for rigid particles. This mismatch can lead to expensive computation if using the same time step to resolve particle collisions and advance fluid flow simultaneously.} 
On the other hand, the stretching collision time algorithm (SCTA) \cite[]{feng2010three,papista2011direct,kempe2012collision,costa2015collision,biegert2017collision,rettinger2022efficient} was proposed as a remedy and has been shown to be successful in reproducing the particle trajectories of particle-wall \cite[]{joseph2001particle,gondret2002bouncing,joseph2004oblique} and particle-particle \cite[]{yang2006dynamics} collisions.
In the SCTA, the collision time $T_c$ was stretched to $N$ times the flow solver time step $\Delta t_f$ (\emph{i.e.}, $T_c=N\Delta t_f$) since the fluid requires time to adapt to the sudden change in particle velocity during collisions. By stretching the collision time, the computational efficiency can be improved compared with the traditional soft-sphere model. The particle stiffness and damping coefficient are determined by the desired restitution coefficient $e_{n,d}$ and collision time $T_c$. However, the SCTA can only retain the particle kinematic properties rather than the mechanical properties as it greatly reduces the physical stiffness. 
Most of the SCTA is based on a linear spring-dashpot system, and the stiffness and damping coefficients were determined by an algebraic expression \cite[]{izard2014simulation,kidanemariam2014interface,costa2015collision,ardekani2016numerical}. 
Other implementations are based on the Hertzian contact theory, which is nonlinear, and the stiffness and damping coefficients were determined by an optimization procedure \cite[]{kempe2012collision,ray2015efficient,biegert2017collision}. 
\cite{rettinger2022efficient} developed an efficient four-way coupled lattice Boltzmann-discrete element method using the stretching collision time algorithm. They found that the collision time $T_c$ is a critical parameter of the method, which has a significant effect on the collision dynamics. The collision time should be stretched as $T_c=2.31D_p/c_s$ to allow the fluid to adapt to the sudden change in particle velocity, where $D_p$ is particle diameter and $c_s$ is the lattice speed of sound.
Recently, different from the SCTA developed based on the soft-sphere model, \cite{jain2019collision} proposed an algorithm based on the hard-sphere model, and the hydrodynamic forces are incorporated during the collision using a semi-implicit IB method \cite[]{tschisgale2018general}.


Although the SCTA can significantly reduce the computational cost compared with the traditional soft-sphere model and reproduce the experimental results of particle-particle and particle-wall collisions successfully \cite[]{joseph2001particle,gondret2002bouncing,joseph2004oblique,yang2006dynamics}, it has several significant drawbacks. 
First, particle stiffness in the SCTA is greatly reduced by stretching the collision time, which is not physical. 
Second, the artificially stretched collision time $T_c$ should be large enough to get converged trajectory \cite[]{rettinger2022efficient}. However, the hydrodynamic force during such an artificially long collision process is usually neglected and significantly underestimated, which may lead to erroneous predictions of turbulence statistics \cite[]{xia2020effects} and can be expected to also affect particle statistics in massive-particle-laden flows, like sediment transport.
To avoid introducing a stretched collision time, \cite{li2020particle} {were} the first to use the physical collision time $T_c$ in particle-resolved simulations. They adopted single non-uniform grid points with $ O \left ( 10^5 \right ) $ grid elements to resolve the particle and lubrication force, yielding excellent agreement with experimental results. However, such an extra-high grid resolution is almost impossible in cases with a large number of particles.

The present study proposed a multiple-time-step integration algorithm (MTSA) to reproduce the experimental results \cite[]{joseph2001particle,gondret2002bouncing,joseph2004oblique,yang2006dynamics} and simulations \cite[]{costa2015collision} using the soft-sphere model with physical collision time on a uniformly coarse grid. The grid resolution is about $D_p/\Delta x=O(10)$, where $\Delta x$ is the grid spacing. To overcome the time scale mismatch problem, three different time steps are employed in the MTSA to resolve fluid flow, fluid-particle interaction and particle motion during particle collisions. The pressure Poisson equation is only solved at the fluid flow time step to reduce computational cost. When time is advanced to the same instant, the fluid and particle information will be updated and synchronized. The MTSA can reduce the computational cost of the traditional soft-sphere model by nearly one order of magnitude and avoids artificially reducing particle stiffness like the SCTA. Meanwhile, we validated the accuracy and convergence of the MTSA with experimental data \cite[]{joseph2001particle,gondret2002bouncing,joseph2004oblique,yang2006dynamics}, numerical data \cite[]{costa2015collision}, and a reference simulation using the traditional soft-sphere model with fine time steps. We also presented guidelines for determining the optimal values of the parameters in the MTSA. The MTSA was then applied to a sediment transport case with thousands of finite-size particles. The results of the MTSA were compared for the first time with those obtained by the SCTA to investigate the effects of particle stiffness in a particle-resolved simulation of sediment transport.

The structure of the paper is arranged as follows. The {governing} equations and physical models for fluid flow and particle motion are given in \S 2. In \S 3, the MTSA is introduced. In \S 4, the influence of the time substep size is systematically analyzed and discussed in particle-wall and particle-particle collisions to yield a universally optimal time substep size. The MTSA is further applied to a sediment transport case with thousands of finite-size particles, {and the effect of particle stiffness is investigated.} Finally, we present the main conclusions of the paper in \S 5.

\section{Models}
\subsection{Governing equations}

The particle-laden flow considered here is governed by the Navier-Stokes equations for the carrier phase and the Newton-Euler equations for the dispersed particulate phase. The motion of an incompressible, Newtonian fluid flow is governed by the following continuity and momentum equations:
\begin{equation}
\nabla\cdot \bm u=0, 
\end{equation}
\begin{equation}
\frac{\partial \bm u}{\partial t}= -\nabla\cdot\left(\bm {uu}\right)-\frac{1}{\rho_f}\nabla p+\nu_f\nabla^2 \bm u+\bm f,
\label{eqn:eq2}
\end{equation}
where $\bm u$ is the fluid velocity, $p$ is the pressure, $\bm f$ is the volume force imposed to take into account the fluid-particle interaction or the IB force, $\rho_f$ is the fluid density, and $\nu_f$ is the fluid kinematic viscosity.

The translational and angular velocities of a particle are advanced by solving the Newton-Euler equations, which for a spherical particle are reduced to
\begin{equation}
\rho_p V_p\frac{{\rm d} \bm u_p}{{\rm d} t}={\oint_{\partial V}{\bm \tau\cdot \bm n_p d A}}+\left(\rho_p-\rho_f\right)V_p\bm g+{\bm F_{p,lub}+\bm F_{p,col}},
\label{eqn:eq3}
\end{equation}
\begin{equation}
I_p\frac{{\rm d}\bm \omega_p}{{\rm d} t}={\oint_{\partial V}{\bm r\times(\bm \tau\cdot \bm n_p)d A}}+{\bm T_{p,col}},
\label{eqn:eq4}
\end{equation}
where $\rho_p$ is the particle density, $V_p$ is the volume of the particle and equal to $(4/3)\pi R_p^3$ for a sphere with radius $R_p$; $I_p$ is the moment of inertia of the particle and equal to $(2/5)\rho_p V_p R_p^2$ for a spherical particle; $\bm u_p$ and $\bm \omega_p$ are the translational and angular particle velocities, respectively; $\bm \tau=-p \bm I+\mu_f (\nabla \bm u+\nabla \bm u^T )$ is the hydrodynamic stress tensor (the superscript $T$ indicates the transposition of a tensor); $\mu_f$ is the fluid dynamic viscosity; $\bm n_p$ is the outward-pointing unit normal vector at the surface $\partial V$ of the particle; $\bm r$ is the relative position vector of the particle surface to the particle center; $\bm g$ is the gravitational acceleration; {$\bm F_{p,lub}$, $\bm F_{p,col}$, and $\bm T_{p,col}$ are the total lubrication force, total collision force, and total collision torque acting on the particle $p$, respectively.} The subscript $p$ refers to the quantities of particle $p$. {The carrier and dispersed phases are coupled by the volume force {$\bm f$}, which is calculated by the multidirect forcing IB method \cite[]{luo2007full,breugem2012second}. The governing equations of the carrier phase are discretized on a uniform staggered Cartesian grid, which is referred to as the Eulerian grid. A particle is discretized by a fixed grid attached to the particle surface, which is referred to as the Lagrangian grid. The quantities between the Eulerian and Lagrangian grid points are transferred through a regularized Dirac delta function $\delta_d$ \cite[]{roma1999adaptive}}.

To avoid the difficulty of directly calculating the surface integrals $\oint_{\partial V}{\bm \tau\cdot \bm n_p d A}$ and $\oint_{\partial V}{\bm r\times(\bm \tau\cdot \bm n_p)d A}$, these surface integrals are converted into volume integrals by means of a momentum balance over the volume occupied by a particle. The volume force $\bm f$ is transferred from the Eulerian grid to the Lagrangian grid. Based on the above operations, the equations of particle motion  (\ref{eqn:eq3}) and (\ref{eqn:eq4}) can be rewritten as \cite[]{uhlmann2005immersed,kempe2012improved,breugem2012second}
\begin{equation}
\rho_p V_p\frac{{\rm d} \bm u_p}{{\rm d} t}=\rho_f\frac{{\rm d}}{{\rm d}t} \left ( \int_{V_p}{\bm udV} \right )-\rho_f\sum_{l=1}^{N_l}\bm F_{p,l} \Delta V_l+\left(\rho_p-\rho_f\right)V_p\bm g+{\bm F_{p,lub}+\bm F_{p,col}},
\end{equation}
\begin{equation}
\bm I_p\frac{{\rm d}\bm \omega_p}{{\rm d} t}=\rho_f\frac{{\rm d}}{{\rm d}t} \left ( \int_{V_p}{\bm r\times \bm udV} \right )-\rho_f\sum_{l=1}^{N_l}\left (\bm r_l\times \bm F_{p,l}  \right ) \Delta V_l+{\bm T_{p,col}},
\end{equation}
{where $N_l$ is the number of Lagrangian grid points, $F_{p,l}$ is the volume force on the Lagrangian grid point calculated by equation (\ref{eqn:f3}) of the IB method, and $\Delta V_l=\pi\Delta x\left ( 12R_p^2+\Delta x^2 \right ) /(3N_l)$ is the volume of a Lagrangian grid cell \cite[]{uhlmann2005immersed}.} The discretization schemes for these governing equations are given in Appendix B with details. According to the approximation of the hydrodynamic force \cite[]{kempe2012collision,costa2015collision,biegert2017collision}, the hydrodynamic force is excluded {during contact} for particles with the Stokes number $St=\rho_pu_{in}D_p/(9\rho_f\nu_f) \geqslant  5$, where $u_{in}$ is the particle impact velocity.

\subsection{Lubrication force}

When particles are close to colliding in particle-resolved simulations, the evaluation of the forces acting on a particle can be performed as follows \cite[]{Breugem2010combined,kempe2012collision,costa2015collision,biegert2017collision}: (a) when $\varepsilon \geqslant \varepsilon_{\Delta x}$, the IB method or a similar approach can be directly used to calculate the hydrodynamic force on the particle, where $\varepsilon=\delta_{n}/R_{p}$ is the non-dimensional gap width, $\delta_{n}$ is the distance between two surfaces in a collision, $R_p$ is particle radius, and $\varepsilon_{\Delta x}$ is a predefined constant. (b) When {$0 \leq \varepsilon<\varepsilon_{\Delta x}$}, the IB method or a similar approach may underpredict the hydrodynamic force due to a lack of spatial grid resolution. As a remedy, a lubrication model based on asymptotic expansions of analytical solutions for the lubrication force in the Stokes regime can be employed \cite[]{brenner1961slow,cox1967slow,cooley1969slow}. (c) When $\varepsilon < 0$, the collision model takes control of the particle, and the hydrodynamic force is neglected. It should be noted that neglecting the hydrodynamic force during a collision is a model assumption.

When a particle approaches another particle or a wall with a finite relative velocity, fluid is squeezed out of the gap and the lubrication model can be employed to calculate the hydrodynamic force with the Stokes flow assumption \cite[]{brenner1961slow,cox1967slow,cooley1969slow}.
{The total lubrication force $\bm F_{p,lub}$ acting on the particle $p$ can be computed as follows:}
\begin{equation}
{
\bm F_{p,lub}= \sum_{p,q \neq p}^{N_{p} } \bm F_{n,pq}^{lub}+\bm F_{n,pw}^{lub},} 
\end{equation}
{where $\bm F_n^{lub}$ is the normal lubrication force for a collision pair, $N_{p}$ is the total particle number, the subscript $n$ indicates the normal direction, and the subscripts $pq$ and $pw$ indicate the contact of particle $p$ with particle $q$ and the wall, respectively. Here, $\bm F_n^{lub}$ is calculated by the following the two-parameter lubrication model proposed by \cite{Breugem2010combined} and \cite{costa2015collision}:} 
\begin{gather}
\bm F_n^{lub}=\begin{cases} -6\pi \mu_f R_{p} \bm u_{cp,n}\left ( \lambda (\varepsilon ) - \lambda (\varepsilon_{\Delta x} ) \right ), \quad \varepsilon_{\sigma }\leqslant \varepsilon<\varepsilon_{\Delta x}   
\\ -6\pi \mu_f R_{p} \bm u_{cp,n}\left ( \lambda (\varepsilon_{\sigma } ) - \lambda (\varepsilon_{\Delta x} ) \right ), \quad 0\leqslant\varepsilon<\varepsilon_{\sigma} 
\\ 0, \quad\quad\quad\quad\quad\quad\quad\quad\quad\quad\quad\quad\quad\ \  \text{otherwise}
\end{cases}, \\
\lambda _{pq}(\varepsilon ) =\frac{1}{2\varepsilon} -\frac{9}{20} \ln{\varepsilon} - \frac{3}{56} \varepsilon \ln{\varepsilon}+O(1),\\
\lambda _{pw}(\varepsilon ) =\frac{1}{\varepsilon} -\frac{1}{5} \ln{\varepsilon} - \frac{1}{21} \varepsilon \ln{\varepsilon}+O(1),
\end{gather}
where $\bm u_{cp,n}$ is the normal component of the relative velocity, as defined in Appendix A; {$\lambda_{pq}$ and $\lambda_{pw}$ are the Stokes amplification factors for the lubrication force in the interactions between particle $p$ and particle $q$ and between particle $p$ and the wall, respectively}; $\varepsilon_{\Delta x}$ is the non-dimensional gap spacing related to the grid size (\emph{i.e.}, $D_p/\Delta x$); $\varepsilon_{\sigma }$ is a fixed minimum gap spacing related to asperities. Following \cite{costa2015collision}, we adopted $\varepsilon_{\Delta x,pw}=0.075$, $\varepsilon_{\Delta x,pq}=0.025$ for $D_p/\Delta x=20$ and $\varepsilon_{\Delta x,pw}=0.05$, $\varepsilon_{\Delta x,pq}=0.025$ for $D_p/\Delta x=30$.  In addition, $\varepsilon_{\sigma,pw}=0.0008$ and $\varepsilon_{\sigma,pp}=0.0001$ were calibrated by matching the trajectories of the particle--wall and particle-particle collisions with the experimental data of \cite{gondret2002bouncing} in the present study. These trajectories are shown in Section 4.1.

\subsection{Collision force}
{The total collision force $\bm F_{p,col}$ and total torque $\bm T_{p,col}$ acting on particle $p$ are computed as follows:}

\begin{equation}
{
\bm F_{p,col}= \sum_{p,q \neq p}^{N_{p} } \left ( \bm F_{n,pq}^{col}+\bm F_{t,pq}^{col}   \right )+\bm F_{n,pw}^{col}+\bm F_{t,pw}^{col},} 
\end{equation}
\begin{equation}
{
\bm T_{p,col}= \sum_{p,q \neq p}^{N_{p} }  R_{p} \bm n_{pq} \times \bm F_{t,pq}^{col}+R_{p} \bm n_{pw} \times \bm F_{t,pw}^{col},} 
\end{equation}
where $\bm F^{col}$ is the collision force for a collision pair, $\bm n$ is the normal unit vector of the contact, and the subscripts $n$ and $t$ indicate the normal and tangential directions, respectively. The collision force $\bm F^{col}$ is calculated by the soft-sphere model. The model is composed of a linear spring--dashpot system in the normal direction, and a linear spring--dashpot system with a Coulomb friction slider in the tangential direction \cite[]{costa2015collision}. The normal collision force $\bm F_{n}^{col}$ and the tangential collision force $\bm F_{t}^{col}$ are determined by
\begin{gather}
\bm F_{n}^{col} =-k_{n}\left | \delta _{n} \right |\bm n-d_{n}\bm u_{cp,n},\\
\bm F_{t}^{col} =-k_{t}\bm \delta _{t}-d_{t}\bm u_{cp,t},     
\end{gather}
where $k_n$ and $k_t$ are the stiffnesses in the normal and tangential directions, respectively; $d_n$ and $d_t$ are the damping coefficients in the normal and tangential directions, respectively; {$\delta_n$ is the distance between two surfaces}; $\bm \delta_{t}$ and $\bm u_{cp,t}$ are the tangential displacement of the spring and the tangential component of the relative velocity, respectively, as defined in Appendix A. 

{The stiffness and damping coefficients can be derived by the collision time and the restitution coefficient \cite[]{costa2015collision}, as
\begin{equation}
k_n=\frac{m_{e,n}\left ( \pi^2+\ln^2e_{n,d}   \right ) }{T_c} ,\quad d_n=-\frac{2m_{e,n}\ln e_{n,d}}{T_c},
\label{eqn:eq33}
\end{equation}
\begin{equation}
k_t=\frac{m_{e,t}\left ( \pi^2+\ln^2e_{t,d}   \right ) }{T_c} ,\quad d_t=-\frac{2m_{e,t}\ln e_{t,d}}{T_c} ,
\end{equation}
where $m_{e,n}=(m_p^{-1}+m_q^{-1})^{-1}$ and $m_{e,t}=(1+1/K^2)^{-1}m_{e,n}$ are the reduced masses of the linear spring--dashpot system; $m_p$ and $m_q$ are the masses of particles $p$ and $q$, respectively; $K$ is the normalized particle radius of gyration with $K^2=2/5$ for a homogeneous spherical particle; $e_{n,d}$ and $e_{t,d}$ are the dry coefficients of restitution in the normal and tangential directions, respectively.}

The stretching collision time model determines the stiffness and damping coefficients through a nonphysical stretched collision time $T_c$, significantly reducing the stiffness and damping coefficients. 
To avoid introducing the stretched collision time, the physical collision time should be used \cite[]{li2020particle}. The physical collision time $T_c$ can be calculated by the physical properties of  particle as \cite[]{zenit1997collisional}
\begin{equation}
T_c=7.894\left ( \frac{m_p^2}{E^{*2}u_{in}D_p}  \right )^{1/5},
\label{eqn:tc}
\end{equation}
where $E^*$ is the equivalent elasticity of the collision system depending on Young's modulus $E$ and Poisson's ratio $\nu$, which is expressed as follows:
\begin{equation}
\frac{1}{E^{*}}=\frac{1}{\pi }  \left ( \frac{1+\nu_p^2 }{E_p}+ \frac{1+\nu_q^2 }{E_q} \right ).
\end{equation}
In the present study, we use $E=200$ GPa and $\nu=0.3$ for steel particles and $E=55$ GPa and $\nu=0.25$ for glass particles. {The ratio between $T_c$ and the particle response time $\tau_p=\rho_pD_p^2/18\mu_f$ is $O(10^{-4} \sim 10^{-3})$ for $10$-mm steel/glass particles with $St=100$ impacting in air or water, respectively.} 

When $\left | \bm F_t^{col} \right |  \geqslant \mu_c \left | \bm F_n^{col} \right |$ ($\mu_c$ is the friction coefficient), a particle starts to slide. The tangential force is controlled by the slider in the tangential direction based on the Coulomb friction law instead of the tangential displacement, as
\begin{equation}
\bm F_t^{col}= \mu_c | \bm F_n^{col} | \bm t.
\end{equation}
where $\bm t$ is the unit vector in the tangential direction.
The tangential displacement needs to be reset in order to comply with Coulomb's friction law, which is
\begin{equation}
\bm \delta _{t}=-\frac{\mu_c | \bm F_n^{col} | \bm t +d_t \bm u_{cp,t}  }{k_t}. 
\end{equation}

\section{The multiple-time-step integration algorithm (MTSA)} 

\subsection{The motivation}
The mismatch problem of the characteristic time scales between particle collision and fluid flow is crucial for particle-resolved simulations. 
In fact, the characteristic time scales for fluid flow, fluid-particle interaction and particle collision are quite different in magnitude.
We have estimated that the particle collision time can be 3 or 4 orders of magnitude smaller than the particle response time, and thus 1 or 2 orders of magnitude smaller than the flow time scale with $St \sim O(100)$.
Using a very fine fluid flow time step $\Delta t_f$ to resolve particle collision can lead to a highly expensive computation.
When employing a single large time step in a particle-resolved simulation, the error of the soft-sphere model is also large and mainly comes from two aspects.
First, the particle motion cannot be well resolved by a large flow time step during a collision. It will result in an excessive overlap between colliding surfaces and overpredicting the rebounding trajectory, as shown in figure 1 of \cite{kempe2012collision}. Second, particle velocity rapidly changes during a collision. The surrounding fluid can hardly follow this rapid change with a large flow time step. It may cause an overestimation of the hydrodynamic force by the IB method and underpredicts the rebounding trajectory, as shown in figure 8 of \cite{costa2015collision}. These two facts prevent the usage of the traditional soft-sphere model with a single large time step.
On the other hand, as we have discussed in the introduction, artificially reducing particle stiffness like the SCTA is nonphysical and may also bring in errors in simulations \cite[]{xia2020effects}. 

Due to the aforementioned drawbacks of the existing methods, we propose the MTSA employing a normal fluid flow time step while inserting additional two-level time substeps to resolve fluid-particle interaction and particle motion during a collision, so that the prediction errors by increasing particle stiffness can be significantly reduced and the computation efficiency is not degraded too much at the same time.

\begin{figure}[ht]
\centering
\hspace{-50pt}
\includegraphics[width=\textwidth]{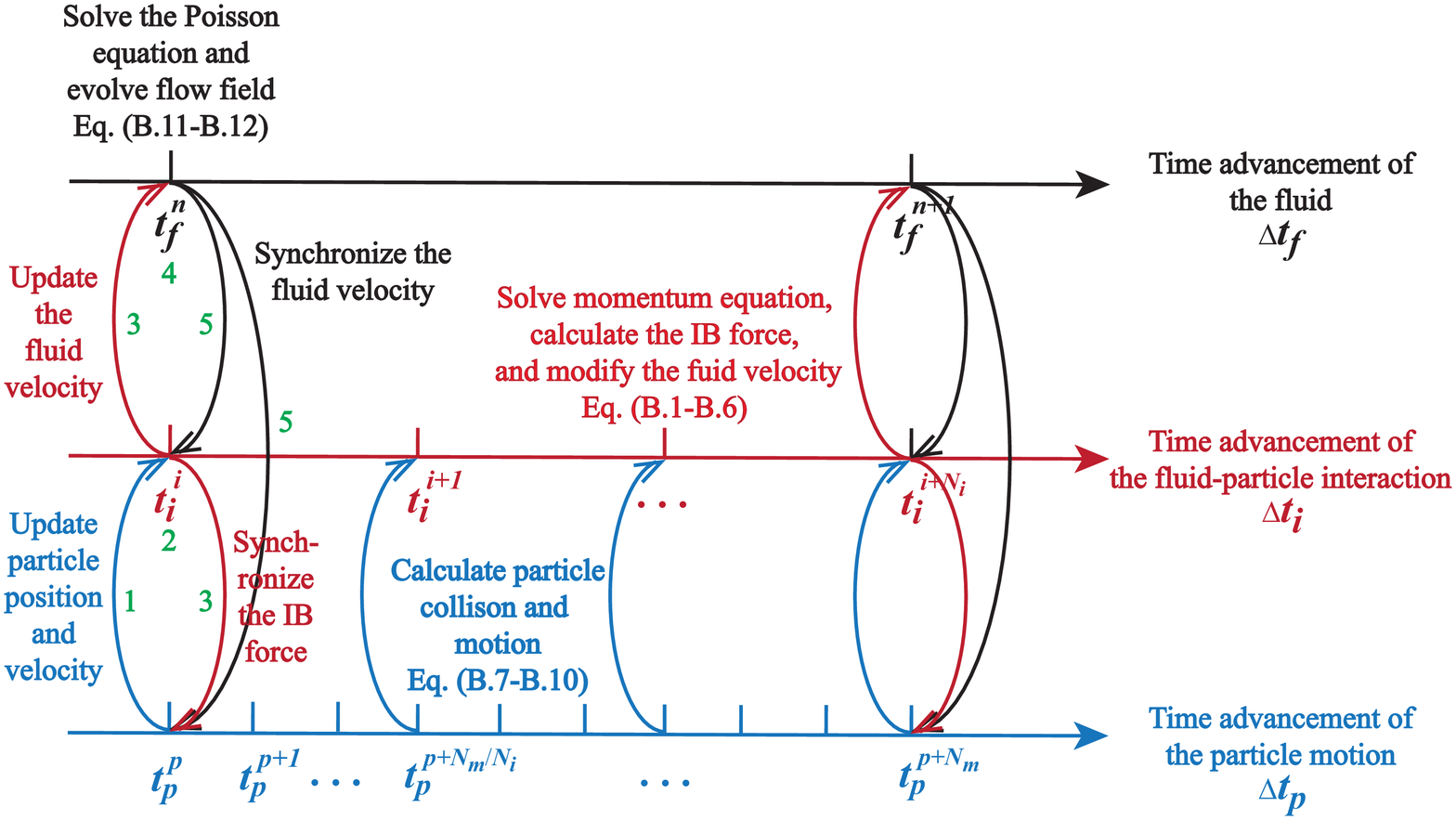}
\caption{Sketch of the multiple-time-step integration algorithm (MTSA).}
\label{fig:flowchart}
\end{figure}

\subsection{The computation procedure}

The MTSA was activated once particle-particle/wall collision happens. A single time step  $\Delta t$ is replaced by three different time steps, namely a fluid flow time step $\Delta t_f$ and additional two-level time substeps (\emph{i.e.}, fluid-particle interaction time substep $\Delta t_i$ and particle motion time substep $\Delta t_p$) hierarchically inserted into it. As shown in figure~\ref{fig:flowchart}, $\Delta t_f=T_c/R_f$ is employed to advance fluid flow, $\Delta t_i=T_c/R_i$ to advance fluid-particle interaction, and $\Delta t_p=T_c/R_m$ to advance particle motion. Here, $R_f$, $R_i$ and $R_m$ are the iteration numbers of fluid flow, fluid-particle interaction and particle motion substeps in one collision time $T_c$, respectively. Based on the above three parameters, $N_i=R_i/R_f$ and $N_m=R_m/R_f$ can be defined as the iteration numbers of fluid-particle interaction and particle motion substeps in one fluid flow time step. It should be noted that $N_i$ and $N_m$ should be integers, which restrict the choices for $R_i$ and $R_m$.

The pressure Poisson equation is only solved at the fluid flow step $t=t_f$, which greatly improves the computational efficiency of the MTSA. Local pressure is assumed to be a constant at the fluid-particle interaction and particle motion substeps in a fluid flow step. At a fluid-particle interaction substep $t=t_i$, we solve the flow momentum equation, calculate the IB force, and modify the fluid velocity by the IB force accordingly. At a particle motion substep $t=t_p$, we calculate particle collision and motion. When the time is advanced to the same instant, the information of the fluid flow and particle will be updated and synchronized. The updating and synchronizing process are demonstrated by the arrow lines in figure~\ref{fig:flowchart}. Particularly, when $t=t_f=t_i=t_p$, the particle position and velocity at $t=t_p$ are updated to the particle-fluid interaction substep before calculating the IB force. Second, the IB force is calculated and used to modify the fluid velocity at $t=t_i$. Third, the calculated IB force at $t=t_i$ is synchronized to the fluid flow step as the input for particle motion, and the modified fluid velocity at $t=t_i$ is also updated as the input for the Poisson equation. Fourth, the Poisson equation and flow field at $t=t_f$ are solved and updated. Finally, the updated fluid velocity at $t=t_f$ is synchronized to $t=t_i$ and $t=t_p$ as the input for the flow momentum equation and particle motion. When $t \neq t_f$, we only update the particle position and velocity.

In the MTSA, an explicit second-order Runge--Kutta (RK2) method is used for the temporal advancement of the fluid flow \cite[]{yang2017numerical,yang2018direct,cui2018sharp,he2022numerical,zhu2022particle}. At each substep of the RK2 method, the fractional-step method of \cite{kim1985application} is applied to ensure that the flow field is divergence-free. In order to insert enough substeps to ensure that the flow can adapt to the change of particle motion during collisions and reduce the computational cost, the explicit Euler method is adopted in the particle-fluid interaction and particle-motion substeps for time advancement. The discretized equations are presented in Appendix B.



\subsection{The advantages}

In summary, the proposed MTSA has the following advantages:

(1) Compared with the traditional soft-sphere model, the MTSA significantly improves computational efficiency as it employs a large fluid flow time step $\Delta t_f$. The pressure Poisson equation is only solved at the fluid flow time step $t=t_f$, which greatly saves the computation cost. The computational time of the MTSA can be reduced to about one order of magnitude less than that of the traditional soft-sphere model with almost the same accuracy, which will be shown in \S 4.

(2) Compared with the SCTA, the MTSA employs physical particle stiffness without reducing particle stiffness. It is expected that the problem of hydrodynamic force underestimation will be alleviated or resolved by the MTSA, and the simulation of sediment transport will not be affected by nonphysically reduced particle stiffness. In addition, the parameters $R_i$ in the MTSA has very good universality for different fluid time steps, which will also be shown in \S 4.


In addition to the aforementioned algorithms proposed in particle-resolved simulations, some numerical algorithms have also been proposed in point particle simulations to address the mismatch of the time scales between particle collision and the fluid flow \cite[]{deen2007review,capecelatro2013euler,finn2016particle}.
They generally employed two-time scales, where a large time step was adopted to advance fluid flow and the inter-phase coupling, while a small time step was employed for particle motion. Some researchers even used a smaller time scale to calculate particle collisions \cite[]{darmana2006parallelization,marshall2009discrete}. However, as will be shown in \S 4, these algorithms are ineffective in particle-resolved simulations, since fluid-particle interaction is not resolved during a collision. 

\section{Results}
In this section, the accuracy and convergence of the traditional soft-sphere model and the MTSA are assessed through simulations of particle-wall and particle-particle collisions. Furthermore, we discuss the impact of $R_f$ and $R_i$ on the prediction results using the MTSA. Finally, a sediment transport case with thousands of finite-size particles is simulated as a practical application of the MTSA and a demonstration of the difference between the MTSA and the SCTA. \textcolor{blue}{It should be noted that a version of SCTA similar to Costa et al. (2015) is adopted here to be compared with the MTSA, which incorporates a two-parameter lubrication model, a linear spring-dashpot system, and substeps for particle collisions and motions.}

Because the hydrodynamic force is neglected in a particle collision, which is totally governed by the spring-dashpot system, $R_m$ is not influenced by the grid size and $St$. We directly determine $R_m$ by normal dry particle-wall and particle-particle collision cases. The simulation results of the restitution coefficient $e_n$ were compared with a given $e_{n,d}=0.97$ for steel and glass particles. For both $R_m=40$ and $80$, it is found that the relative error between $e_n$ and $e_{n,d}$ is less than $0.1\%$, which is sufficiently small. Therefore, the impact of $R_m$ is not studied in this section. And $R_m=40$ is used in the following cases, and $R_m=80$ is only used when $R_i=16$.

\subsection{Accuracy and convergence of the traditional soft-sphere model}

{Cases REF, S1 and S2 in table~\ref{table:cases} are designed to verify the accuracy and convergence of the traditional soft-sphere model as the benchmark result. These three cases can be regarded as using the MTSA without the fluid-particle interaction substep (\emph{i.e.}, $N_i=R_i/R_f=1$), which is similar to the algorithm proposed for the point-particle simulations \cite[]{deen2007review}.} Therefore, there are only two-time steps in cases REF, S1 and S2. 
Particularly, the Navier-Stokes and fluid-particle interaction equations are solved with the flow time step $\Delta t_f$, while the Newton-Euler equations are advanced with the particle motion time step $\Delta t_p$. {Case REF is regarded as the benchmark case. In case S1, $\Delta t_f$ was smaller than that in case REF, which is used to verify the accuracy and convergence of case REF. In case S2, $\Delta t_f$ is larger than that in case REF, which is used to compare with the results of the MTSA with the same $\Delta t_f$ to demonstrate the improvement by inserting additional fluid-particle interaction substeps.}

\begin{table}[H]
{
   \centering
   \caption{The MTSA parameters $R_f$, $R_i$, and $R_m$ used in different cases of particle-wall and particle-particle collisions. $N_i=R_i/R_f=1$ indicates that the traditional soft-sphere model is used without the fluid-particle interaction (FPI) substep.}
   \setlength{\tabcolsep}{1.5mm}{
   \begin{tabular}{ccccc}
   \toprule
   Case & $R_f$ & $R_i$ & $R_m$ & Notes\\
   \midrule
   REF & 8   & 8   & 40 & Benchmark case \\
   S1  & 16  & 16  & 80 & Reduce $\Delta t_f$, without FPI substep \\
   S2  & 2   & 2   & 40 & Increase $\Delta t_f$, without FPI substep \\
   MTSA  & 0.5,1,2 & 2,4,8,16 & 40,80 & With FPI substep\\ \bottomrule
   \end{tabular}}
   \label{table:cases}}
\end{table}

First, the bouncing motion of a single particle in a viscous fluid with different $St$ and $D_p/\Delta x$ values were simulated to assess the accuracy and convergence of the traditional soft-sphere model in particle-wall collisions. 
The physical parameters in the simulation were comparable to the experiment of \cite{gondret2002bouncing}: $\rho_p=7800$ kg/m$^3$ and $e_{n,d}=0.97$, and other parameters are listed in table~\ref{table:pw}.
The computational domain was $12.8D_p \times 25.6D_p \times 12.8D_p$. Two grid resolutions, $D_p/\Delta x=20$ and $D_p/\Delta x=30$, were implemented in this simulation. A periodic boundary condition was imposed in the horizontal directions, and a no-slip boundary condition was imposed on both the top and bottom surfaces. A particle was initially placed at $x=L_x/2$, $y=L_y-0.75D_p$, and $z=L_z/2$. The falling velocity was prescribed following \cite{biegert2017collision}, where it accelerated smoothly and $u_{in}$ matched the Stokes number in the previously reported experiment \cite[]{gondret2002bouncing} before the collision, as
\begin{equation}
   u_p(t)=u_{in}\left ( e^{-40t}-1  \right ),\quad \text{if} \quad \delta _{n}>R_{p}.
\label{eqn:uin}
\end{equation}
Once the particle reached a distance of $\delta_n=R_p$, we turned off the prescribed velocity. Then, the particle moved under hydrodynamic, gravitational, buoyant, and collision forces. The normal coefficient of restitution is defined as $e_n=u_{out}^*/u_{in}^*$. Following \cite{costa2015collision}, $u_{in}^*$ and $u_{out}^*$ are defined at the instants $t-t_c=\mp f_s^{-1}$, where $t_c$ is the instant of collision and $f_s$ is the sampling frequency in measurements, which is 500 Hz in the experiment of particle-wall collision \cite[]{joseph2001particle} and 100 Hz in the experiment of particle-particle collision \cite[]{yang2006dynamics}.

\begin{table}[htbp]
   \centering
   \caption{Parameters used in the simulation of particle-wall collisions.}
   \setlength{\tabcolsep}{3mm}{
   \begin{tabular}{cccccc}
   \toprule
   $St$ & $D_p$ (mm) & $\rho_f$ (kg/m$^{3}$) & $\mu_f$ (cP) &$u_{in}$ (m/s) & $T_c$ (s)\\
   \midrule
   6   & 3 & 965 & 100 & 0.231 & $2.3\times 10^{-5}$ \\
   15  & 6 & 965 & 100 & 0.288 & $4.4\times 10^{-5}$ \\
   27  & 6 & 965 & 100 & 0.519 & $3.9\times 10^{-5}$ \\
   60  & 3 & 953 & 20  & 0.462 & $2.0\times 10^{-5}$ \\
   100 & 4 & 953 & 20  & 0.577 & $2.6\times 10^{-5}$ \\
   152 & 3 & 935 & 10  & 0.585 & $1.9\times 10^{-5}$ \\
   193 & 6 & 953 & 20  & 0.742 & $3.6\times 10^{-5}$ \\
   742 & 5 & 920 & 5   & 0.856 & $3.0\times 10^{-5}$ \\ \bottomrule
   \end{tabular}}
   \label{table:pw}
\end{table}


{The resulting normal coefficients of restitution $e_n$ are compared with the experimental data of \cite{joseph2001particle} and \cite{gondret2002bouncing} here. The simulation results of the coefficient of restitution $e_n$ for normal particle-wall collisions by the traditional soft-sphere model are shown in figure~\ref{fig:pw_en_ssm}. The resulting $e_n$ of case REF agrees well with the experimental results over the entire range of the particle Stokes number. This verifies the accuracy of the traditional soft-sphere model. The resulting $e_n$ values are almost the same when refining the grid (\emph{i.e.}, $D_p/\Delta x=30$) at $St=15, 27, 152, 742$ or refining the flow time step $\Delta t_f$ (\emph{i.e.}, case S1) at $St=27, 152$. This verifies the convergence of the traditional soft-sphere model. However, $e_n$ is evidently underestimated with a much larger $\Delta t_f$ (case S2). This is because $\Delta t_f$ is too large to resolve the fluid-particle interaction so that the surrounding fluid cannot follow the rapid change of the particle velocity during the collision, and the velocity difference between the particle and fluid after the collision is increased. This leads to an overestimation of the hydrodynamic force by the IB method. The underestimated $e_n$ in case S2 can cause a remarkable discrepancy in the particle bouncing trajectories, as shown in figure~\ref{fig:pw_tra_ssm}. It is seen that the trajectory of case S1 is almost the same as that of case REF, because the values of $e_n$ are close in the two cases. The trajectory of the refined grid (\emph{i.e.}, $D_p/\Delta x=30$) is not shown here, as the corresponding $e_n$ and trajectory are very close to those of cases REF and S1.}

\begin{figure}[htbp]
\centering
\includegraphics[width=0.8\textwidth]{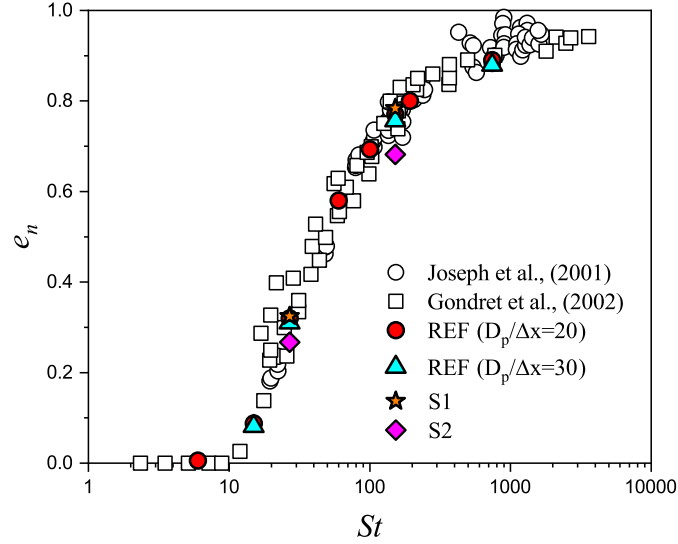}
\hspace{0in}
\caption{Wet coefficient of restitution for normal particle-wall collisions by the traditional soft-sphere model.}
\label{fig:pw_en_ssm}
\end{figure}

\begin{figure}[htbp]
\centering
\hspace{-16mm}
\includegraphics[width=8cm]{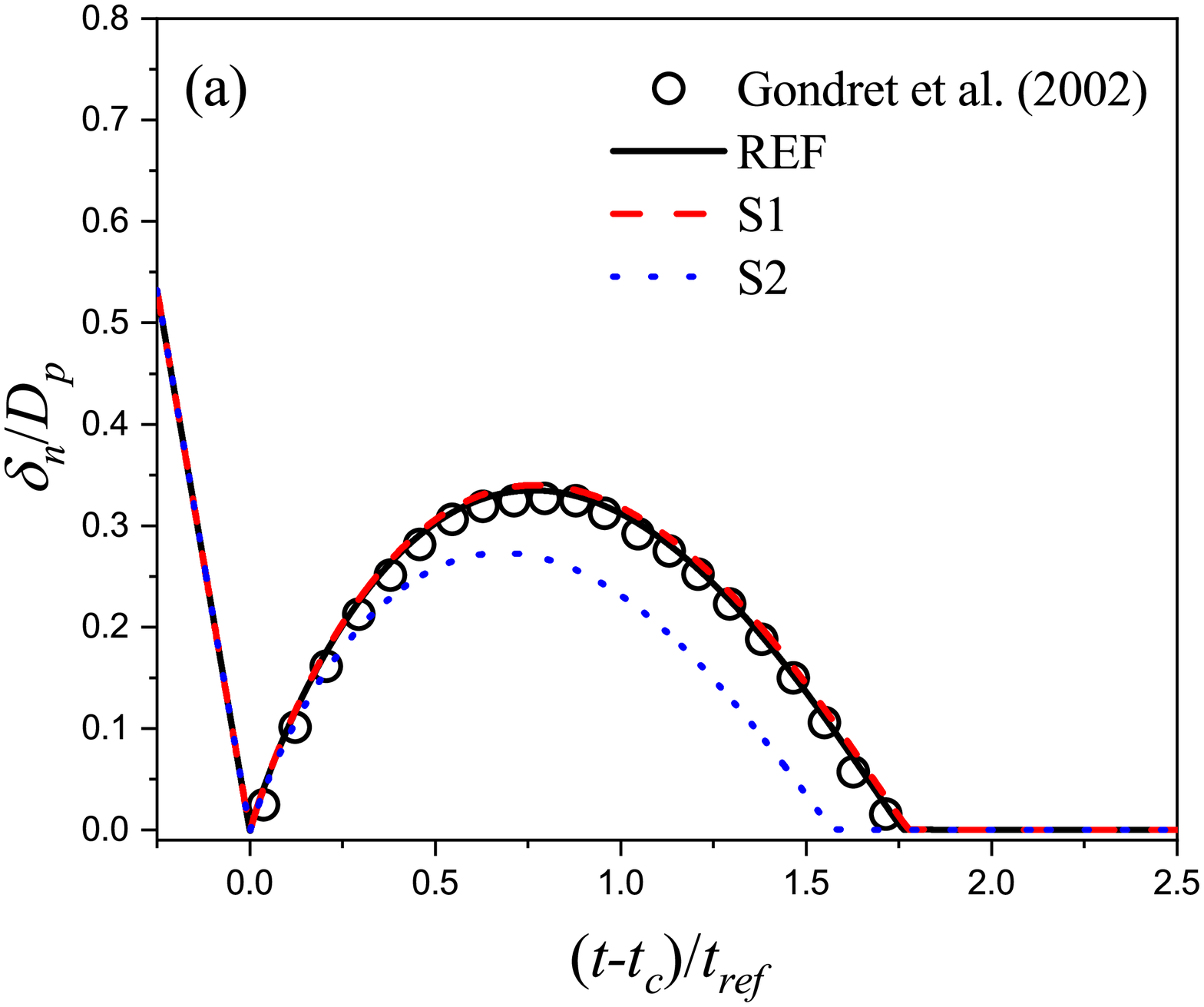}
\hspace{-10mm}
\includegraphics[width=8cm]{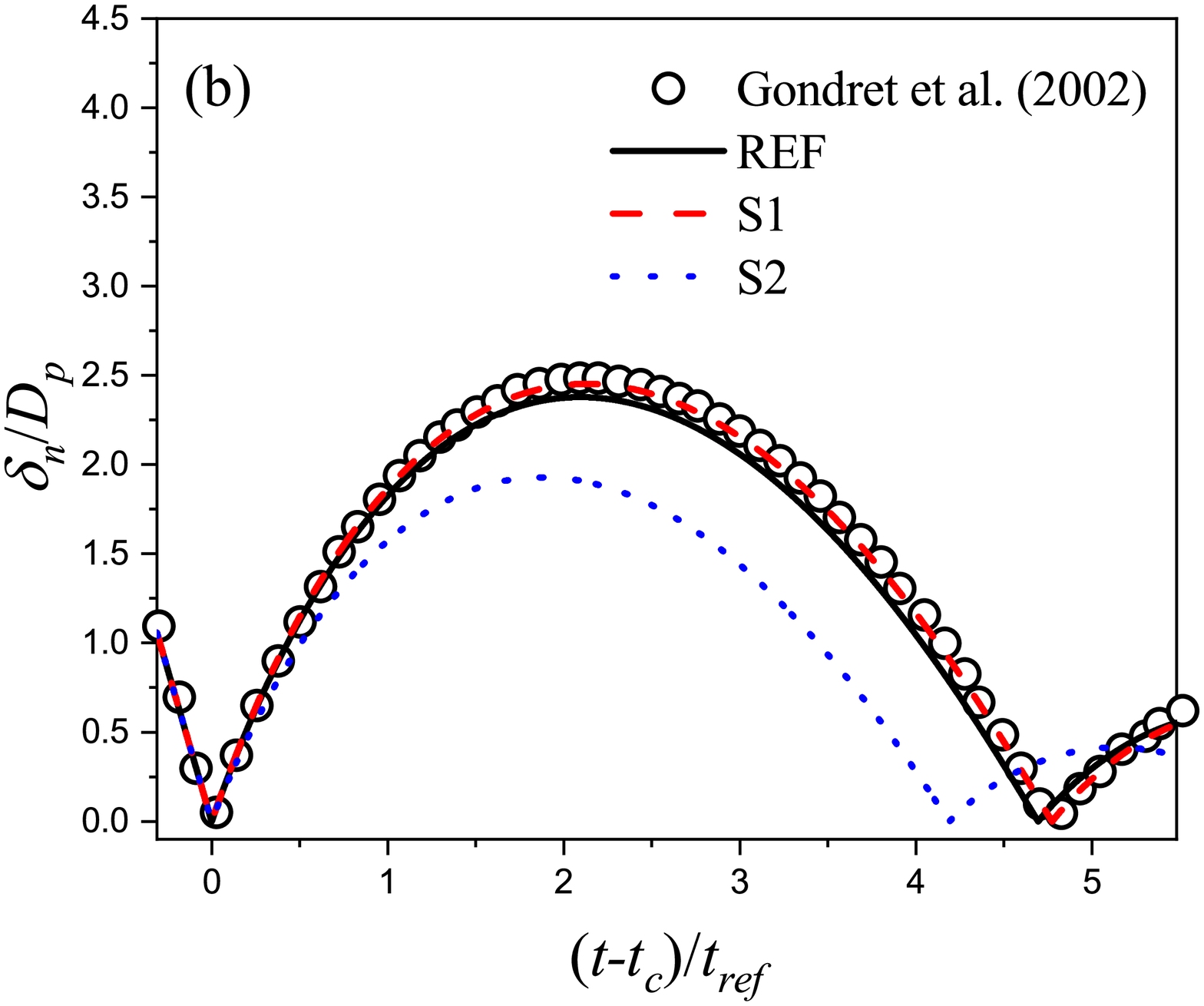}
\hspace{-16mm}
\caption{Comparison of the particle trajectories between the experiment and the simulation results by the traditional soft-sphere model with different Stokes numbers and time steps: (a) $St=27$ and (b) $St=152$. $t_c=t\mid_{\delta_n=0}$ is the instant of collision and $t_{ref}=\sqrt{D_p/\left | \bm g \right |}$ is the reference time scale.}
\label{fig:pw_tra_ssm}
\end{figure}

Next, a moving particle colliding with a steady particle in a viscous fluid with different $St$ and $D_p/\Delta x$ values was simulated to assess the accuracy and convergence of the traditional soft-sphere model in particle-particle collisions. 
The physical parameters in the simulations were comparable to the experiments of \cite{yang2006dynamics}: $D_p=12.7$ mm, $\rho_p=7780$ kg/m$^{3}$, $\rho_f=1125$ kg/m$^{3}$, $\mu_f=45$ cP and $e_{n,d}=0.97$, and other parameters are listed in table~\ref{table:pp}.
The computational domain was $6D_p \times 24D_p \times 6D_p$. Two grid resolutions of $D_p/\Delta x=20$ and $30$ were implemented in the simulations. The periodic boundary conditions were applied in all three directions. The steady and moving particles were initially placed at $y=19D_p$ and $2D_p$, respectively, centered at $x=L_x/2$ and $z=L_z/2$. The velocity of the moving particle was prescribed by the negative form of equation (\ref{eqn:uin}) since it was moving upward. The gravitational and buoyancy forces were not considered. 

\begin{table}[htbp]
   \centering
   \caption{Parameters used in the simulation of particle-particle collisions.}
   \setlength{\tabcolsep}{8mm}{
   \begin{tabular}{ccc}
   \toprule
   $St$ & $u_{in}$ (m/s$^{-1}$) & $T_c$ (s)\\
   \midrule
   12.7  & 0.059 & $1.3\times 10^{-4}$ \\
   21.5  & 0.099 & $1.2\times 10^{-4}$ \\
   34.3  & 0.158 & $1.1\times 10^{-4}$ \\
   52.7  & 0.243 & $9.6\times 10^{-5}$ \\
   135.2 & 0.623 & $8.0\times 10^{-5}$ \\
   345   & 1.591 & $6.6\times 10^{-5}$ \\ \bottomrule
   \end{tabular}}
   \label{table:pp}
\end{table}


In figure~\ref{fig:pp_en_ssm} (a), the resulting normal coefficients of restitution $e_n$ in the particle-particle collisions are compared with the experimental data of \cite{yang2006dynamics} and the numerical results of \cite{costa2015collision}. The conclusions for the particle-particle collisions are similar to those for particle-wall collisions. The resulting $e_n$ of case REF agrees well with the experimental and numerical results over the entire range of the Stokes number $St$. The resulting $e_n$ is almost unchanged by refining the grid ($D_p/\Delta x=30$) at $St=12, 34, 135, 345$ or reducing the flow time step $\Delta t_f$ (case S1) at $St=34$, which verifies the accuracy and convergence of the traditional soft-sphere model. Although the resulting $e_n$ by increasing $\Delta t_f$ (case S2) is within the range of the widely scattered experimental results, there still was a visible deviation compared with the results of case REF due to the insufficient temporal resolution for fluid-particle interactions. Figure~\ref{fig:pp_en_ssm} (b) further shows the contact point trajectories of particles at $St=34$. The trajectories in cases REF and S1 have good agreement with those of \cite{costa2015collision}. It is seen that the underestimated $e_n$ of case S2 causes a visible discrepancy in the trajectories of the contact point compared with case REF.

{The results of the particle-wall and particle-particle collisions indicate that the traditional soft sphere model can be used to reproduce experimental results only if the flow time step $\Delta t_f$ is sufficiently small, and increasing $\Delta t_f$ would result in an underestimation of $e_n$ and non-negligible errors in the particle trajectories. However, the computational cost of using such a small $\Delta t_f$ is quite high in particle-resolved simulations with a large number of particles. Hence, the MTSA can be a good method to reduce the computational cost of the traditional soft-sphere model while maintaining its accuracy. In the following, the results of case REF will be employed as a benchmark to test the accuracy of the MTSA.}

\begin{figure}[htbp]
\centering
\hspace{-16mm}
\includegraphics[width=8cm]{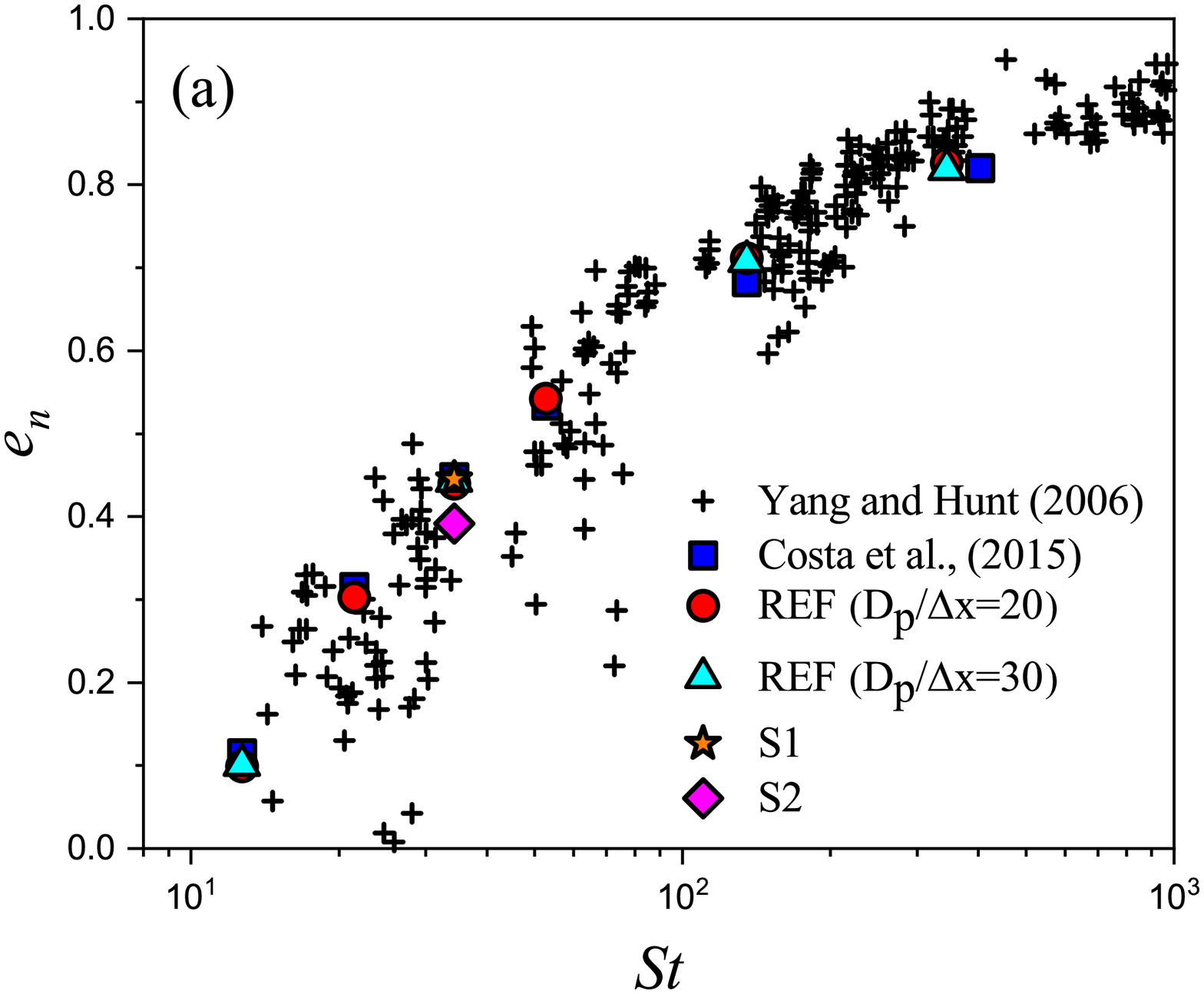}
\hspace{-10mm}
\includegraphics[width=8cm]{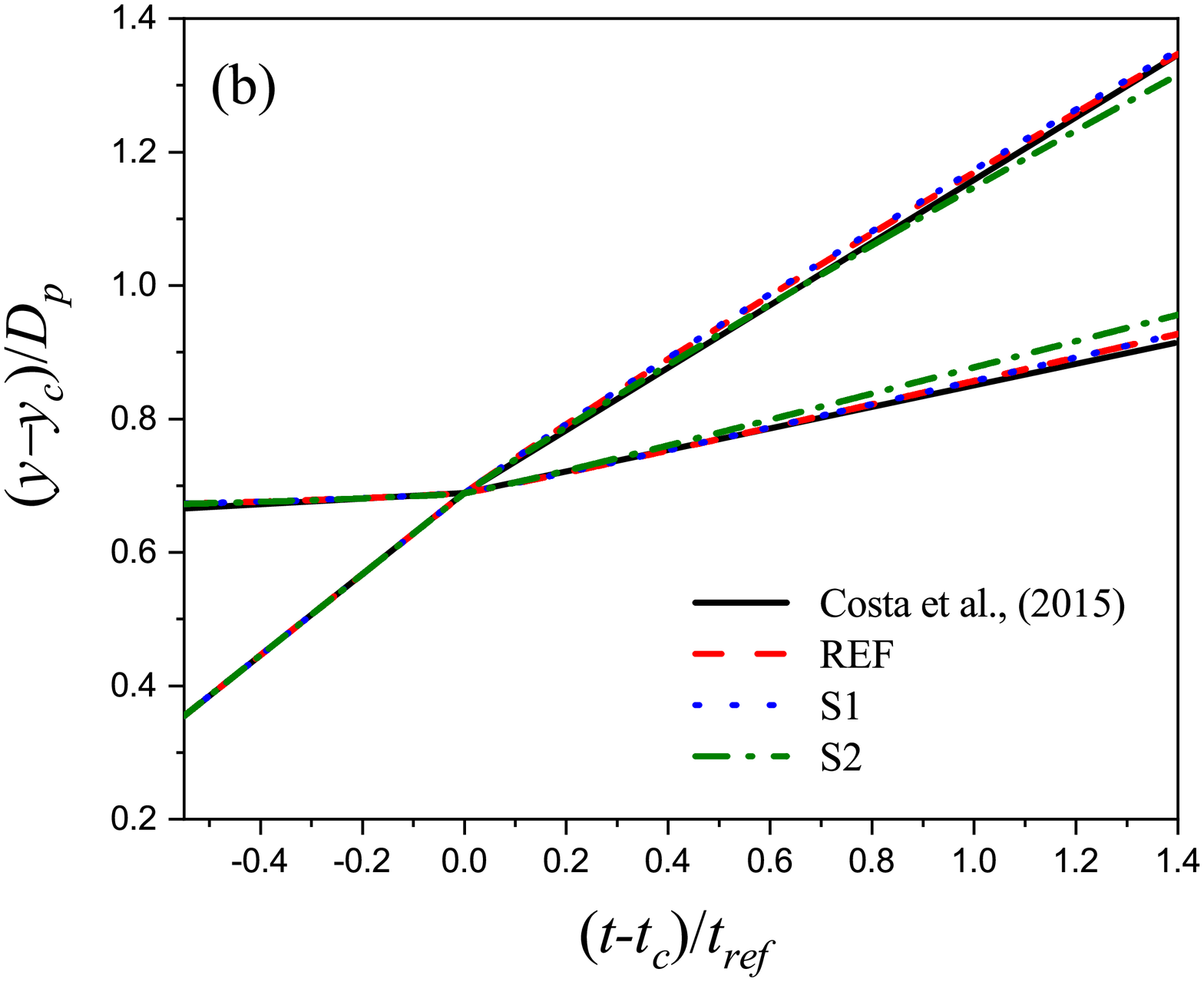}
\hspace{-16mm}
\caption{Simulation results of the particle-particle collisions by the traditional soft-sphere model: (a) wet coefficient of restitution in normal particle-particle collisions and (b) trajectories of particle contact points at $St=34$. $t_c=t\mid_{\delta_n=0}$ is the instant of collision, and $t_{ref}=\sqrt{D_p/\left | \bm g \right |}$ is the reference time scale.}
\label{fig:pp_en_ssm}
\end{figure}

\subsection{Assessment of the MTSA}
{Based on the above results that a large flow time step $\Delta t_f$ cannot be used to resolve fluid-particle interaction during a collision, we employ a large $\Delta t_f$ for fluid flow advancement while inserting additional time substeps for fluid-particle interaction in the MTSA. In this subsection, we investigate the influence of the substep iteration number on the simulation results with different particle Stokes number $St$ and grid resolution $D_p/\Delta x$.} 

\subsubsection{Particle-wall collision}

For particle-wall collision, we use the same simulation setup and parameters as in \S 4.1. 
The resulting normal coefficient of restitution $e_n$ using the MTSA is compared with that of case REF. Figure~\ref{fig:pw} shows the error of $e_n$ between the MTSA and case REF with different $St$ and $D_p/\Delta x$ values. The left and right columns show the errors with grid resolutions of $D_p/\Delta x=20$ and $D_p/\Delta x=30$, respectively. The dashed line corresponds to zero error. 

It can be seen that the general trends of the error curves with $R_i$ in each subfigure are similar in figures~\ref{fig:pw} (a)-(f). {As discussed before, $e_n$ is underestimated ($e_{n,MTSA}-e_{n,REF}<0$) without including additional fluid-particle interaction substeps, since the fluid flow surrounding the particle cannot adapt to the rapid change of the particle velocity. Therefore, the error curves are mostly under the zero-error line at small $R_i$ values. If more fluid-particle interaction substeps are inserted (increasing $R_i$), the absolute value of $e_{n,MTSA}-e_{n,REF}$ decreases first. However, as the pressure Poisson equation is not solved in the particle-fluid interaction substeps, the intermediate flow velocities are nonphysical, which leads to an overestimation of $e_n$ at larger $R_i$ values. Therefore, the overall trend of the absolute error of $e_{n,MTSA}-e_{n,REF}$ decreases first then increases with $R_i$, as shown in figures~\ref{fig:pw} (a)-(f).}


\begin{figure}
\vspace{-60pt}
\centering
   \begin{minipage}{0.51\linewidth}
       \centerline{\includegraphics[width=6.5cm]{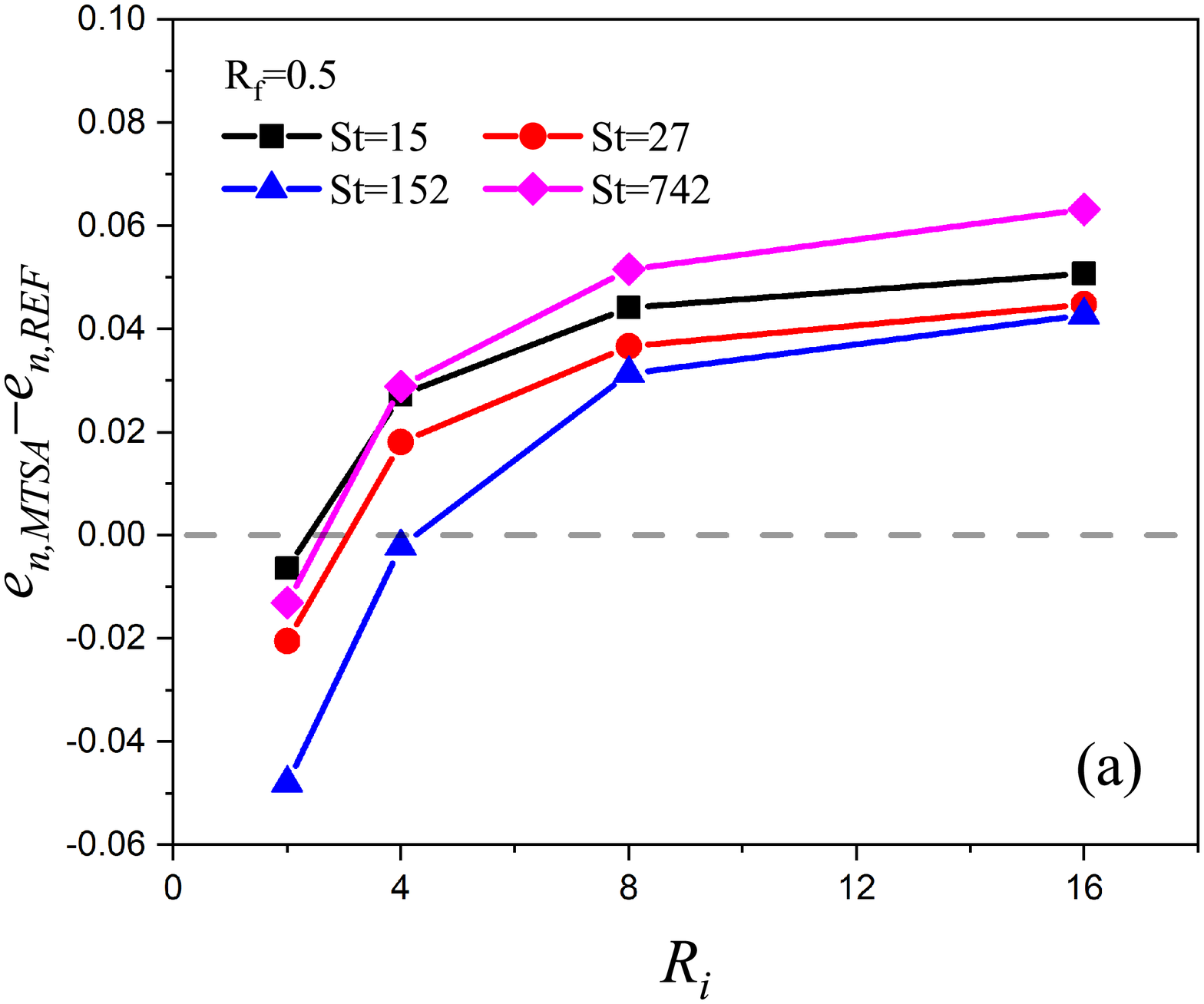}}
       \vspace{-10pt}
       \centerline{\includegraphics[width=6.5cm]{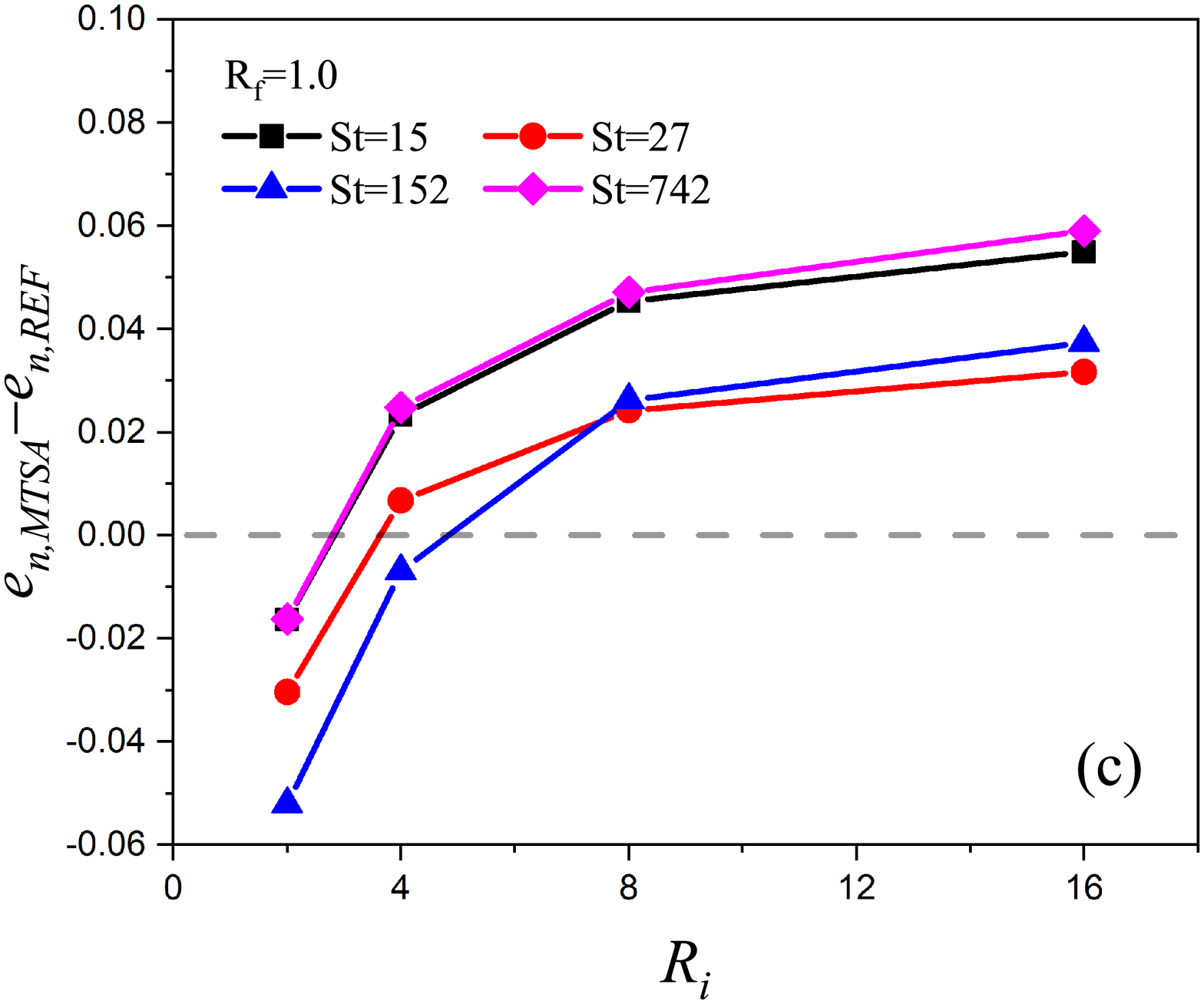}}
       \vspace{-10pt}
       \centerline{\includegraphics[width=6.5cm]{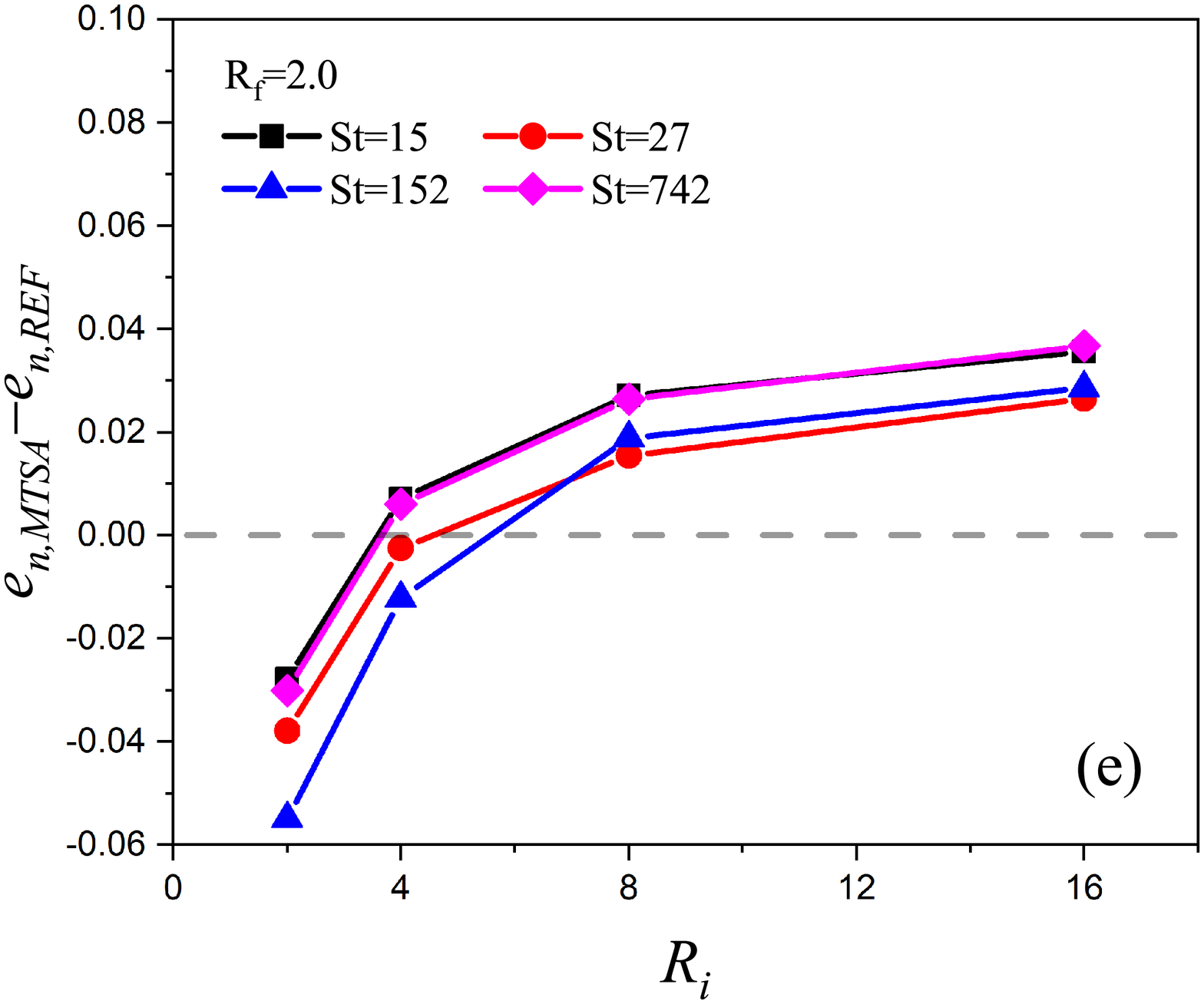}}
       \vspace{-10pt}
       \centerline{\includegraphics[width=6.5cm]{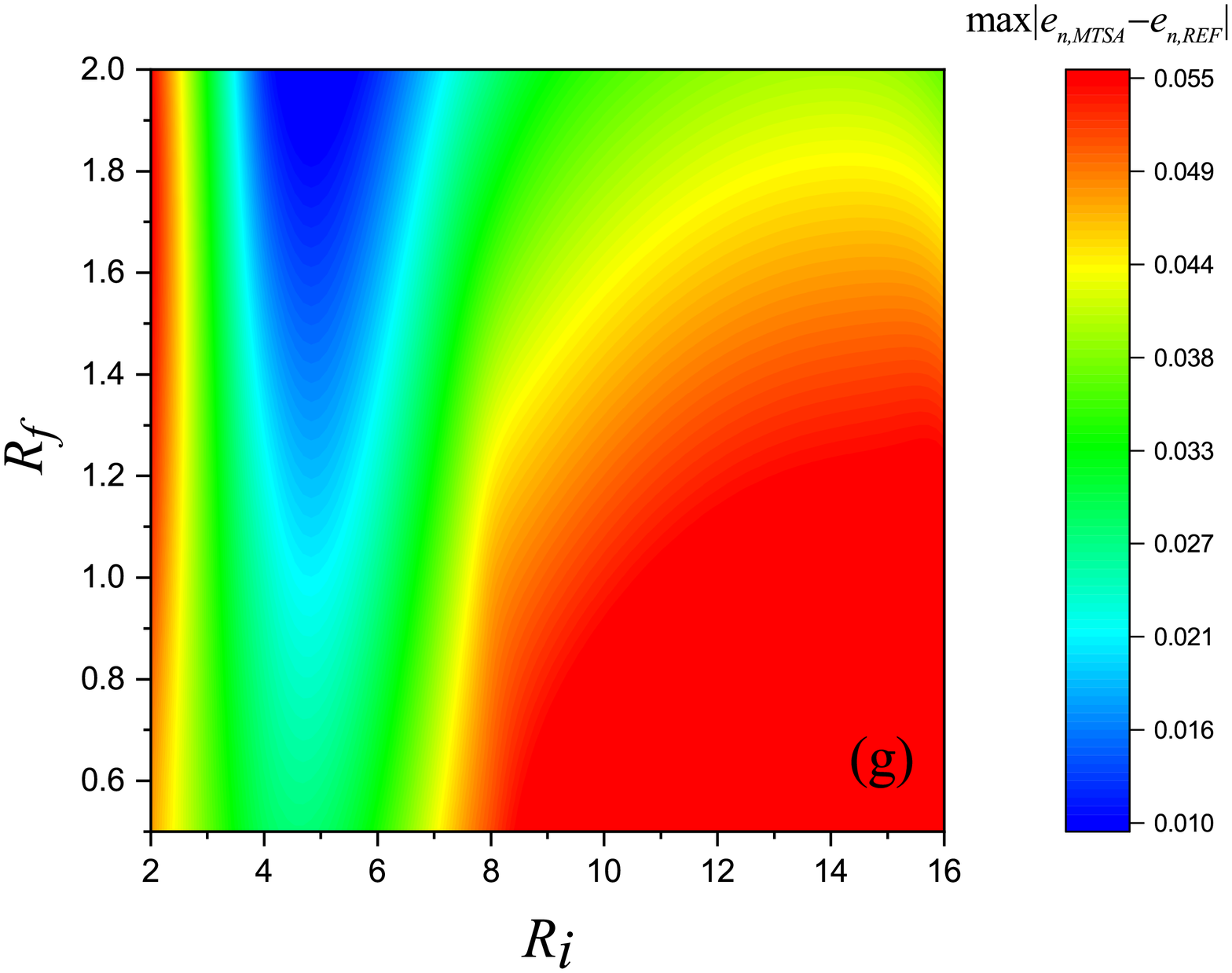}}
       \centerline{$D_p/\Delta x=20$}
   \end{minipage}
      \begin{minipage}{0.48\linewidth}
       \centerline{\includegraphics[width=6.5cm]{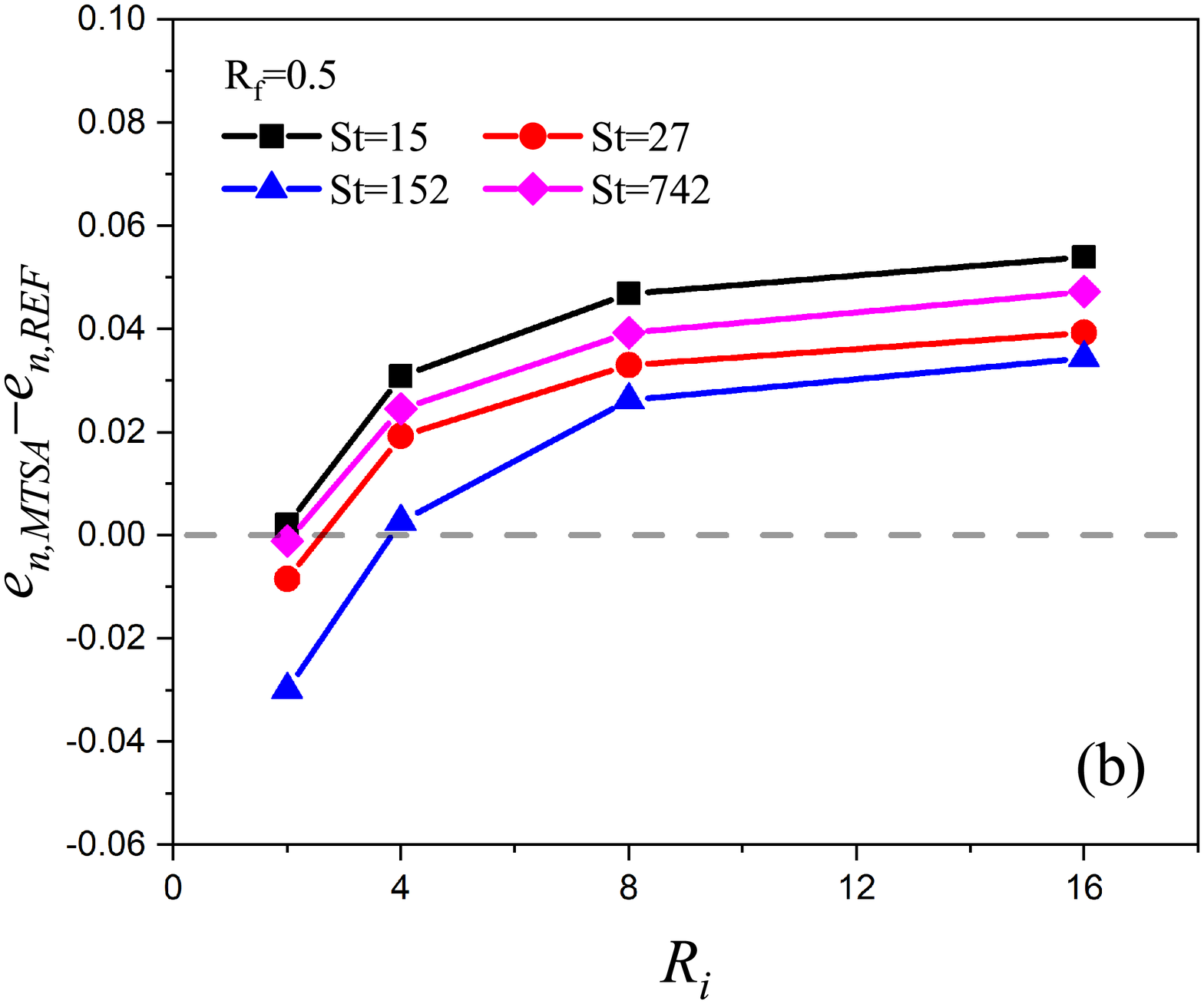}}
       \vspace{-10pt}
       \centerline{\includegraphics[width=6.5cm]{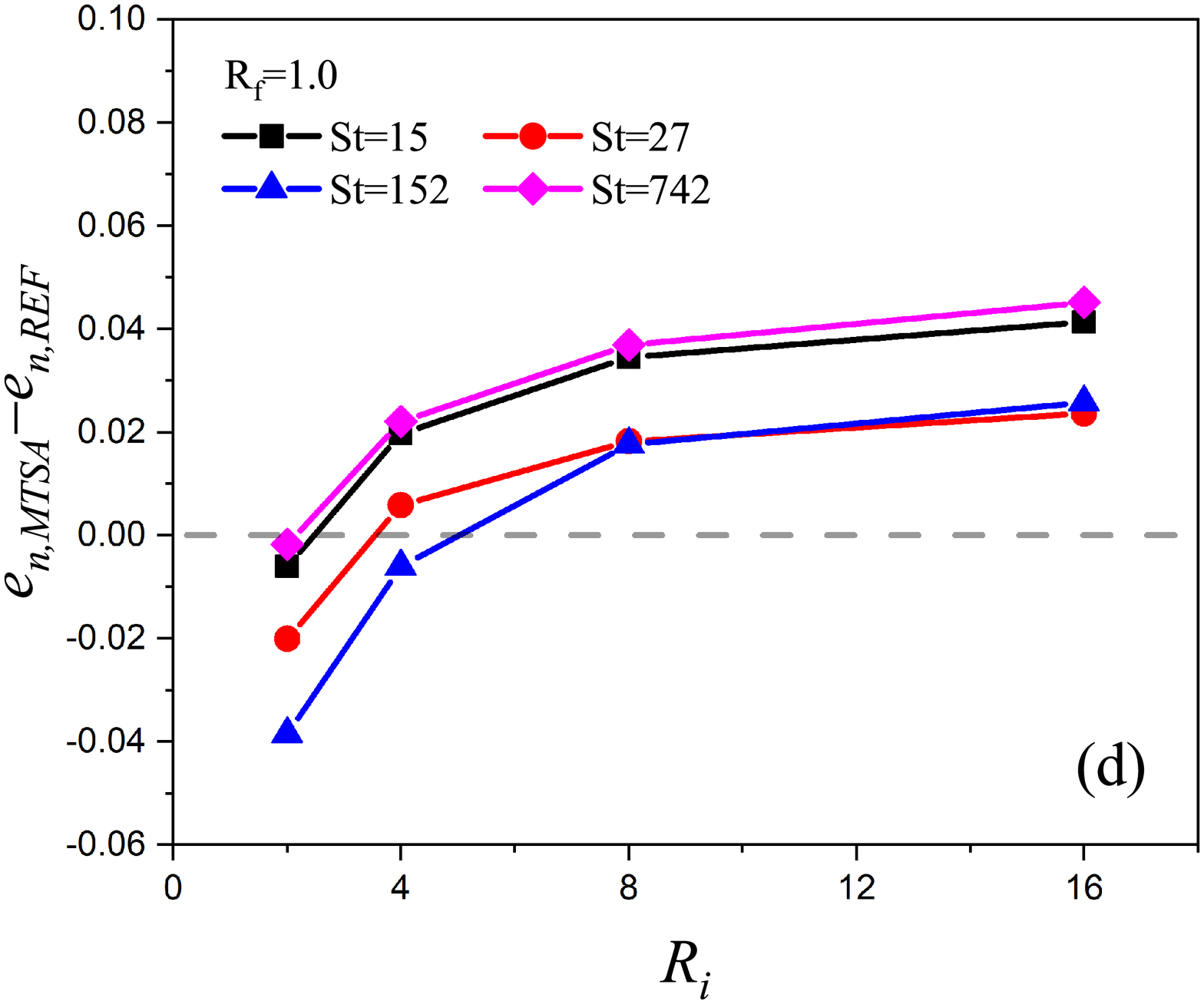}}
       \vspace{-10pt}
       \centerline{\includegraphics[width=6.5cm]{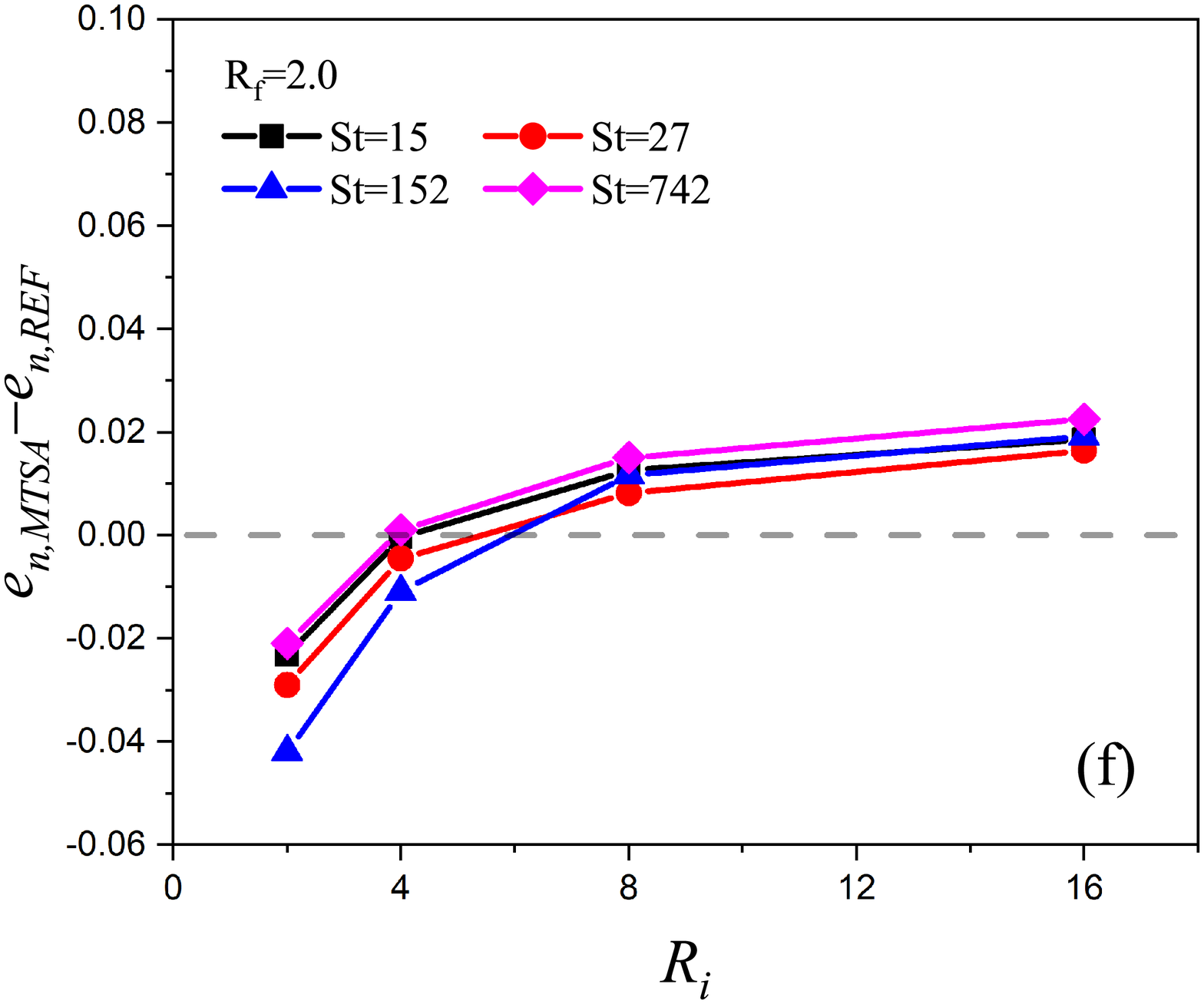}}
       \vspace{-10pt}
       \centerline{\includegraphics[width=6.5cm]{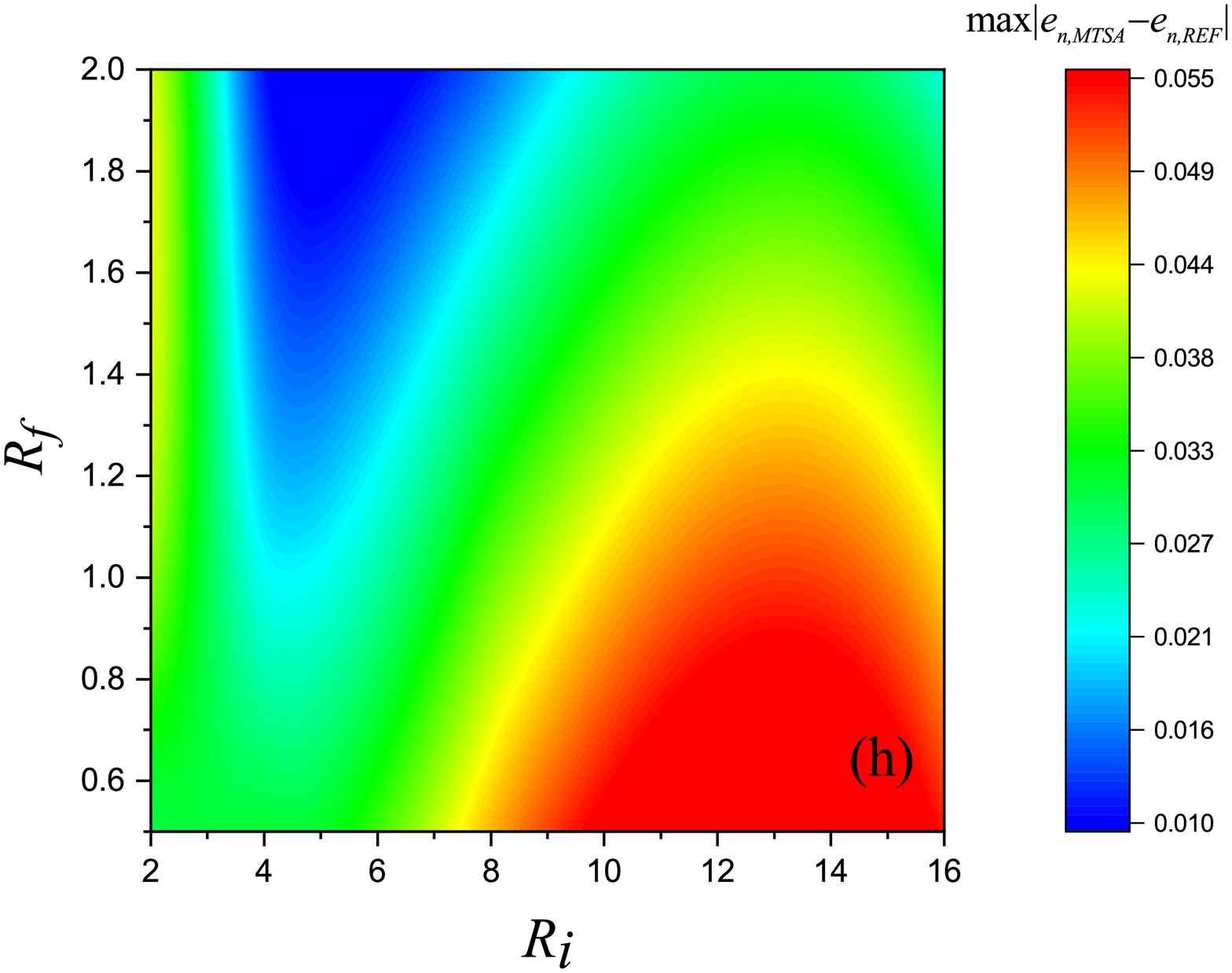}}
       \centerline{$D_p/\Delta x=30$}
   \end{minipage}
   \caption{Prediction error of the normal coefficient of restitution $e_n$ between the MTSA and case REF in particle--wall collisions: (a,b) $R_f=0.5$, (c,d) $R_f=1.0$, (e,f) $R_f=2.0$, and (g,h) the contours of the maximum error $\left | e_{n,MTSA}-e_{n,REF} \right |$ with $St=15, 27, 152, 742$.} (a,c,d,g) $D_p/\Delta x=20$ and (b,d,f,h) $D_p/\Delta x=30$. The dashed line corresponds to zero error.
   \label{fig:pw}
\end{figure}

{
We can elucidate the increase in the error of $e_{n,MTSA}-e_{n,REF}$ with $R_i$ as follows. Based on the Navier-Stokes equations discretized by the explicit Euler method, the intermediate flow velocities at the $i$th fluid-particle interaction substep during a collision can be expressed as
\begin{align}
	\bm u^i &=\bm u^{i-1}+\Delta t_i\left (\bm H^{i-1}+ \bm f^{i-1}-\frac{1}{\rho_f}\nabla p^{i-1}\right ), \\
	\widetilde{\bm u}^i &=\widetilde{\bm u}^{i-1}+\Delta t_i\left (\widetilde{\bm H}^{i-1}+\widetilde{\bm f}^{i-1}\right ), 
\end{align}
where $\bm u$ and $\widetilde{\bm u}$ are the physical (by solving the pressure Poisson equation) and nonphysical (without solving the pressure Poisson equation) intermediate flow velocities, respectively; $\bm H$ and $\widetilde{\bm H}$ are the sums of the convective and viscous terms with physical and nonphysical intermediate flow velocities, respectively; $\bm f$ and $\widetilde{\bm f}$ are the volume forces with physical and nonphysical intermediate flow velocities, respectively. Assuming that the initial flow velocity before a collision is $\bm u^0$, the flow velocity after the collision ($i=R_i$) can be expressed as
\begin{align}
	\bm u^{R_i} &=\bm u^0+\sum_{i=1}^{R_i}\Delta t_i\left ( \bm H^{i-1}+\bm f^{i-1}-\frac{1}{\rho_f}\nabla p^{i-1}\right ), \\
	\widetilde{\bm u}^{R_i} &=\bm u^0+\sum_{i=1}^{R_i}\Delta t_i\left (\widetilde{\bm H}^{i-1}+\widetilde{\bm f}^{i-1}\right ).
\end{align}
The quantities at Eulerian grid points can be transferred to Lagrangian grid points through a regularized Dirac delta function $\delta_d$ \cite[]{roma1999adaptive}. The flow velocity on the particle surface after the collision is
\begin{align}
	\bm U_l^{R_i} &=\bm U_l^0+\sum_{i=1}^{R_i}\Delta t_i\left ( \bm H_l^{i-1}+\bm F_{p,l}^{i-1}-\frac{1}{\rho_f}\nabla P_l^{i-1}\right ) \label{eqn:error1}, \\
	\widetilde{\bm U}_l^{R_i} &=\bm U_l^0+\sum_{i=1}^{R_i}\Delta t_i\left (\widetilde{\bm H}_{l}^{i-1}+ \widetilde{\bm F}_{p,l}^{i-1}\right ), \label{eqn:error2}
\end{align}
where $\bm U_l$, $\bm F_{p,l}$, $\bm H_l$, and $P_l$ are the quantities of $\bm u$, $\bm f$, $\bm H$, and $p$ at Eulerian grid points interpolated at Lagrangian grid points, respectively; and $\widetilde{\bm U}_l$, $\widetilde{\bm F}_{p,l}$, and $\widetilde{\bm H}_l$ are the quantities of $\widetilde{\bm u}$, $\widetilde{\bm f}$, and $\widetilde{\bm H}$ on Eulerian grid points interpolated on Lagrangian grid points, respectively. During a collision, the particle motion is only determined by the collision model and not affected by the fluid flow \cite[]{kempe2012collision,costa2015collision,biegert2017collision}, thus the particle velocities with physical and nonphysical intermediate flow velocities are the same ($\bm U_p^{R_i}=\widetilde{\bm U}_p^{R_i}$). Based on the IB method as in equation (\ref{eqn:f3}), the differences in the hydrodynamic force with physical and nonphysical intermediate flow velocities after a collision can be expressed as
\begin{align}
	\Delta \bm F_{p,l}^{R_i} &=\widetilde{\bm F}_{p,l}^{R_i}-\bm F_{p,l}^{R_i} \nonumber \\
	&=\frac{\widetilde{\bm U}_p^{R_i}-\bm U_p^{R_i}+\bm U_l^{R_i}-\widetilde{\bm U}_l^{R_i}}{\Delta t_i} \label{eqn:error3} \\
	&=\frac{\bm U_l^{R_i}-\widetilde{\bm U}_l^{R_i}}{\Delta t_i}. \nonumber
\end{align}
Substituting equations (\ref{eqn:error1}) and (\ref{eqn:error2}) into equation (\ref{eqn:error3}) yields
\begin{equation}
	\Delta \bm F_{p,l}^{R_i}=\sum_{i=1}^{R_i}(\Delta \bm H_l^{i-1}-\Delta \bm F_{p,l}^{i-1}-\frac{1}{\rho_f}\nabla P_l^{i-1}). \label{eqn:error4}
\end{equation}
Equation (\ref{eqn:error4}) includes three terms: $\Delta \bm H_l^{i-1}=\widetilde{\bm H}_{l}^{i-1}-\bm H_l^{i-1}$, $\Delta \bm F_{p,l}^{i-1}=\widetilde{\bm F}_{p,l}^{i-1}-\bm F_{p,l}^{i-1}$, and $\nabla P_l^{i-1}$, representing the flow velocity error, the hydrodynamic force error, and the pressure error at each fluid-particle interaction substep, respectively. If $\Delta \bm H_l^{i-1}-\Delta \bm F_{p,l}^{i-1}-\nabla P_l^{i-1}/\rho_f \ne 0$, the error of $\Delta \bm F_{p,l}^{R_i}$ increases as $R_i$ increases, which leads to the error of $e_n$ increasing with $R_i$. As shown in figures~\ref{fig:pw} (a-f) and figures~\ref{fig:pp} (a-f), $e_{n,MTSA}$ increases with $R_i$, which indicates that the MTSA underestimates the hydrodynamic force as $R_i$ increased and $\Delta \bm H_l^{i-1}-\Delta \bm F_{p,l}^{i-1}-\nabla P_l^{i-1}/\rho_f<0$.
}

{Since insufficient fluid-particle interaction substeps can cause an underestimation of $e_n$, and too many substeps can cause an overestimation of $e_n$. Therefore, we may find an optimal value of $R_i$ to minimize the error of $e_n$.} {Figures~\ref{fig:pw} (g) and (h) show the contours of $\max\left | e_{n,MTSA}-e_{n,REF} \right |$ for different $St$ and $D_p/\Delta x$ values. It can be found that the maximum error is small for $R_i\in\left \{ 4,5,6 \right \}$ with different $St$ and $D_p/\Delta x$ values, which indicates that the MTSA has good accuracy with $R_i\in\left \{ 4,5,6 \right \}$ in the particle-wall collisions. Based on the maximum error in the particle-particle collisions shown in figures~\ref{fig:pp} (g) and (h), $R_i=4$ is determined to be the optimal value. The maximum error for $R_i=4$ is less than 0.031 in the particle-wall collisions, which is small enough.}

{After determining that $R_i=4$ is optimal, we employed the MTSA with $R_f=1, R_i=4, R_m=40$ to calculate $e_n$ for particle-wall collisions in the entire range of $St$. As shown in figure~\ref{fig:pw_mtsa}, the simulation results of the MTSA agree well with those of case REF and the experimental results over the entire range of $St$. Therefore, after inserting proper additional fluid-particle interaction substeps, the prediction results by the MTSA are significantly improved compared with those of case S2 in particle-wall collisions. 
The yellow error bars in figure~\ref{fig:pw_mtsa} indicate the prediction uncertainty of $e_n$ with $R_f=0.5, 1, 2$. The short lengths of the error bars imply that the MTSA has good convergence with these $R_f$ values.}
A consistent conclusion can be further drawn from the particle trajectories shown in figure~\ref{fig:pw_tra_mtsa}. It can be seen that the particle trajectories predicted by the MTSA show good agreements with those of case REF, and they are significantly improved compared with those of case S2. The grey area is the particle trajectory range with $R_f=0.5, 1, 2$. {Such small areas demonstrate that the particle trajectories in the particle-wall collisions will not be much affected by $\Delta t_f$ if it is sufficiently small ($R_f\geqslant 0.5$).}

\begin{figure}[htbp]
\centering
\includegraphics[width=0.8\textwidth]{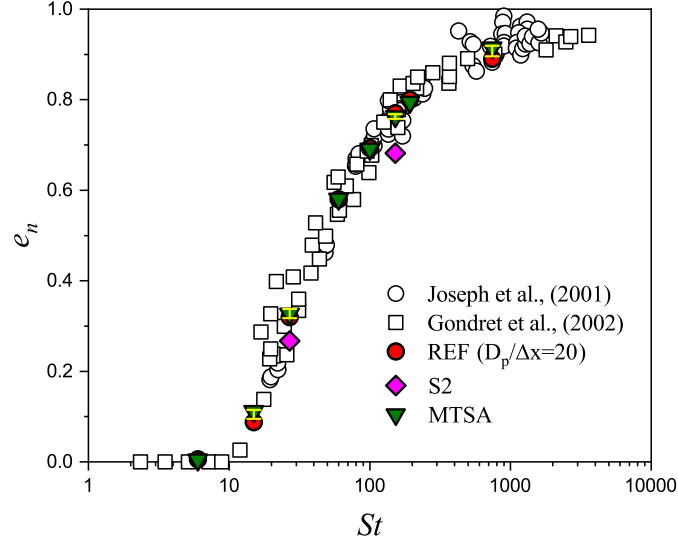}
\hspace{0in}
\caption{Wet coefficient of restitution in normal particle-wall collisions predicted by the MTSA ($R_f=1$, $R_i=4$, $R_m=40$). The yellow error bar indicates the prediction uncertainty of $e_n$ with $R_f=0.5, 1, 2$.}
\label{fig:pw_mtsa}
\end{figure}

\begin{figure}[htbp]
\centering
\hspace{-16mm}
\includegraphics[width=8cm]{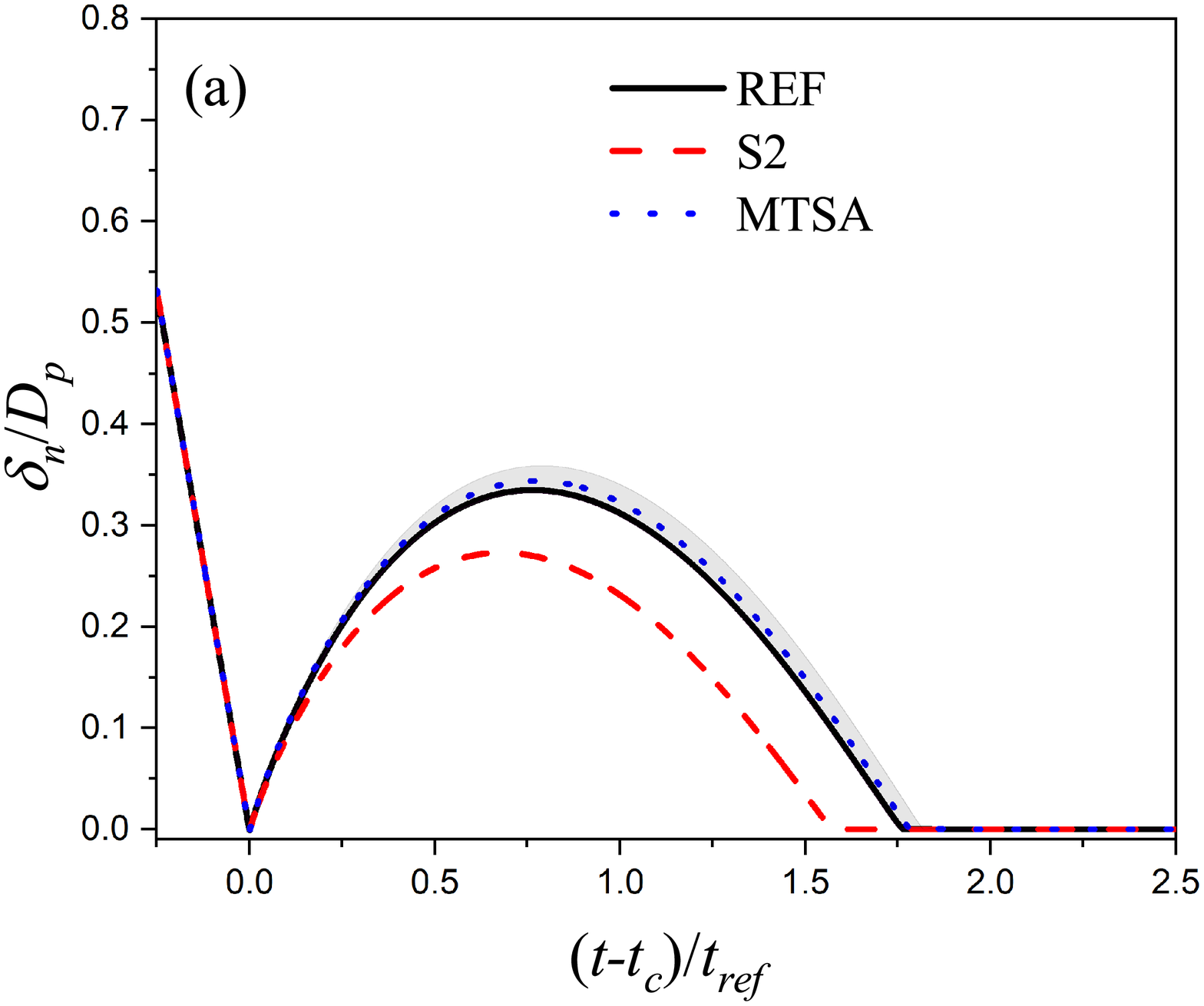}
\hspace{-10mm}
\includegraphics[width=8cm]{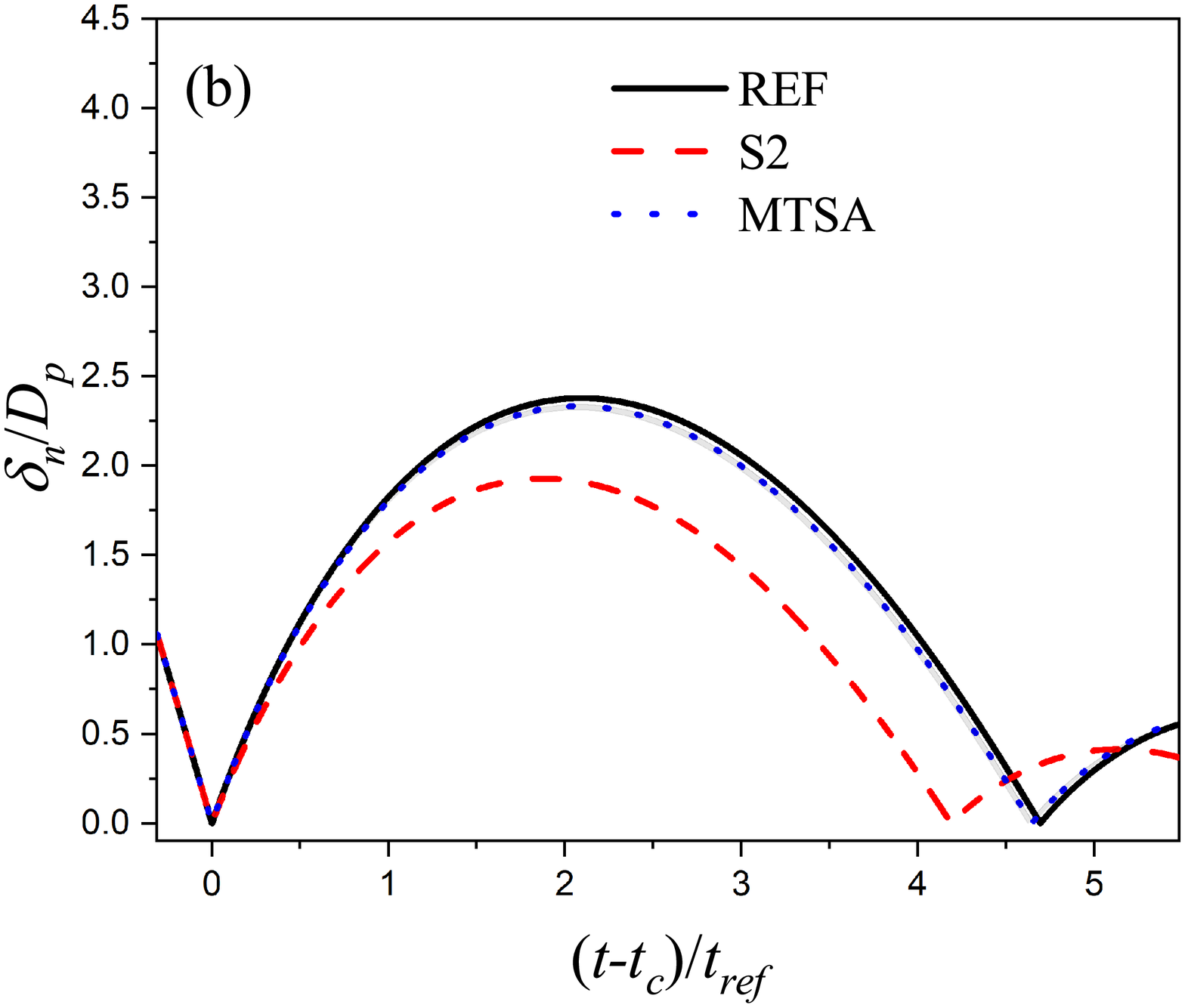}
\hspace{-16mm}
\caption{Particle trajectories with different Stokes numbers: (a) $St=27$ and (b) $St=152$. The grey area is the particle trajectory range with $R_f=0.5, 1, 2$. $t_c=t\mid_{\delta_n=0}$ is the instant of collision, and $t_{ref}=\sqrt{D_p/\left | \bm g \right |}$ is the reference time scale.}
\label{fig:pw_tra_mtsa}
\end{figure}

\subsubsection{Particle-particle collision}

For particle-particle collisions, we adopted the same simulation setup as that in \S 4.1. 
Figure~\ref{fig:pp} shows the error of the normal coefficient of restitution $e_n$ between the MTSA and case REF. The error curves in each subfigure exhibit the same trend as those in particle-wall collisions, for the same reason as that in particle-wall collisions discussed above. 

\begin{figure}
\vspace{-60pt}
\centering
   \begin{minipage}{0.51\linewidth} 
       \centerline{\includegraphics[width=6.5cm]{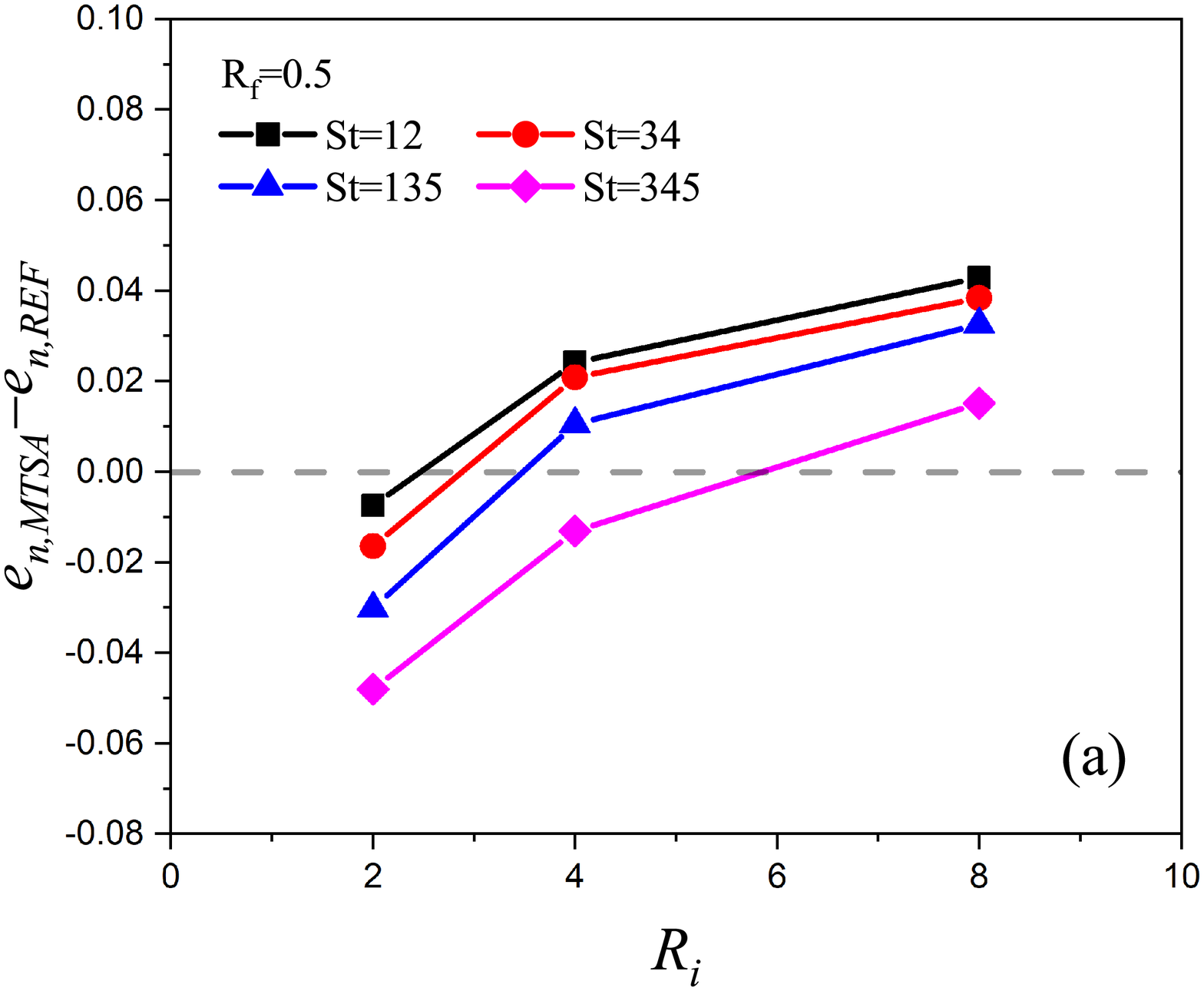}}
       \vspace{-10pt}
       \centerline{\includegraphics[width=6.5cm]{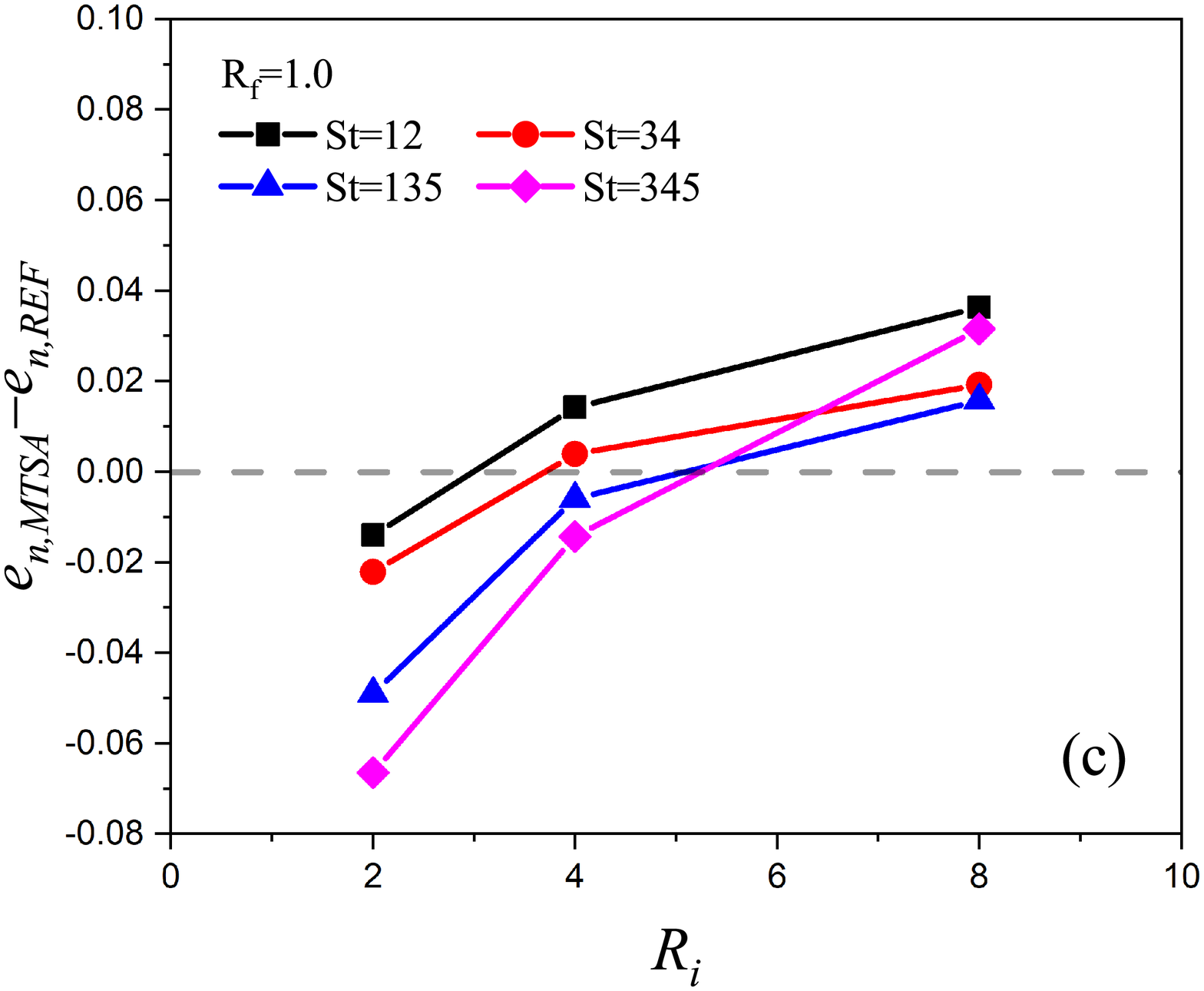}}
       \vspace{-10pt}
       \centerline{\includegraphics[width=6.5cm]{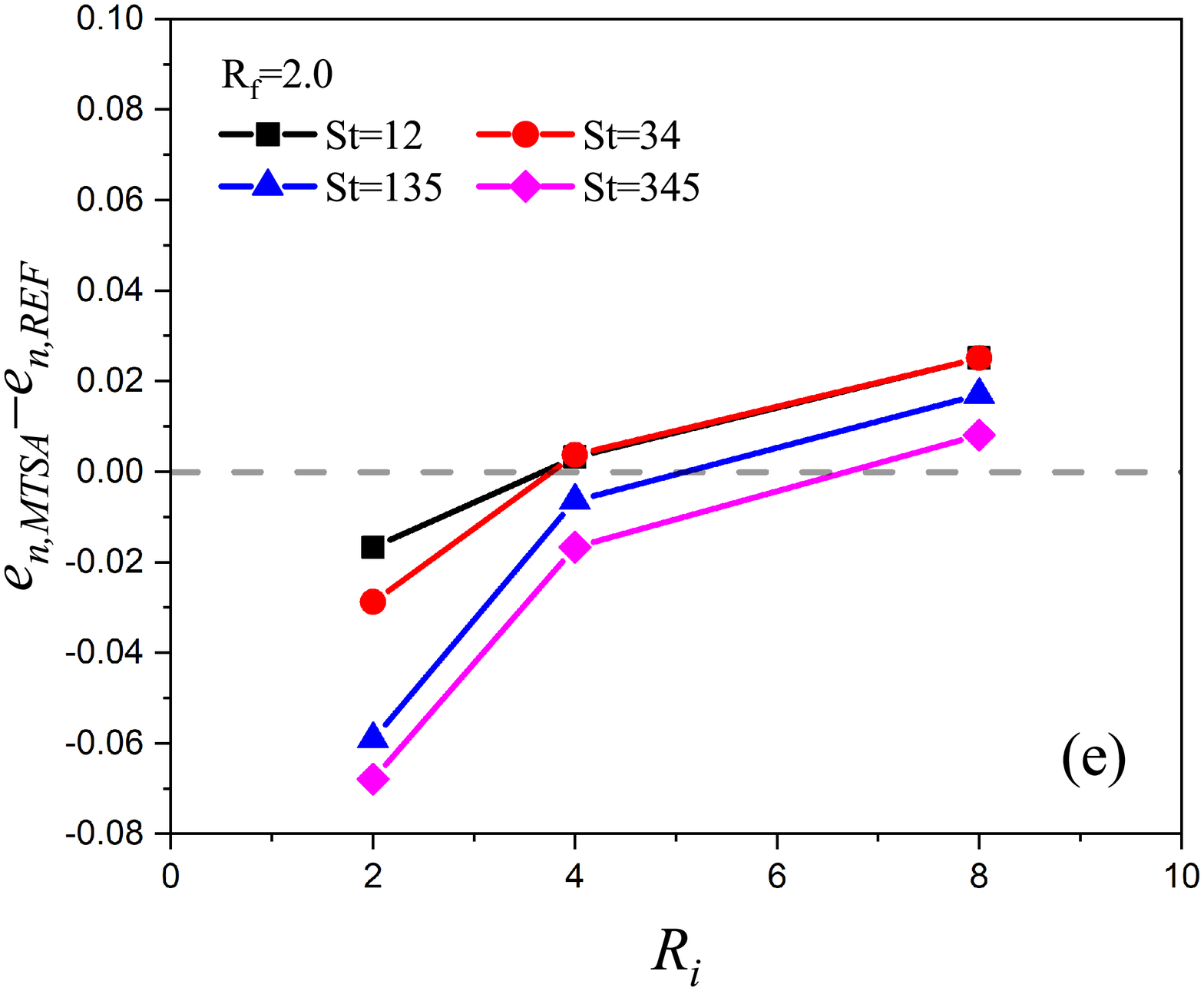}}
       \vspace{-10pt}
       \centerline{\includegraphics[width=6.5cm]{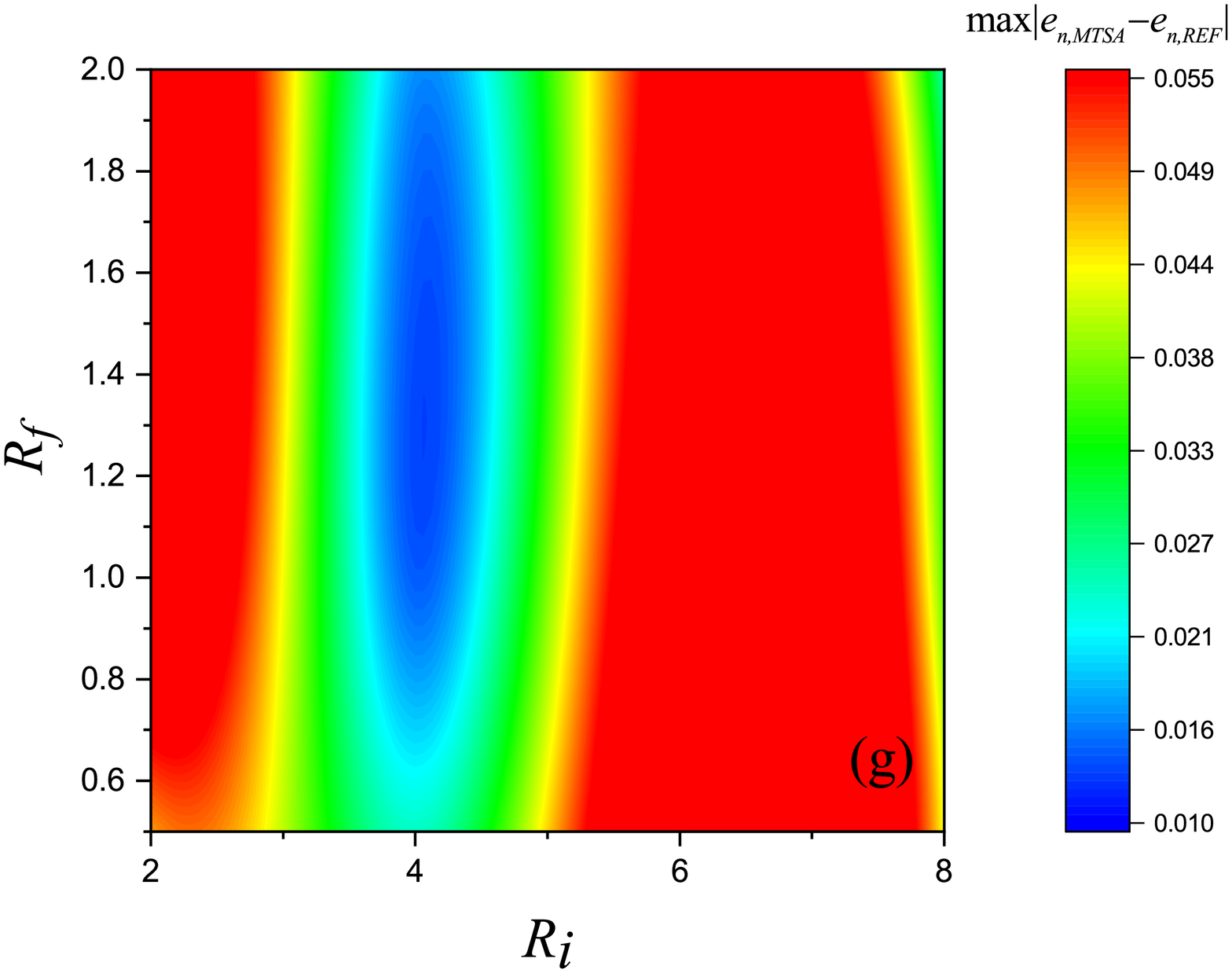}}
       \centerline{$D_p/\Delta x=20$}
   \end{minipage}
      \begin{minipage}{0.48\linewidth}
       \centerline{\includegraphics[width=6.5cm]{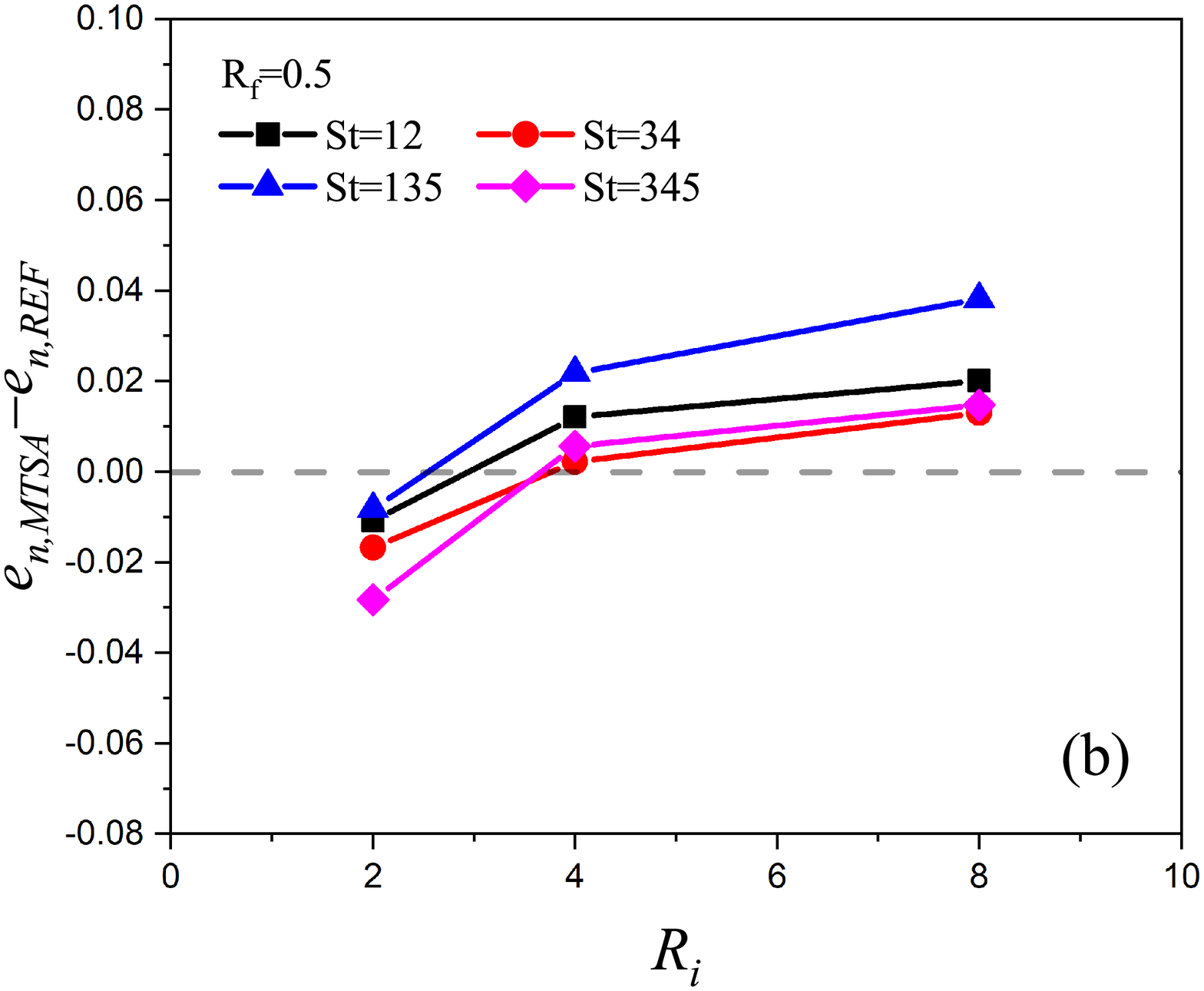}}
       \vspace{-10pt}
       \centerline{\includegraphics[width=6.5cm]{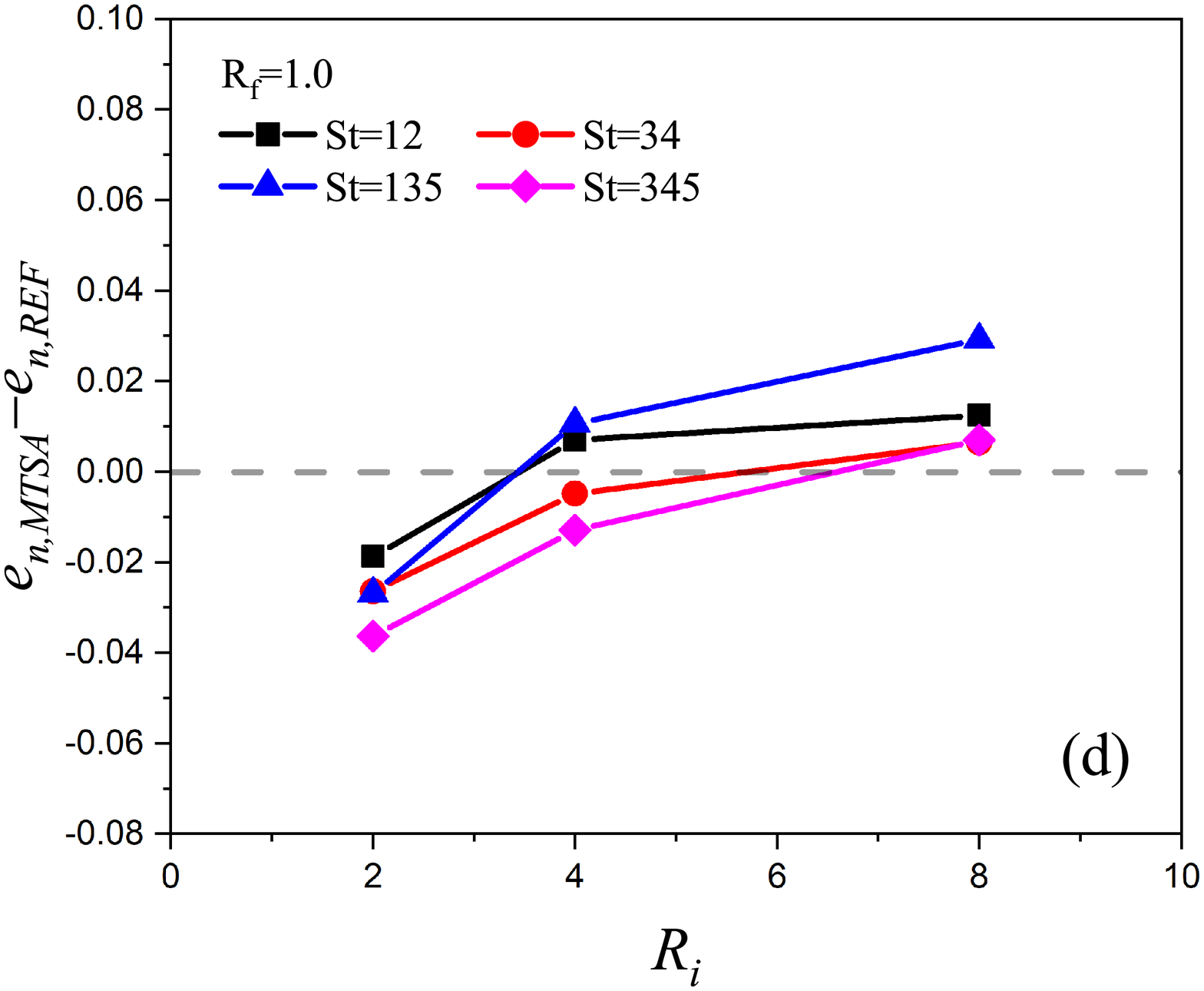}}
       \vspace{-10pt}
       \centerline{\includegraphics[width=6.5cm]{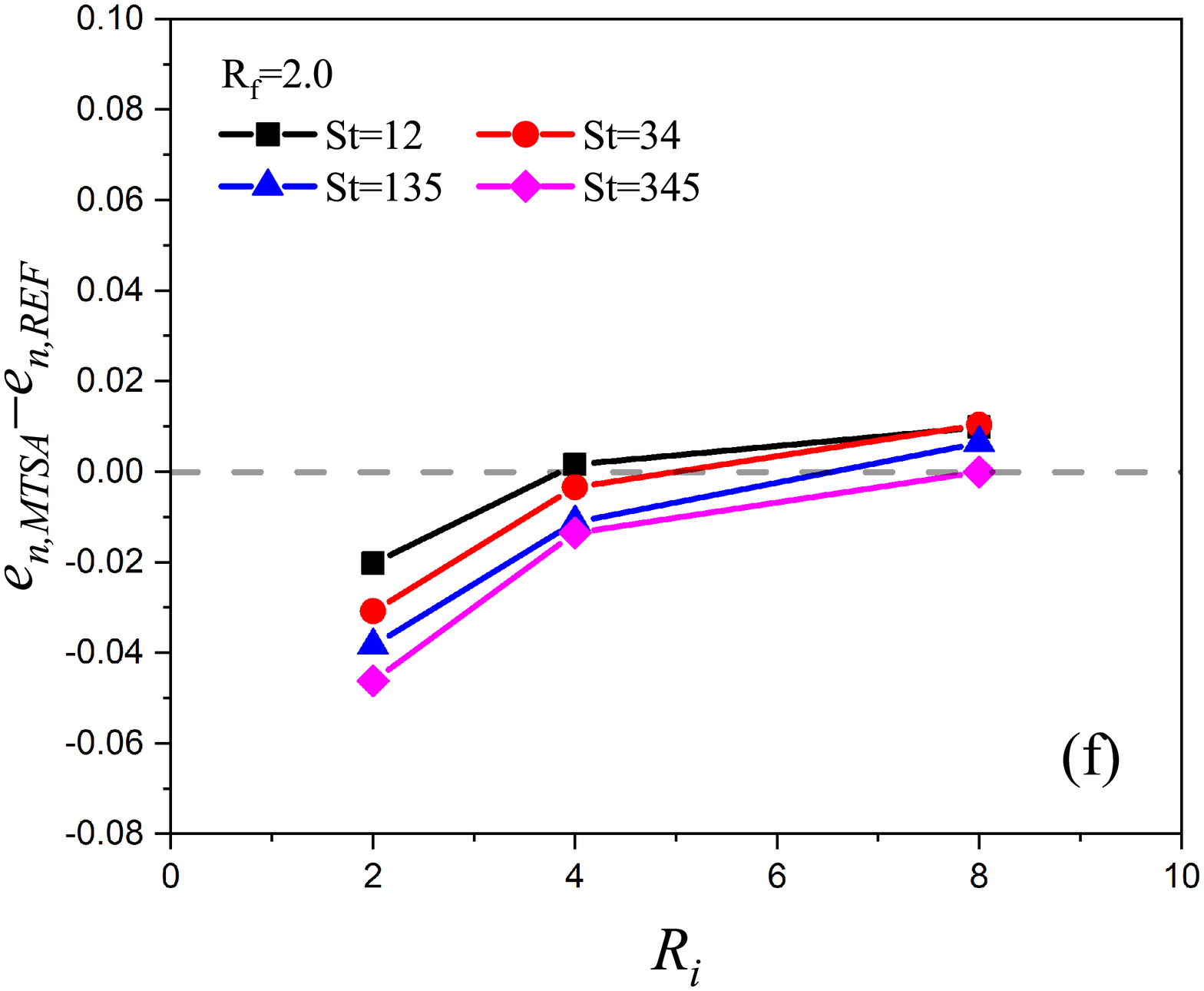}}
       \vspace{-10pt}
       \centerline{\includegraphics[width=6.5cm]{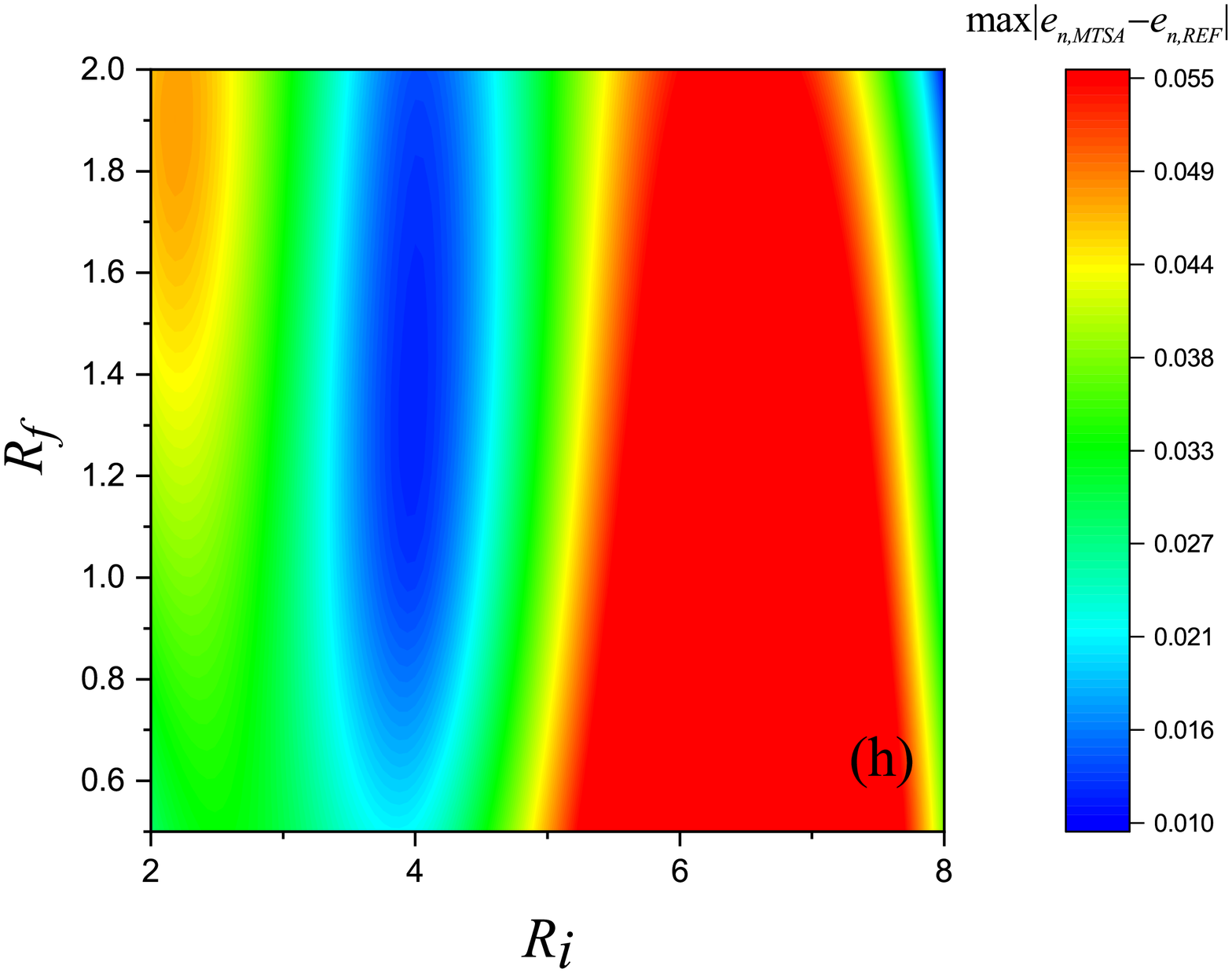}}       
       \centerline{$D_p/\Delta x=30$}
   \end{minipage}
   \caption{Prediction error of the normal coefficient of restitution $e_n$ between the MTSA and case REF in particle-particle collisions: (a,b) $R_f=0.5$, (c,d) $R_f=1.0$, (e,f) $R_f=2.0$, and (g,h) the contours of the maximum error $\left | e_{n,MTSA}-e_{n,REF} \right |$ with $St=12, 34, 135, 345$.} (a,c,d,g) $D_p/\Delta x=20$ and (b,d,f,h) $D_p/\Delta x=30$. The dashed line corresponds to zero error.
   \label{fig:pp}
\end{figure}

{Figures~\ref{fig:pp} (g) and (h) show the contours of $\max\left | e_{n,MTSA}-e_{n,REF} \right |$ with different $St$ and $D_p/\Delta x$ values. It is seen that the maximum error is small for $R_i\in\left \{ 4 \right \}$ with different $St$ and $D_p/\Delta x$ values, which indicates that the MTSA has good accuracy with $R_i=4$ in particle-particle collisions. Based on the combined results of the particle-wall and particle-particle collisions, $R_i=4$ is determined to be the optimal value and will be used in the following. The absolute maximum error for $R_i=4$ is less than 0.024 in the particle-particle collisions, which is also small enough.}

We further employed the MTSA with $R_f=1, R_i=4, R_m=40$ to calculate $e_n$ for particle-particle collisions over the entire range of $St$. As shown in figure~\ref{fig:pp_mtsa} (a), the simulation results of the MTSA agree well with those of case REF and the experimental results. Its predictions are much better than those of case S2. The yellow error bars represent the prediction uncertainty of $e_n$ with $R_f=0.5, 1, 2$. The short lengths of the error bars demonstrate that the MTSA has good convergence with these $R_f$ values. Figure~\ref{fig:pp_mtsa} (b) shows the particle trajectories with $St=34$. The particle trajectories predicted by the MTSA have excellent agreement with those of case REF and are much better than those of case S2. The grey area is the particle trajectory range with $R_f=0.5, 1, 2$. {The small grey areas indicate that the predicted particle trajectories in the particle-particle collisions are not much affected by $\Delta t_f$ if it is sufficiently small ($R_f\geqslant 0.5$).}

\begin{figure}[htbp]
\centering
\hspace{-16mm}
\includegraphics[width=8cm]{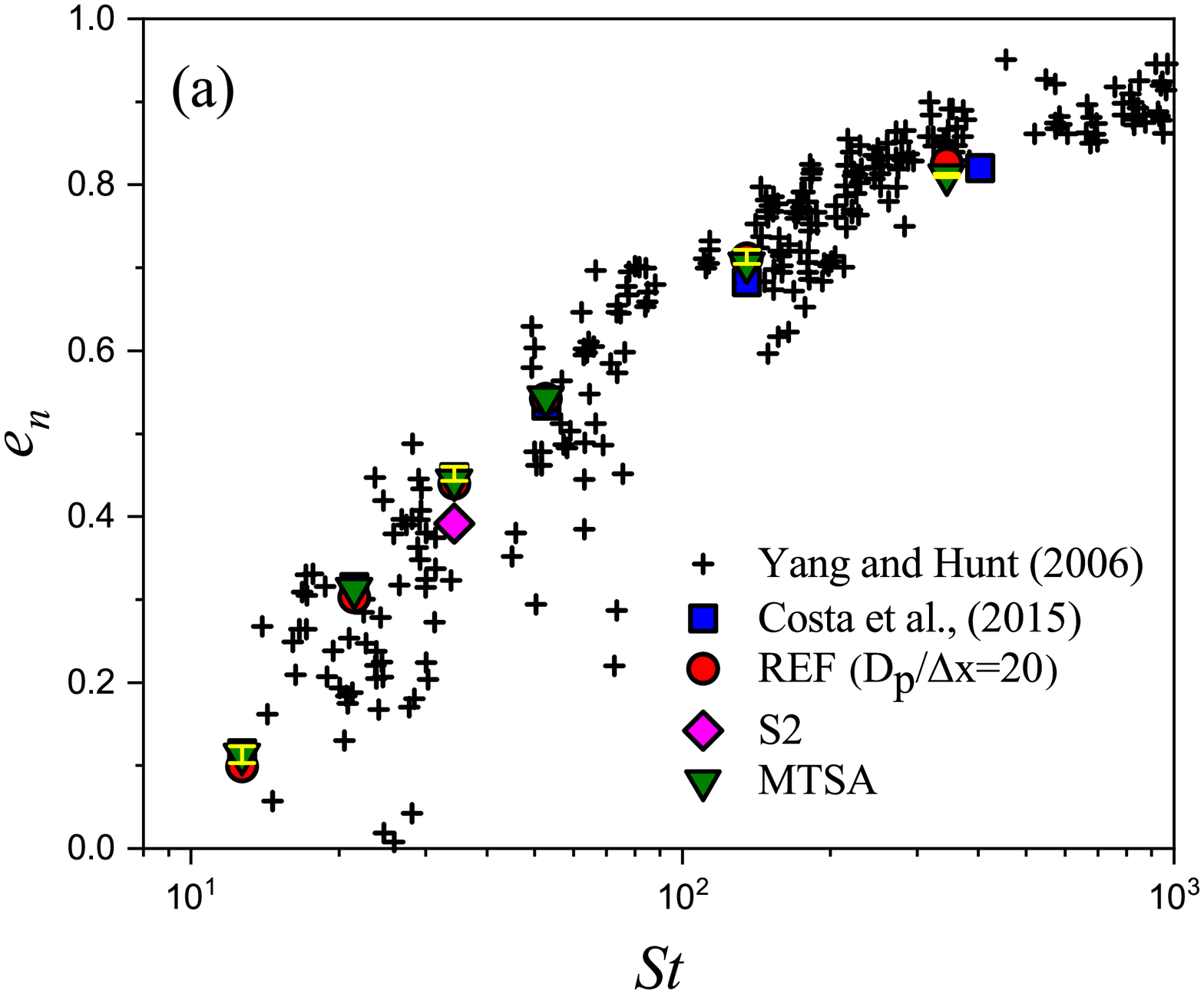}
\hspace{-10mm}
\includegraphics[width=8cm]{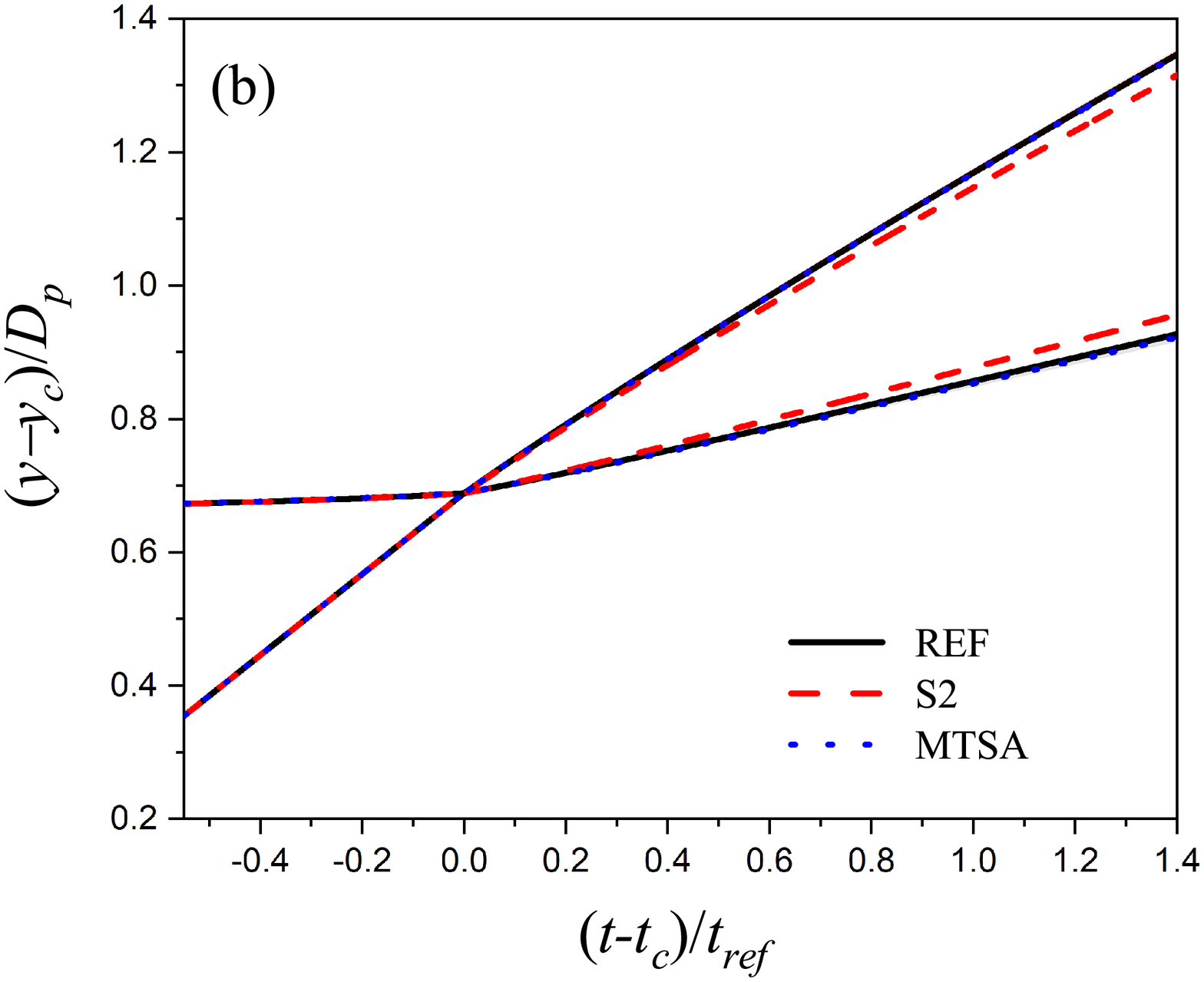}
\hspace{-16mm}
\caption{Simulation results of particle-particle collisions by the MTSA: (a) wet coefficient of restitution in normal particle-particle collisions, where the yellow error bars indicate the prediction uncertainty of $e_n$ with $R_f=0.5, 1, 2$; (b) the predicted trajectories of the particle contact points with $St=34$, where the grey area is the particle trajectory range with $R_f=0.5, 1, 2$. $t_c=t\mid_{\delta_n=0}$ is the instant of collision, and $t_{ref}=\sqrt{D_p/\left | \bm g \right |}$ is the reference time scale.}
\label{fig:pp_mtsa}
\end{figure}

{To summarize, by considering the results of particle-wall and particle-particle collisions together, $R_i=4$ is determined to be optimal for collisions with different $St$ and $D_p/\Delta x$ values. The prediction results by the MTSA with $R_i=4$ have been shown to be in good agreement with the experimental measurements and the benchmark results (case REF) and made significant improvements over the results without inserting additional fluid-particle interaction substeps (case S2).}

\subsubsection{Oblique particle-wall collision}

Here we assessed the MTSA ($R_f=1$, $R_i=4$, $R_m=40$) for oblique particle-wall collisions and compared the results with the experimental data of \cite{joseph2004oblique}. 
The physical parameters in the simulations are close to those in the experiment of \cite{yang2006dynamics}: $D_p=2.5$ mm, $\rho_f=998$ kg/m$^{3}$, $\mu_f=1$ cP and $e_{n,d}=0.97$, and other parameters are listed in table~\ref{table:oblique}.
The effective angles of incidence and rebounding are defined as $\Psi_{in}=u_{in,t}/u_{in,n}$ and $\Psi_{out}=u_{out,t}/u_{in,n}$, respectively, where $u_{in,n}$ and $u_{in,t}$ are the normal and tangential components of the particle impact velocity, respectively, and $u_{out,t}$ is the tangential component of the particle rebounding velocity. The computational domain and grid resolution are the same as those used for the previous particle-wall collisions in \S 4.1. Following the work of \cite{costa2015collision}, the particle falling velocity was controlled by an oblique acceleration with the direction vector $\bm e_g=-\sin(\phi_{in}) \bm e_y-\cos(\phi_{in}) \bm e_x$ to yield the desired incidence angle $\Psi_{in}$. The magnitude of the particle acceleration was set to $g=98.1$ m/s$^2$. 

\begin{table}[H]
   \centering
   \caption{Parameters used in the simulation of oblique particle-wall collisions. $\mu_{c,wet}$ is the wet friction coefficient.}
   \setlength{\tabcolsep}{5.5mm}{
   \begin{tabular}{ccccccc}
   \toprule
   Material & $\rho_p$ (kg/m$^{3}$) & $e_{n,d}$ & $e_{t,d}$ & $\mu_{c}$ & $\mu_{c,wet}$ \\
   \midrule
   Steel & 7800 & 0.97 & 0.34 & 0.11 & 0.02 \\
   Glass & 2540 & 0.97 & 0.39 & 0.10 & 0.15 \\ \bottomrule
   \end{tabular}}
   \label{table:oblique}
\end{table}


Comparison of the normalized incidence and the rebounding angle of the oblique collision between the simulation results obtained by the MTSA and the experimental data for steel and glass particles are presented in figure~\ref{fig:oblique}. It is seen that the simulation results by the MTSA are in good agreement with the experimental data of \cite{joseph2004oblique} in the entire range of incidence angles.

\begin{figure}[htbp]
\centering
\hspace{-16mm}
\includegraphics[width=8cm]{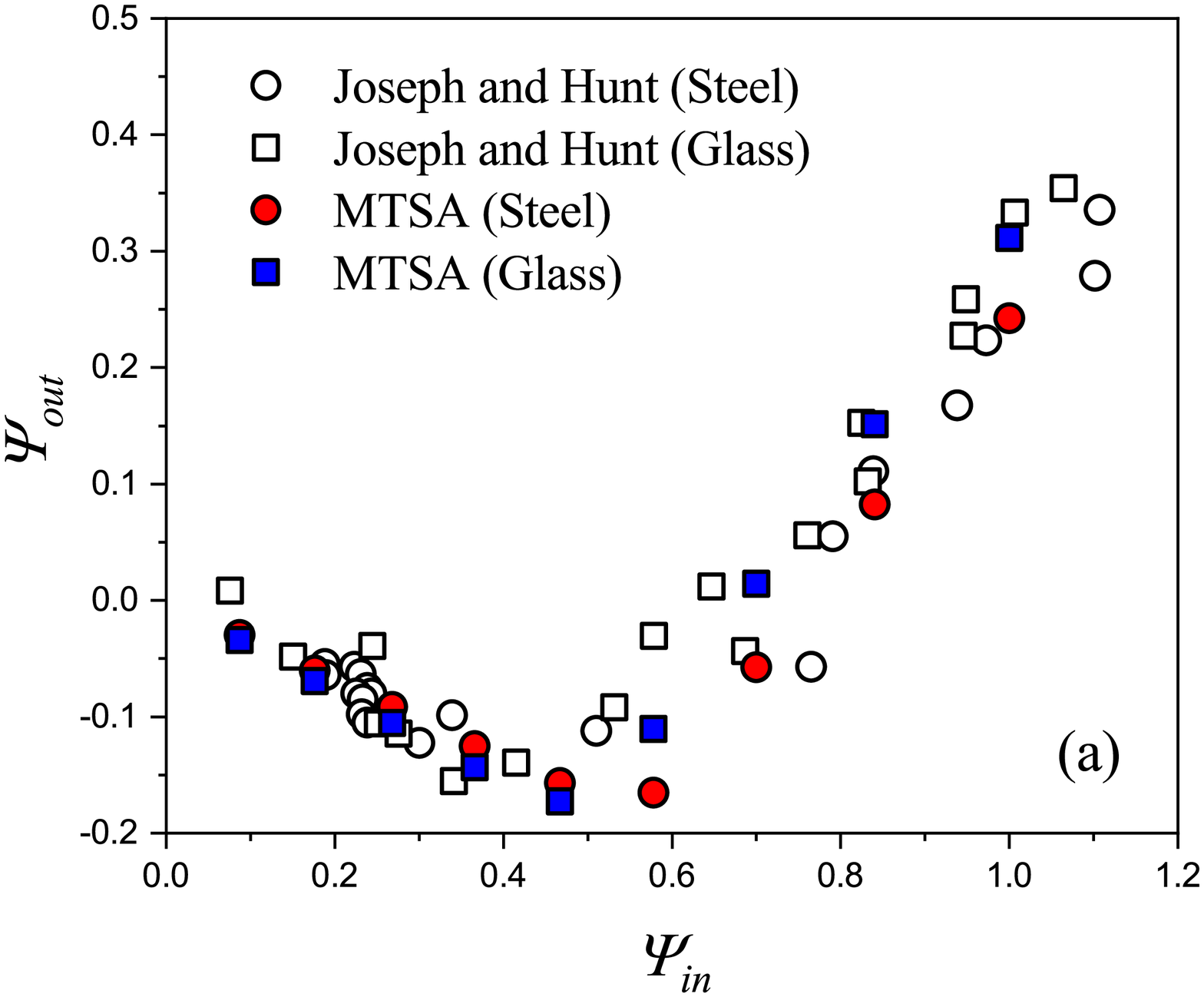}
\hspace{-10mm}
\includegraphics[width=8cm]{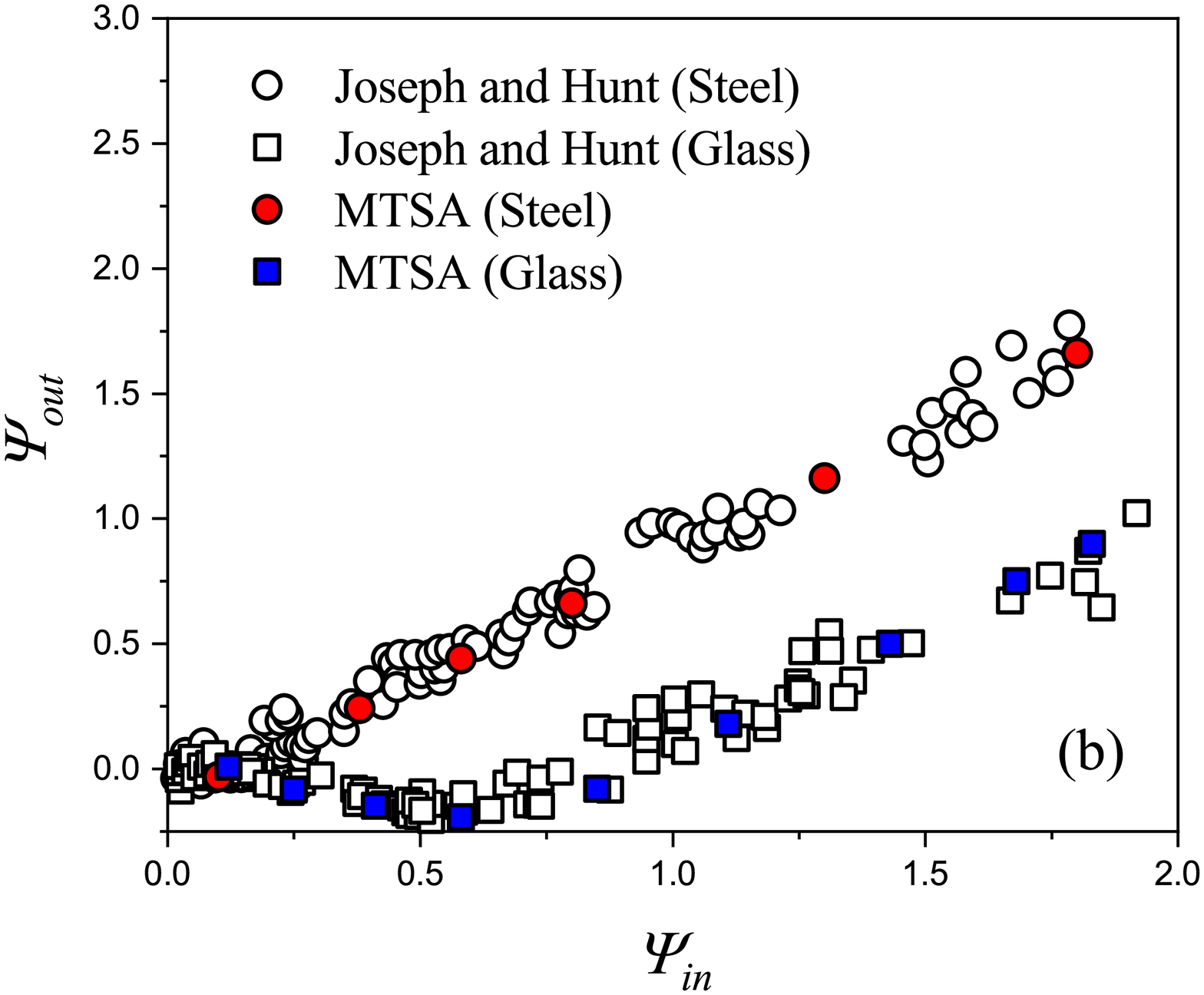}
\hspace{-16mm}
\caption{Dependence between the effective angles of the incidence and rebounding: (a) without viscous fluid and (b) with viscous fluid.}
\label{fig:oblique}
\end{figure}

\subsection{Simulation of turbulent flow over an erodible sediment bed}

Now we applied the MTSA with $R_f=0.5, R_i=4, R_m=40$ to simulate a turbulent flow over an erodible sediment bed. The turbulent flow was driven by a constant pressure gradient that is balanced by the shear stress on the sediment bed. Periodic boundary conditions were imposed in the streamwise and spanwise boundaries of the computational domain, no-slip boundary conditions were applied on the bottom surface of the domain and the particle surfaces, and free-slip boundary conditions were imposed on the top surface of the domain. The sediment bed was laid on the smooth bottom wall of the channel under gravity. The sediment bed consisted of $N_p=7200$ particles with four layers, as shown in Figure~\ref{fig:channel}. In the present study, the sediment bed was generated by a sedimentation simulation for particles settling under gravity while turning off the hydrodynamic force, similar to \cite{kidanemariam2014interface}.

\begin{figure}[htbp]
\centering
\includegraphics[width=0.6\textwidth]{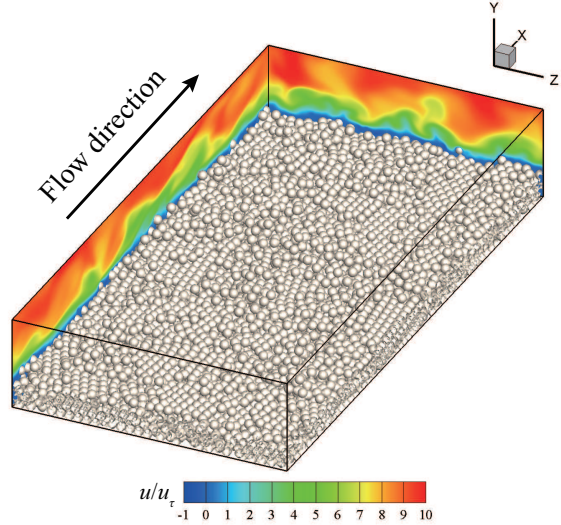}
\caption{Instantaneous snapshot of the turbulent flow over the erodible sediment bed. The flow field is colored by the value of the non-dimensional streamwise velocity $u/u_{\tau}$.}
\label{fig:channel}
\end{figure}

{For the parameters of the carrier phase, the size of the computational domain was $L_x \times L_y \times L_z=(6 \times 1 \times 3)H$ with a uniform Cartesian grid of $N_x \times N_y \times N_z=1200 \times 200 \times 600$ grid numbers, where $H$ is the computational domain height in the $y$ direction. Here, $x$, $y$, and $z$ denote the streamwise, wall-normal and spanwise directions, respectively.
The friction Reynolds number is defined as $Re_\tau=u_{\tau}H_e/\nu=94$, where $u_{\tau}=\sqrt{\tau_b/\rho_f}$ is the friction velocity, $\tau_b$ is the mean shear stress at an effective sediment-bed height $y_b$, and $H_e=H-y_b$ is the effective boundary-layer height. The bulk Reynolds number $Re_b=U_b H_e/\nu$ are 569, 582, and 614, for case MTSA, SCTA1, and SCTA2, respectively, where $U_b=(1/H)\int_{0}^{H}\left \langle u \right \rangle dy$ is the bulk velocity.} The effective bed height $y_b$ was determined at the height where the mean particle porosity $\left \langle\phi_p\right \rangle=0.1$  \cite[]{kidanemariam2014direct}, which yielded $y_b=0.35H$. The Shields number is ${\it \Theta} =u_{\tau}^2/((\rho_p / \rho_f- 1)gD_p)=0.12$, which is above the critical Shields number ${\it \Theta_c}=0.047$ \cite[]{meyer1948formulas,wong2006reanalysis}, and the corresponding Galileo number is $Ga=\sqrt{(\rho_p/ \rho_f-1)gD_p^3}/\nu =42$. The gravitational acceleration is $g=9.81$ m/s$^2$.

For the parameters of the dispersed phase, the density ratio between the particle and the fluid is $\rho_p/\rho_f=2.65$. The dimensionless particle diameter is $D_p/H=0.1$. 
Referring to the recent particle-resolved sediment transport simulations \cite[]{jain2021impact,kidanemariam2022open}, the grid resolution was chosen as $D_p/\Delta x=20$, which is sufficient for sediment transport. The corresponding Eulerian grid size is $\Delta x^+=\Delta y^+=\Delta z^+=0.47<1$, which is able to fully resolve the turbulent channel flow. The superscript $+$ indicates quantities normalized in viscous units (by $u_\tau$ and $\nu_f$).
The particle diameter in wall units was $D_p^+= D_p/(\nu_f/u_\tau) = 14.5$. The mean volume fraction of the dispersed phase was 21\%. 

For the parameters in the collision model, the normal restitution coefficient is $e_{n,d}=0.97$, the tangential restitution coefficient is $e_{t,d}=0.39$, and the friction coefficient is $\mu _c=0.15$; these were chosen according to the material properties of sand particles \cite[]{joseph2004oblique}.

The simulations were carried out on the Tianhe-2A supercomputer with 50 nodes (1200 cores). Each node has two Intel Xeon E5-2692 cores and 64 GB of memory.

\subsubsection{Effect of particle stiffness}

Cases MTSA, SCTA1, and SCTA2 with different particle stiffness listed in table~\ref{table:cases_bed} were designed to investigate the effect of particle stiffness on the turbulence and particle statistics of the sediment transport simulation. For the MTSA, $\Delta t_f=T_{c,min}/R_f=7.6\times 10^{-6}$ s was chosen to ensure that it can resolve mostly all collisions with different impact velocities, where $T_{c,min}$ is the minimum physical collision time obtained by equation (\ref{eqn:tc}) with the maximum impact velocity $u_{in,max}=8u_{\tau}$ based on a similar particle-resolved simulation by \cite{vowinckel2014fluid}. The physical stiffness $E=55$ GPa for the glass particle was applied in the MTSA, while the particle stiffness was reduced in cases SCTA1 and SCTA2 by stretching the collision time with the stretching coefficient $N=10$ \cite[]{kempe2012collision}. In case SCTA2, the particle stiffness was further reduced by increasing $T_c$ compared with case SCTA1. 

\begin{table}[H]
{
   \centering
   \caption{Parameters used in the simulations of the turbulent flow over the erodible sediment bed. The CFL number was calculated by $CFL=\max_{i=1,3}\left | u_i \right | \Delta t/\Delta x$.}
   \setlength{\tabcolsep}{1.6mm}{
   \begin{tabular}{cccccc}
   \toprule
   Case & $\Delta t_f$ (s) & $T_c$ (s) & $E$ (GPa) & CFL & Notes\\
   \midrule
   MTSA & $7.6\times 10^{-6}$   &  Eq.(\ref{eqn:tc}) & 55 & 0.113 & Benchmark case \\
   SCTA1  & $7.6\times 10^{-6}$  & $10\Delta t_f$  & 0.03 & 0.115 & Same $\Delta t_f$ as case MTSA \\
   SCTA2  & $7.6\times 10^{-5}$   & $10\Delta t_f$  & $10^{-4}$ & 1.21 & Same $N$ as case SCTA1  \\ \bottomrule
   \end{tabular}}
   \label{table:cases_bed}}
\end{table}

In our simulations, once the volume flux of the turbulent flow was statistically stationary, an averaging procedure described in Appendix C was applied to the flow and particle variables. In total, 250 snapshots of the flow field and 2500 snapshots of the particles during a time period of $15H/u_{\tau}$ were collected for time averaging. In order to distinguish whether the error came from reducing particle stiffness or statistical uncertainty, the time period $15H/u_{\tau}$ was equally divided into two periods of $7.5H/u_{\tau}$. The results averaged over these two time periods were defined as the lower and upper boundaries (statistical uncertainty range) of the simulation, as shown by the shaded region in figures \ref{fig:phi} to \ref{fig:pu}. The lines/symbols in the following figures are the results averaged over a total time period of $15H/u_{\tau}$. In the following, the turbulence and particle statistics are compared between cases MTSA, SCTA1 and SCTA2. The wall-normal coordinate $Y=y-y_b$ is adopted and normalized by $H_e$.

Figure~\ref{fig:phi} shows the profiles of the mean particle porosity (planar density) obtained in each $x-z$ plane. The operator $\left \langle \cdot \right \rangle$ indicates an average over $x-z$ plane (not a slab) and time. With this averaging operation, the theoretical maximum limit of particle porosity in dense packing is 0.906 (the largest planar density in the face-centered cubic structure), which is much larger than 0.74 of the largest particle volume fraction in three-dimensional space. 
Following \cite{kidanemariam2014interface}, the mean particle porosity was calculated by equation (\ref{eqn:b1}) in Appendix C. The four peaks under the grey line indicate that the sediment bed consists of four layers of particles. The peak heights correspond to the mean center height of each layer of particles \cite[]{vowinckel2014fluid}. 
{It should be noted that the particle porosity \cite[]{kidanemariam2014interface,vowinckel2014fluid}, which is the particle fraction in the $x-z$ plane rather than the volume fraction \cite[]{rao2019coarse}, was employed since the particles are larger than the grid. If we chose a $y$-slab with a height $y_s$ larger than $D_p$, the mean volume fraction $\Phi$ can be calculated through the integration of the particle porosity as $\Phi=(1/y_s)\int_{y_0}^{y_0+y_s}\left \langle \phi _p \right \rangle dy$, where $y_0$ is the lower boundary of the $y$-slab. The mean volume fraction of the entire sediment bed is 0.58 with $y_0=-y_b/H_e$ and $y_s=y_b/H_e$ in the present simulation, which is lower than the theoretical maximum limit of dense packing of 0.74 \cite[]{rao2019coarse}. The maximum particle porosity is 0.832, which is also lower than the theoretical maximum limit of 0.906 for dense packing.}  As shown in figure~\ref{fig:phi}, the particle porosity above $Y/H_e=0.15$ is approximately 0.01, which indicates that particles seldom jumped above $Y/H_e=0.15$. Therefore, particle samples above $Y/H_e=0.15$ within the averaging time window may not be sufficiently large to obtain convergent statistics. This problem can also be seen in figures 9 and 10 of \cite{ji2014saltation}. Therefore, we only focus on the near-wall region ($0<Y/H_e<0.15$) in the following analysis.

\begin{figure}[htbp]
\centering
\includegraphics[width=0.7\textwidth]{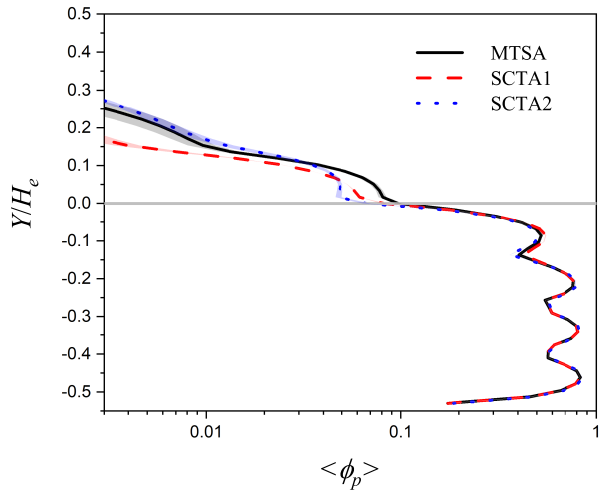}
\caption{Profiles of the mean particle porosity. The grey line indicates the effective height of the sediment bed. {The shaded regions of different colors indicate the ranges of the statistical errors in different cases.}}
\label{fig:phi}
\end{figure}

\textcolor{black}{For comparing the present simulation results with available experimental or numerical results, we cannot find any particle-resolved simulation studies or experimental measurements under similar conditions. In many experiments, both the size of particles and the Reynolds number of turbulent flow are much larger than the present one. In addition, as the particle volume fraction is usually very high near a sediment bed, it is very challenging to distinguish tracers and inertial particles in experiment \cite[]{brandt2022particle}. As a result, near-wall turbulence statistics are almost impossible to be extracted. We cannot find any experimental results of near-sediment-bed turbulence statistics in the literature also. A measurable quantity of sediment transport rate can be compared,  which is defined as $q=(\pi D_p^3/6) \textstyle \left(\sum_{N_p} \langle{u}_p \rangle \right)/\left ( L_x L_y L_z \sqrt{(\rho_p/\rho_f-1)gD_p^3} \right )$ in a non-dimensional form, where $\langle{u}_p\rangle$ is mean particle velocity. The non-dimensional sediment transport rates in cases MTSA, SCTA1, and SCTA2 are $q_{MTSA}=0.0461$, $q_{SCTA1}=0.0568$, and $q_{SCTA2}=0.1396$, respectively. The well-known empirical formula of \cite{wong2006reanalysis} gives $q^*=4.93\left ( \it \Theta-\it \Theta_c \right ) ^{1.6}$, where $\it \Theta_c$ is the critical Shields number. Following \cite{wong2006reanalysis}, if ${\it \Theta_c}=0.047$ is employed, we can get $q^*=0.0748$. It seems that the prediction by SCTA1 is the most accurate, but we argue that the formula of \cite{wong2006reanalysis} was fitted with data at much larger flow-submerge ($Y/D_p\sim 40$) and flow Reynolds numbers. The particle stability increases with decreasing flow depth under constant shear velocity, which is likely related to the suppression of the energetic large-scale turbulent motions due to the limited separation between flow depth and roughness length scales (smaller flow Reynolds number) \cite[]{cameron2020entrainment}. Therefore, we believe that the actual sediment transport rate under the present simulation condition should be much smaller than $q^*=0.0748$.}

In the near-wall region, the statistical uncertainties of the particle porosity are negligible compared with the errors caused by reducing particle stiffness. The present results show that when particle stiffness is reduced, the mean particle porosity is underestimated, and the underestimation increases as particle stiffness decreased, as displayed in figure~\ref{fig:phi}. The relative difference between the results of case SCTA1 and case MTSA and between case SCTA2 and case MTSA are 24\% and 40\% at $Y/H_e=0.02$, respectively. Thus, fewer particles are entrained in cases SCTA1 and SCTA2. 
This may be caused by excluding the hydrodynamic force during the collisions, which significantly ignores the effect of turbulence on particle entrainment with a time scale smaller than $T_c$. \cite{vowinckel2016entrainment} found that 96.5\% of the entrained particles were triggered by particle collisions. Therefore, the effects of turbulence may be substantially underestimated if $T_c$ is greatly stretched, and fewer particles would be entrained in the near-wall region.}

Figure~\ref{fig:u} (a) shows the streamwise mean flow velocity profile. In the near-wall region, the statistical uncertainties of the streamwise mean flow velocity are negligible. The results show that when particle stiffness is reduced by the SCTA, the streamwise mean flow velocity is overestimated compared with the MTSA, and the difference increases as particle stiffness decreases. The relative difference between the results of case SCTA1 and case MTSA and the results of case SCTA2 and case MTSA are 6\% and 35\% at $Y/H_e=0.02$, respectively. This is probably because fewer particles are entrained from the bed in cases SCTA1 and SCTA2, which exert a smaller drag force on the flow. 
The profiles of the root-mean-squared (r.m.s) fluctuating flow velocities in the three directions are shown in figure~\ref{fig:u} (b). In the near-wall region, it can be seen that the wall-normal and spanwise fluctuating flow velocities are close between cases SCTA1, SCTA2 and MTSA. For the streamwise fluctuating flow velocity, the statistical uncertainty is smaller than the uncertainty by the collision model. In fact, the r.m.s streamwise fluctuating flow velocity is underestimated by the SCTA compared with the MTSA, and the underestimation increases as particle stiffness decreases. This may be because particles are more difficult to be entrained in cases SCTA1 and SCTA2, which imposes fewer disturbances to the fluid in the near-wall region. 
Figure~\ref{fig:u} (c) shows the profiles of the total shear stress $\left ( -\left \langle u^\prime v^\prime \right \rangle+ \nu_f {\rm d}\langle{u}\rangle/{\rm d}y  \right )/u_{\tau}^2 $, where $u^\prime$ and $v^\prime$ are the streamwise and wall-normal flow velocity fluctuations, respectively. The total shear stress nicely follows a linear profile in the outer flow indicating that the turbulent flow has been fully developed. The total shear stress is underestimated by the SCTA compared with the MTSA in the near-wall region, and the underestimation increases as particle stiffness decreases.

\begin{figure}[htbp]
\centering
\hspace{-16mm}
\includegraphics[width=8cm]{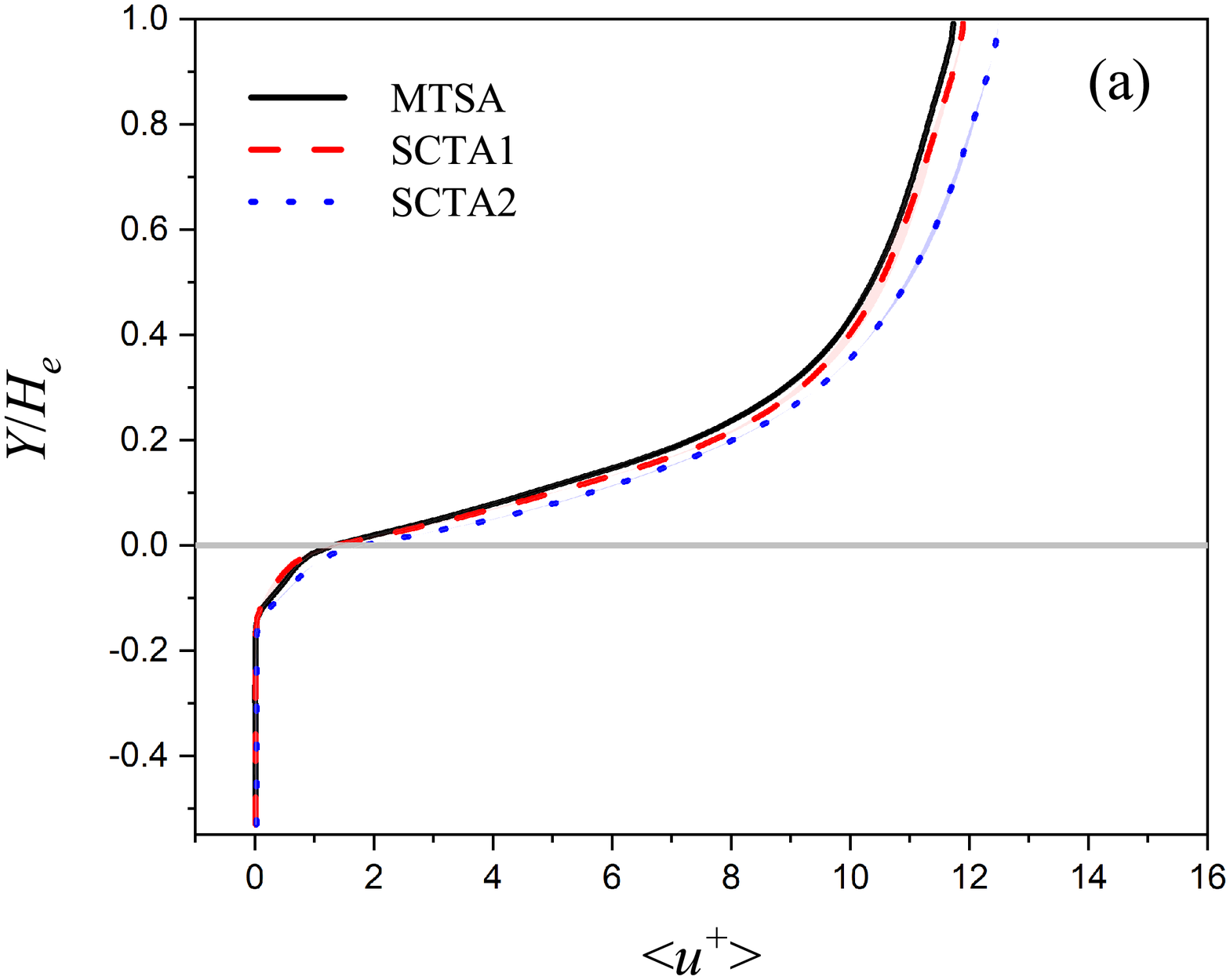}
\hspace{-10mm}
\includegraphics[width=8cm]{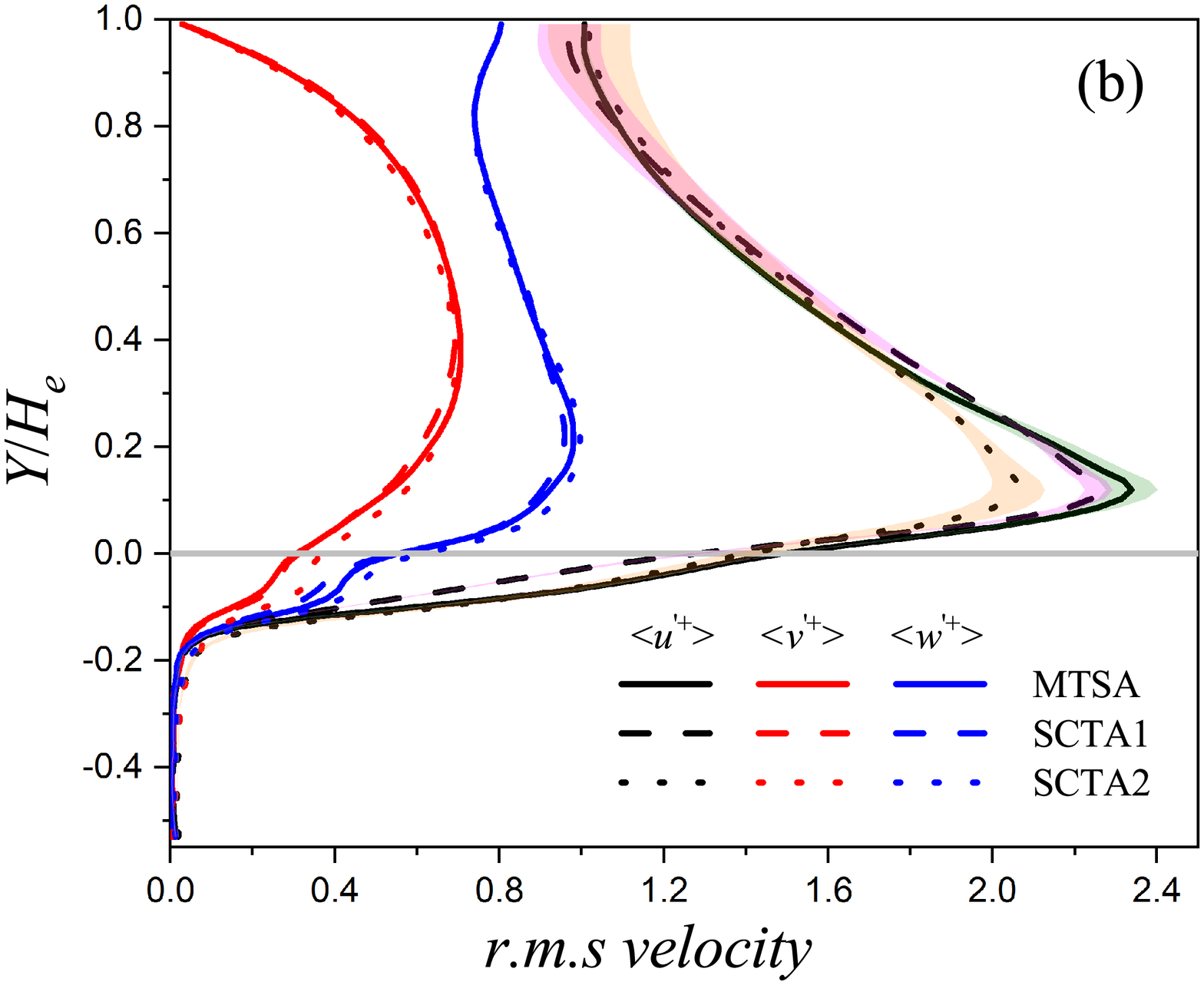}
\hspace{-16mm}

\centering
\includegraphics[width=8cm]{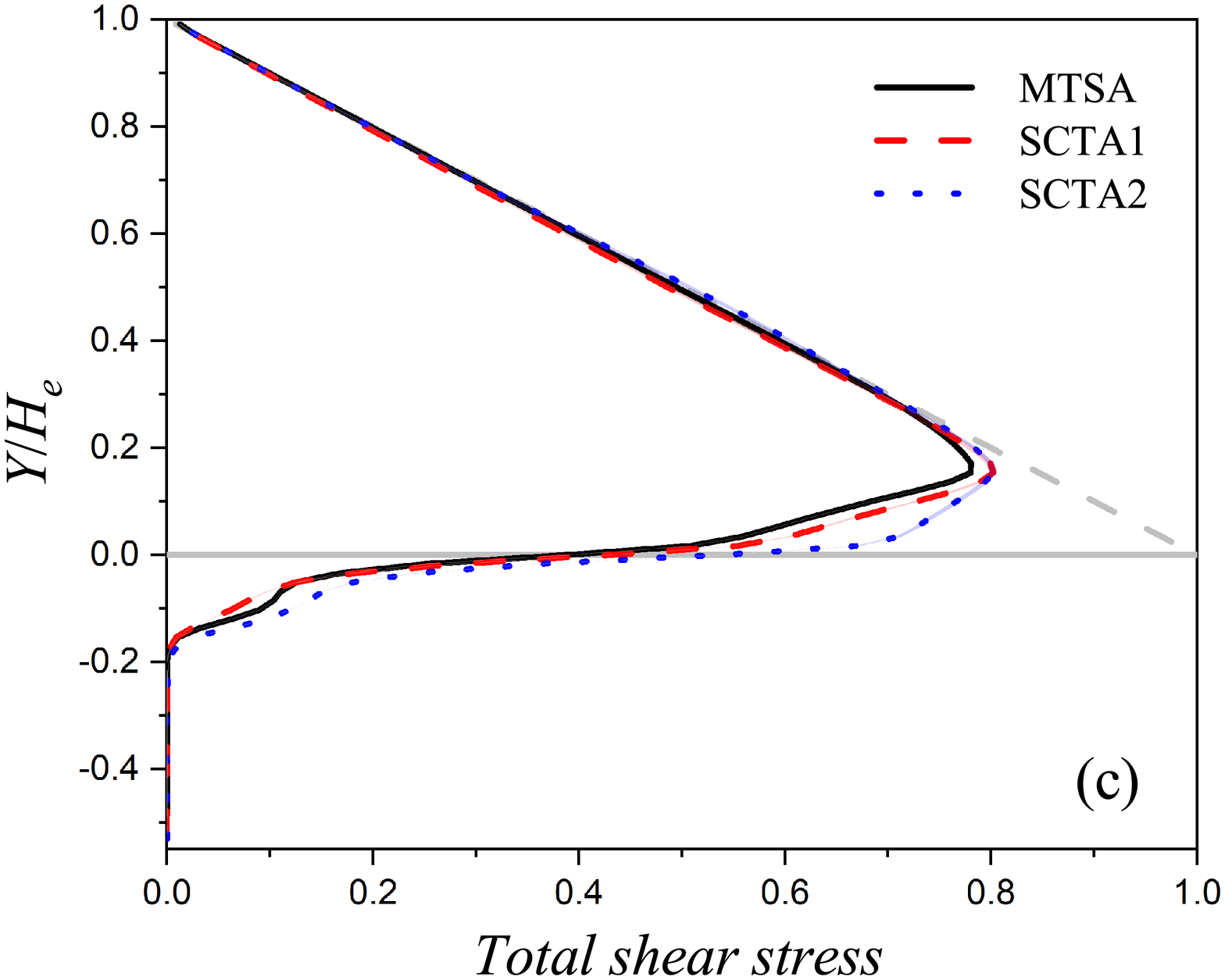}
\caption{Turbulence statistics in sediment transport simulations: (a) the streamwise mean flow velocity and (b) the root-mean-squared fluctuating flow velocities, and (c) the total shear stress $\left ( -\left \langle u^\prime v^\prime \right \rangle + \nu_f {\rm d}\langle{u}\rangle/{\rm d}y  \right )/u_{\tau}^2$. The grey line indicates the effective height of the sediment bed. The grey dash line indicates $1-Y/H_e$. The shaded regions indicate the ranges of statistical uncertainties.}
\label{fig:u}
\end{figure}

{Figure~\ref{fig:pu} (a) shows the profiles of the mean particle velocities. It is seen that the profiles of the wall-normal and spanwise mean particle velocities are quite close between cases SCTA1, SCTA2 and MTSA. In the near-wall region, the statistical uncertainties of the streamwise mean particle velocity are much smaller than those by reducing particle stiffness. This result definitely shows that when particle stiffness is artificially reduced, the streamwise mean particle velocity will be overestimated, and the difference increases as particle stiffness decreases. The relative difference of the streamwise mean particle velocity between cases SCTA1 and MTSA and between cases SCTA2 and MTSA can reach up to 87\% and 325\% at $Y/H_e=0.02$, respectively. This may be because the mean flow velocity of cases SCTA1 and SCTA2 are higher than that of case MTSA, thus the drag force acting on the entrained particles increases to accelerate particles to higher speeds in cases SCTA1 and SCTA2. Figure~\ref{fig:pu} (b) displays the profiles of the mean particle angular velocities. It can be seen that the profiles of the streamwise and wall-normal mean particle angular velocities are close between cases SCTA1, SCTA2 and MTSA. In the near-wall region, the statistical uncertainties of the spanwise mean particle angular velocity are much smaller than those by reducing particle stiffness. The results in the near-wall region demonstrate that when particle stiffness is reduced by the SCTA, the spanwise mean particle angular velocity may be overestimated, and the difference increases as particle stiffness decreases. The relative difference of the spanwise mean particle angular velocity between cases SCTA1 and MTSA and between cases SCTA2 and MTSA can reach up to 62\% and 150\% at $Y/H_e=0.02$, respectively.
This may be because the mean flow velocity gradient is larger in cases SCTA1 and SCTA2.

\textcolor{black}{It should be noted that a version of SCTA similar to \cite{costa2015collision} is adopted here to be compared with the MTSA, which incorporates a two-parameter lubrication model, a linear spring-dashpot system, and substeps for particle collisions and motions. It is found that there are evident differences between the results of the sediment transport case using the SCTA and MTSA for the statistics especially particle statistics if the particle stiffness is greatly reduced. The differences increase as particle stiffness decreases (stretched collision time increases). Using different versions of SCTA with different model details like the lubrication model, linear vs. non-linear spring-dashpot, time integration among others may affect the results. However, we believe that the assumption of excluding the hydrodynamic force during particle collisions and the much longer collision time in the SCTA are the major reasons for the differences between the results using the SCTA and MTSA, as the other modeling and numerical aspects are the same in our code.}

\begin{figure}[H]
\centering
\hspace{-16mm}
\includegraphics[width=8cm]{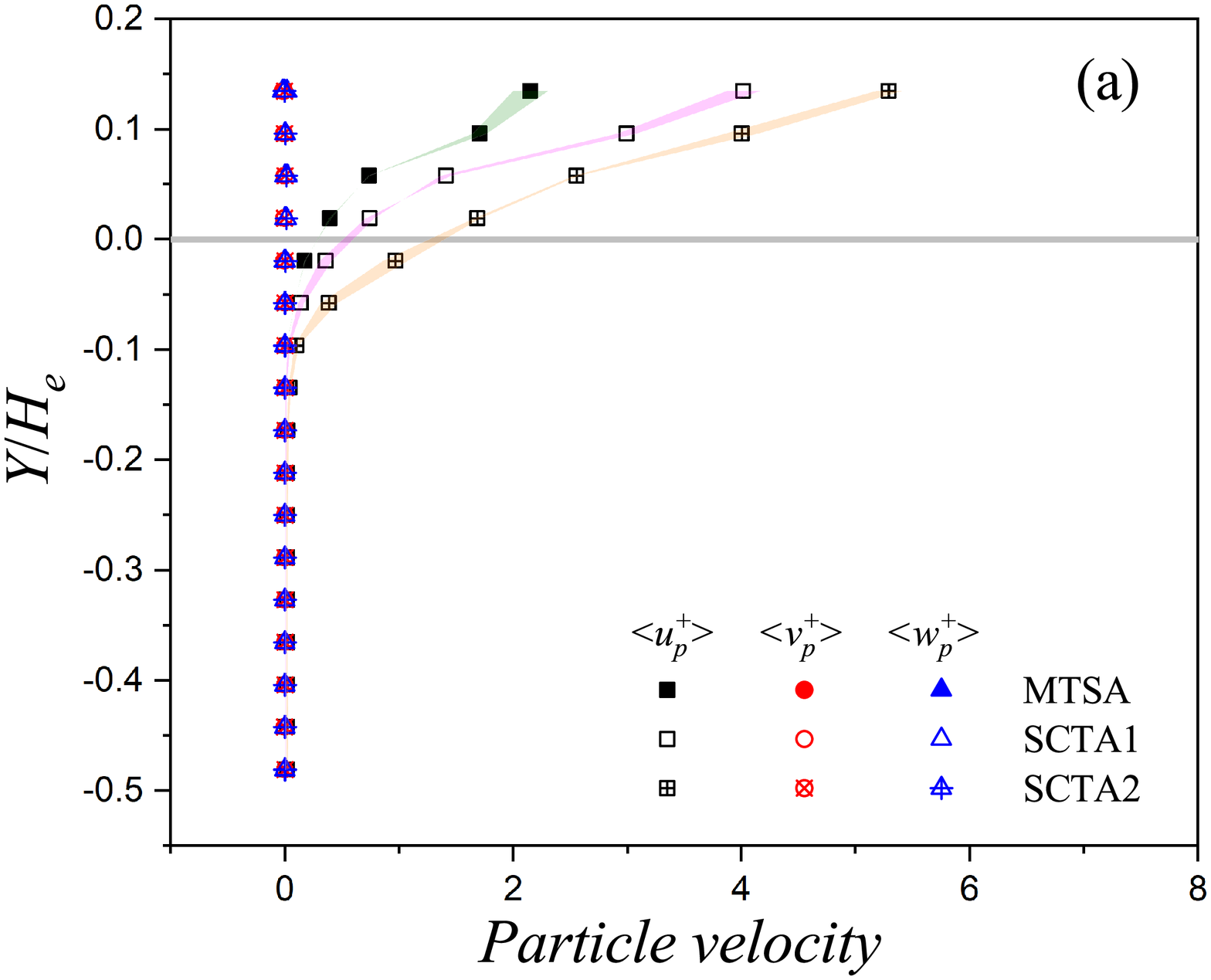}
\hspace{-10mm}
\includegraphics[width=8cm]{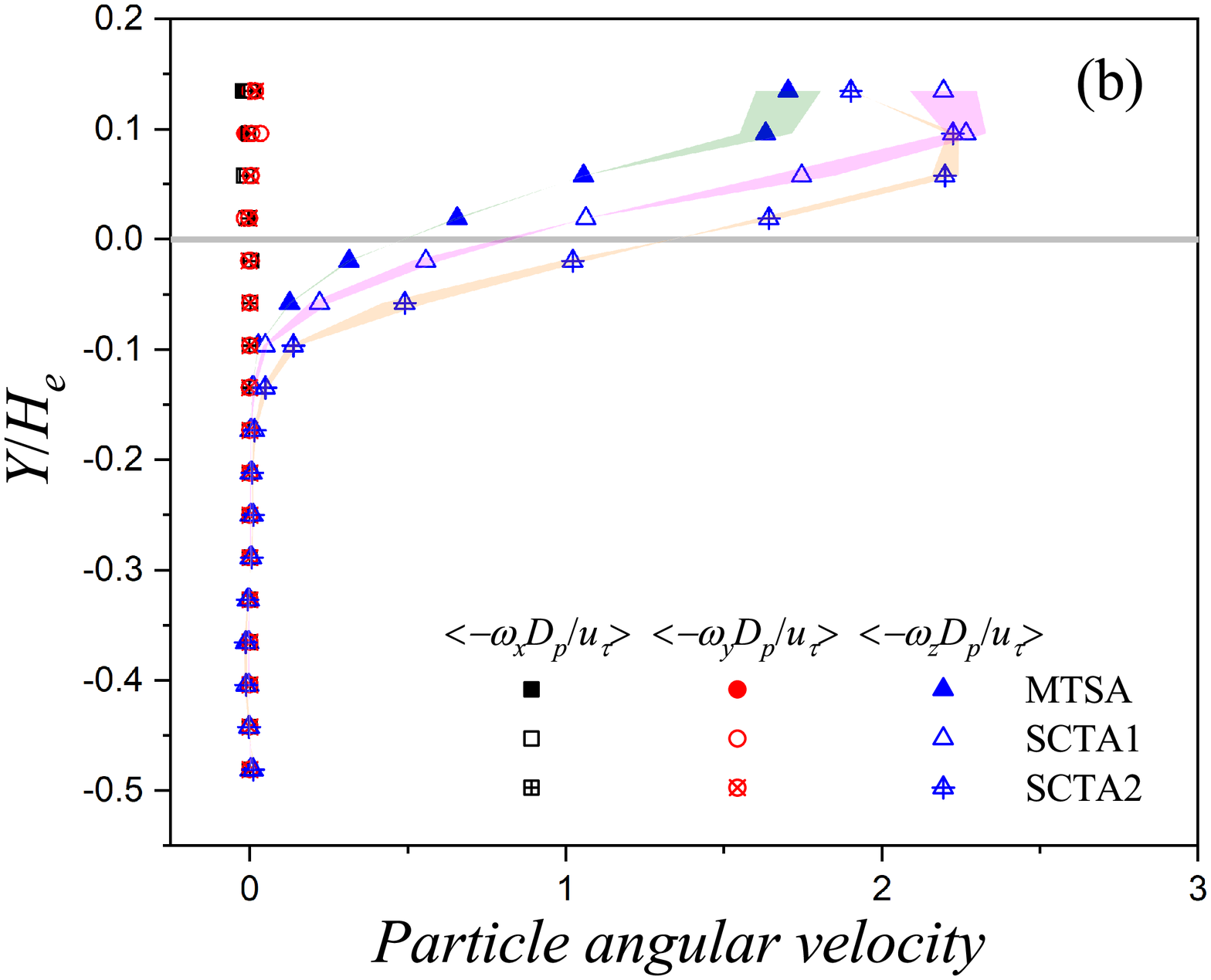}
\hspace{-16mm}
\caption{Particle statistics in sediment transport simulations: (a) the mean particle velocity and (b) the mean particle angular velocity. The grey line indicates the effective height of the sediment bed. {The shaded regions indicate the ranges of the statistical uncertainties.}}
\label{fig:pu}
\end{figure}

\subsubsection{The computational cost of the MTSA}
The computational cost of a complete simulation of sediment transport using the traditional soft-sphere model with a fine time step is very high. Therefore, we only ran 1000 time steps in case REF to obtain its computation performance. 
Here we compare the ratio $(T/\Delta t_{f})/(T_{REF}/\Delta t_{f,REF})$ between different algorithms, which represents the computational time ratio for advancing a unit flow time in the simulations, where $T$ is the elapsed clock time spent over 1000 flow time steps. Furthermore, $\Delta t_{f}/\Delta t_{f,REF}=16$ (\emph{i.e.}, $R_f=0.5$) was used in the following simulations, and the computational time ratios of $\Delta t_{f}/\Delta t_{f,REF}=4$ and 8 were also tested. As shown in figure~\ref{fig:time}, the computational time ratio of the MTSA is close to the ideal ratio $\Delta t_{f,REF}/\Delta t_{f}$. The computational efficiency can be increased by an order of magnitude using the MTSA with $R_f=0.5, R_i=4, R_m=40$ compared with the traditional soft-sphere model. This is because the pressure Poisson equation is not solved in the inserted substeps, which significantly reduces the computational cost. {It is also seen that the computational efficiency of the MTSA is not degraded too much with that of the SCTA.}

\begin{figure}[H]
\centering
\includegraphics[width=0.7\textwidth]{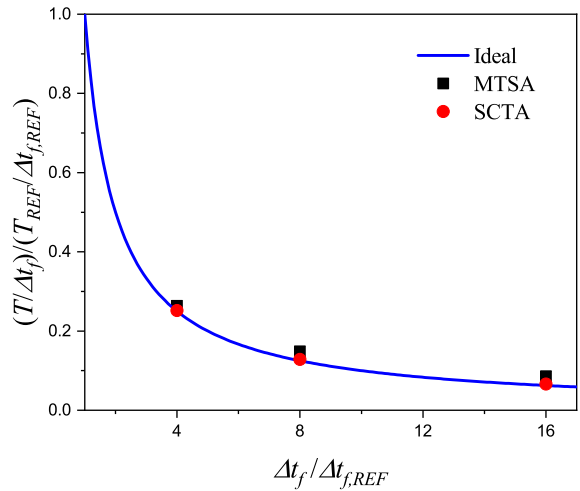}
\hspace{0in}
\caption{Variation of the computational time ratio with the time step ratio. The blue line is the ideal computational time ratio, which equals $\Delta t_{f,REF}/\Delta t_{f}$. }
\label{fig:time}
\end{figure}

\section{Conclusions}

In the present study, a multiple-time-step integration algorithm (MTSA) has been proposed for particle-resolved simulations with physical collision time. The MTSA adopts three different time steps to resolve fluid flow, fluid-particle interaction and particle collision. We have successfully demonstrated the accuracy and convergence of the MTSA in particle-wall and particle-particle collision problems. It is shown that the MTSA can yield excellent agreement with experiments and benchmark simulations using the physical collision time and on a coarse grid ($D_p/\Delta x=O(10)$). {It also overcomes the main drawback of a version of SCTA similar to \cite{costa2015collision}, in which an artificially stretched collision time is introduced, which may lead to the collision time being nonphysical and an underestimation of the hydrodynamic force and particle entrainment in sediment transport.}

{The impact of the parameters $R_f$, $R_i$, and $R_m$ in the MTSA has been systematically analyzed and discussed. We found that $R_i=4$ and $R_m=40$ are universally suitable for particle-wall and particle-particle collisions over a wide range of particle Stokes numbers. The absolute error of $e_n$ can be controlled to be within about 0.03 for different $R_f$ values. After determining the optimal parameters of the MTSA, we applied the MTSA to simulate a turbulent flow over an erodible sediment bed. The statistics of turbulence and particles by the MTSA and a version of SCTA similar to \cite{costa2015collision} were compared to investigate the effect of particle stiffness. The results demonstrated that if particle stiffness is artificially reduced by the SCTA, the number of entrained particles over the erodible sediment bed will be evidently underestimated compared with those by the MTSA, which will mainly impact the streamwise mean flow velocity, streamwise fluctuating flow velocity, streamwise mean particle velocity and spanwise mean particle angular velocity in the near-wall region. The prediction differences increase as particle stiffness decreases. 
Furthermore, the computation efficiencies between the MTSA and the traditional soft-sphere model were compared. The results indicated that the MTSA can substantially reduce the computational cost by up to one order of magnitude, and make the computational cost acceptable for particle-resolved simulations with physical collision time. As such, the proposed MTSA can be a valuable algorithm for generating high-fidelity and accurate data efficiently in particle-resolved simulations.}

For future studies, we plan to perform more intensive particle-resolved simulations of sediment transport, including higher Reynolds numbers and particle/fluid density ratios.
For example, the Reynolds number $Re_\tau$ is an important non-dimensional parameter that may also affect the prediction difference between the simulations using the SCTA and MTSA, thus simulations at higher $Re_\tau$ should be valuable. Simulating sediment transport using the adaptive mesh refinement (AMR) technique to reduce computational costs \cite[]{zeng2022subcycling,zeng2022parallel} is also within our future plans.

\section{Acknowledgments}
This work was financially supported by grants from the National Natural Science Foundation of China (Nos. 92052202 and 11972175) and the National Numerical Windtunnel project. The computation was performed at the Tianhe-2A supercomputer of the National Supercomputer Center in Guangzhou. The authors report no conflict of interest.

\section*{Appendix A. Definitions for particle-particle and particle-wall collision}

The variables used in particle collision are defined below. Some definitions depend on whether the interaction is between a particle $p$ and a wall (P-W) or between a particle $p$ and a particle $q$ (P-P). The variables are defined as follows:

$\bm{n}$ - normal unit vector of contact
\begin{alignat}{2}
&\bm n=\frac{\bm x_q-\bm x_p}{\left | \bm x_q-\bm x_p \right | },  \quad &&\text{(P-P)}, \tag{A.1}\\
&\bm n=\frac{\bm x_w-\bm x_p}{\left | \bm x_w-\bm x_p \right | }, &&\text{(P-W)}, \tag{A.2}
\end{alignat}

$\delta_n$ - distance between two surfaces
\begin{alignat}{2}
&\delta _{n}=\left | \bm x_q- \bm x_p \right |-R_p-R_q,  \quad &&\text{(P-P)}, \tag{A.3}\\
&\delta _{n}=\left | \bm x_w- \bm x_p \right |-R_p, &&\text{(P-W)}, \tag{A.4}
\end{alignat}

$\bm u_{cp}$ - relative velocity of contact point
\begin{alignat}{2}
&\bm u_{cp}=\bm u_p-\bm u_q+ R_{p} \bm \omega_p \times \bm n +R_{q} \bm \omega_q \times \bm n, \quad &&\text{(P-P)}, \tag{A.5}\\
&\bm u_{cp}=\bm u_p+R_{p} \bm \omega_p \times \bm n, &&\text{(P-W)}, \tag{A.6}
\end{alignat}

$\bm u_{cp,n}$ - normal component of $\bm u_{cp}$
\begin{alignat}{1}
&\bm u_{cp,n}=(\bm u_{cp} \cdot \bm n) \bm n, \tag{A.7}
\end{alignat}

$\bm u_{cp,t}$ - tangential component of $\bm u_{cp}$
\begin{alignat}{1}
&\bm u_{cp,t}=\bm u_{cp}-\bm u_{cp,n}, \tag{A.8}
\end{alignat}
and $\bm \delta_t$ is the tangential displacement of the contact point.

The direction of the tangential unit vector changes in every time step. Therefore, we need to rotate the displacement from the previous time step onto a plane tangent to $\bm n$. Then $\bm \delta_t$ is calculated to be the same as \cite{biegert2017collision}:
\begin{alignat}{3}
&\widetilde{\bm \delta _{t}} =\bm \delta _{t}^{k-1}-(\bm \delta _{t}^{k-1} \cdot \bm n)\bm n, \tag{A.9}\\
&\widehat{\bm \delta _{t}} =\frac{\left |\bm \delta _{t}^{k-1} \right | }{\left |\widetilde{\bm \delta _{t}} \right |} \widetilde{\bm \delta _{t}},  \tag{A.10}\\
&\bm \delta _{t}^{k}=\widehat{\bm \delta _{t}}+\Delta t \bm u_{cp,t}. \tag{A.11}
\end{alignat}

\section*{Appendix B. Discretization schemes}

The discretization schemes adopted in the MTSA are as follows: an explicit second-order Runge--Kutta (RK2) method is used for fluid flow advancement \cite[]{yang2017numerical,yang2018direct,cui2018sharp,he2022numerical}. At each substep of the RK2 method, the fractional-step method of \cite{kim1985application} is applied to ensure that the flow field is divergence-free. In order to insert enough substeps to ensure that the flow can adapt to the change of the particle motion during collisions and reduce the computational cost, the explicit Euler method is adopted in the substeps for time advancement. The fluid and particle are coupled by the multidirect forcing IB method \cite[]{luo2007full,breugem2012second}. The discretized equations of fluid flow and particle motion in the $k$th Runge--Kutta step are given as follows:

\begin{align}
& \text{do}\ k=1,2 \nonumber \\
& \ \ \hat{\bm u}^{k-1,0}={\bm u}^{k-1}, \nonumber \\
& \ \ \textcolor{red}{do\ i=1,N_i} \nonumber \\
& \ \ \ \ \hat{\bm u}^{k-1,i}=\hat{\bm u}^{k-1,i-1}+\textcolor{red}{\Delta t_i}\left [ \alpha _{k} \hat{\bm H}^{k-1,i-1}-\beta _{k}\left ( \hat{\bm H}^{k-2,i-1}-\frac{1}{\rho }\nabla p^{k-2}\right )\right ], \tag{B.1} \label{eqn:f1}\\
& \ \ \ \ \bm u^{*,0}=\hat{\bm u}^{k-1,i}, \nonumber \\
& \ \ \ \ \bm F_{p,l}^{k-1,i,0}=0, \nonumber \\
& \ \ \ \ \text{do}\ s=1,N_s \nonumber \\
& \ \ \ \ \ \ \bm U_{l}^{*,s-1}=\sum_{ijk}\bm u_{ijk}^{*,s-1}\delta _{d}\left (\bm x_{ijk}-\bm X_{l}^{k-1,i-1,N_m}  \right )\Delta x\Delta y\Delta z, \tag{B.2} \label{eqn:f2}\\
& \ \ \ \ \ \ \bm F_{p,l}^{*,s-1} = \frac{\bm U_p{\left ( \bm X_{l}^{k-1,i-1,N_m}  \right )-\bm U_{l}^{*,s-1} } }{\textcolor{red}{\Delta t_i}}, 
\tag{B.3} \label{eqn:f3}\\
& \ \ \ \ \ \ \bm f_{ijk}^{*,s-1}=\sum_{l=1}^{N_l}\bm F_{p,l}^{*,s-1}\delta _{d}\left (\bm x_{ijk}-\bm X_{l}^{k-1,i-1,N_m}  \right )\Delta V_l, \tag{B.4}\\
& \ \ \ \ \ \ \bm u^{*,s}=\bm u^{*,s-1}+\left(\alpha_k-\beta_k\right)\textcolor{red}{\Delta t_i}\bm f^{*,s-1}, \tag{B.5} \label{eqn:m1}\\
& \ \ \ \ \ \ \bm F_{p,l}^{k-1,i,s}=\bm F_{p,l}^{k-1,i,s-1}+\left(\alpha_k-\beta_k\right)\bm F_{p,l}^{*,s-1}, \tag{B.6} \label{eqn:f4}\\
& \ \ \ \ \text{enddo} \nonumber \\
& \ \ \ \ \hat{\bm u}^{k-1,i}=\bm u^{*,N_s}, \nonumber\\
& \ \ \ \ \bm u_{p}^{k-1,i,0}=\bm u_{p}^{k-1,i-1,N_m}, \nonumber \\
& \ \ \ \ \bm x_p^{k-1,i,0}=\bm x_p^{k-1,i-1,N_m}, \nonumber \\
& \ \ \ \ \bm \omega_{p}^{k-1,i,0}=\bm \omega_{p}^{k-1,i-1,N_m}, \nonumber \\
& \ \ \ \ \textcolor{blue}{do\ p=1,N_m} \nonumber \end{align}
\begin{align}
& \ \ \ \ \ \ \bm u_{p}^{k-1,i,p}= \bm u_{p}^{k-1,i,p-1}+\left(\alpha_k-\beta_k\right) \Bigg\{ -\frac{\textcolor{blue}{\Delta  t_p}}{V_p}\frac{\rho_f}{\rho_p}\sum_{l=1}^{N_l}\bm F_{p,l}^{k-1,i,N_s}\Delta V_l+ \nonumber \\
& \ \ \ \ \ \ \frac{1}{V_p} \frac{\rho_f}{\rho_p}\left ( \left\{\int_{V_p}\bm udV\right\}^N-\left\{\int_{V_p}\bm udV\right\}^{N-1}  \right )+\textcolor{blue}{\Delta t_p}\left ( 1-\frac{\rho_f}{\rho_p}  \right )\bm g+ \nonumber \\
& \ \ \ \ \ \ \textcolor{blue}{\Delta t_p} \frac{\bm F_{p,lub}^{k-1,i,p-1}}{\rho_p V_p}+\textcolor{blue}{\Delta t_p} \frac{\bm F_{p,col}^{k-1,i,p-1}}{\rho_p V_p},  \Bigg\} \tag{B.7} \label{eqn:f5}\\
& \ \ \ \ \ \ \bm x_p^{k-1,i,p}=\bm x_p^{k-1,i,p-1}+\left(\alpha_k-\beta_k\right)\textcolor{blue}{\Delta t_p}\bm u_p^{k-1,i,p}, \tag{B.8} \\
& \ \ \ \ \ \ \bm \omega_{p}^{k-1,i,p}= \bm \omega_{p}^{k-1,i,p-1}+\left(\alpha_k-\beta_k\right) \Bigg\{ -\textcolor{blue}{\Delta t_p}\frac{\rho_f}{I_p}\sum_{l=1}^{N_l}\bm r_l\times \bm F_{p,l}^{k-1,i,N_s}\Delta V_l+ \nonumber \\
& \ \ \ \ \ \ \frac{\rho_f}{I_p}\left ( \left\{\int_{V_p}\bm r\times \bm udV\right\}^N-\left\{\int_{V_p}\bm r\times \bm udV\right\}^{N-1}  \right )+\textcolor{blue}{\Delta t_p} \frac{\bm T_{p,col}^{k-1,i,p-1}}{I_p} \Bigg\}, \tag{B.9} \label{eqn:f6}
\end{align}
\begin{align}
& \ \ \ \ \textcolor{blue}{enddo} \nonumber \\
& \ \ \ \ \bm U_p\left(\bm X_l^{k-1,i,N_m}\right)=\bm u_p^{k-1,i,N_m}+\bm \omega_p^{k-1,i,N_m}\times(\bm X_l^{k-1,i,N_m}-\bm x_p^{k-1,i,N_m}), \tag{B.10} \label{eqn:f7}\\
& \ \ \textcolor{red}{enddo} \nonumber \\
& \ \ \nabla^2 p^{k-1}=\frac{\rho_f}{\alpha_k \Delta t_f} \nabla \cdot \hat{\bm u}^{k-1,N_i}, \tag{B.11} \label{eqn:f8}\\
& \ \ \bm u^{k}=\hat{\bm u}^{k-1,N_i}-\frac{\alpha_k \Delta t_f}{\rho_f}\nabla p^{k-1}, \tag{B.12} \label{eqn:f9}\\
& \text{enddo} \nonumber
\end{align}
where the superscripts $N$, $k$, $i$, $p$, and $s$ are the indices of the flow time step, the Runge--Kutta substep, the fluid--particle interaction substep, the particle motion substep, and the IB method iteration substep, respectively. The coefficients in the RK2 scheme are $\alpha_1=1$, $\beta_1=0$, and $\alpha_2=\beta_2=0.5$. 
$N_s$ is the number of iterations to calculate the IB force in the multidirect forcing immersed boundary method\cite[]{luo2007full,breugem2012second}. The idea of this method is to iteratively determine the IB force to make the fluid velocity modified by the IB force better satisfy the no-slip boundary condition on the particle surface. $N_s=3$ is employed following the suggestion of \cite{breugem2012second}.
The subscripts $ijk$ refer to the quantities on an Eulerian grid with indices $(i,j,k)$, and the subscript $l$ refers to the quantity on a Lagrangian point with index $l$. $\hat{\bm u}$ is the fluid velocity in f fluid-particle interaction substep, $\hat{\bm H}=-\nabla\cdot\left(\bm {\hat u \hat u}\right)+\nu_f\nabla^2 \bm{ \hat u}$ is the sum of the convection and viscous terms. $\bm u^*$, $\bm f^*$ and $\bm x$ are the intermediate fluid velocity, volume force, and coordinate on the Eulerian grid, respectively, and $\bm U^*$, $\bm F^*$, $\bm X$ are the corresponding quantities on the Lagrangian grid. The quantities between the Eulerian and Lagrangian grid points are transferred through a regularized Dirac delta function $\delta_d$. Here, we use the three-point regularized Dirac delta function of \cite{roma1999adaptive}. The volume integrals $\int_{V_p}{\bm udV}$ and $ \int_{V_p}{\bm r\times \bm udV}$ are numerically calculated according to \cite{kempe2012improved}. $\Delta V_l$ is the volume of the Lagrangian grid cell, and $\bm U_p$ is the velocity on the particle surface.

\section*{Appendix C. Averaging for flow and particle variables}

\subsection*{C.1. Averaging for flow variables}

Before averaging the flow variables, an indicator function needs to be defined as $\phi _f(\bm x,t)$ to distinguish an Eulerian grid point at a position $\bm x$ that is inside or outside of a particle, following \cite{kidanemariam2014interface}, which is
\begin{equation}
\phi _f(\bm x,t)= \begin{cases} 
1,  & \text{ if } \bm x \ \text{is outside a particle,} \\
0,  & \text{ otherwise.}
\end{cases}
\tag{C.1}
\end{equation}
Based on the indicator function $\phi _f(\bm x,t)$, only the flow data outside of the particle are accounted for averaging as follows:
\begin{equation}
n_f(y_j)=\sum \limits_{n=1}^{N_t}\sum \limits_{i=1}^{N_x}\sum \limits_{k=1}^{N_z}\phi _f(\bm x_{ijk},t^n),
\tag{C.2}
\end{equation}
where $n_f(y_j)$ is the total number of grids in the $x-z$ plane over $N_t$ time steps for the flow statistics at a given height $y_j$. Therefore, the ensemble average of the flow variables $\xi_f(\bm x,t)$ can be defined as
\begin{equation}
\left \langle\xi_f\right \rangle (y_j)=\frac{1}{n_f(y_j)}\sum \limits_{n=1}^{N_t}\sum \limits_{i=1}^{N_x}\sum \limits_{k=1}^{N_z} \xi_f(\bm x_{ijk},t^n)\phi _f(\bm x_{ijk},t^n),
\tag{C.3}
\end{equation}
where the operator $\left \langle \cdot \right \rangle$ indicates the average over an $x-z$ plane and time. The particle porosity $\left \langle\phi_p\right \rangle (y_j)$ can be deduced from $\phi _f(\bm x,t)$:
\begin{equation}
\left \langle\phi_p\right \rangle (y_j)=\frac{1}{N_tN_xN_z}\sum \limits_{n=1}^{N_t}\sum \limits_{i=1}^{N_x}\sum \limits_{k=1}^{N_z} (1-\phi _f(\bm x_{ijk},t^n)),
\tag{C.4}
\label{eqn:b1}
\end{equation}

\subsection*{C.2. Averaging for particle variables}
Particle variables are averaged over all particles within each slab. A slab is generated by dividing $H$ by a thickness of $\Delta h$. An indicator function $\varphi_{p}^j(y_p,t)$ is defined to distinguish the center height $y_p$ of a particle inside or outside of a slab with index $j$ as follows:
\begin{equation}
\varphi _{p}^j(y_p,t)= \begin{cases} 
1,  & \text{ if } (j-1)\Delta h \leqslant y<j\Delta h, \\
0,  & \text{ otherwise.}
\tag{C.5}
\end{cases}
\end{equation}
Based on the indicator function $\varphi  _{p}^j(y_p,t)$, the particle number in each slab can be calculated as
\begin{equation}
n_p^j=\sum \limits_{n=1}^{N_t}\sum \limits_{l=1}^{N_p}\varphi  _{p}^j(y_p^l,t^n),
\tag{C.6}
\end{equation}
where $n_p^j$ is the total particle number in the $j$ slab over $N_t$ time steps. Therefore, the averaged particle variable $\xi_p$ can be defined as
\begin{equation}
\left \langle\xi_p(y^j)\right \rangle=\frac{1}{n_p^j}\sum \limits_{n=1}^{N_t}\sum \limits_{l=1}^{N_p}\varphi  _{p}^j(y_p^l,t^n)\xi_p^l(t^n).
\tag{C.7}
\end{equation}
A slab thickness of $\Delta h=D_p/4$ was selected in the present study.

\bibliographystyle{elsarticle-harv}
\bibliography{ref}

\end{document}